# A COGNITIVE AGENT COMPUTING-BASED MODEL FOR THE PRIMARY SCHOOL STUDENT MIGRATION PROBLEM USING A DESCRIPTIVE AGENT-BASED APPROACH

Virtual University

**By**

*MUHAMMAD TAUSIF*

**MS120400102**

**DEPARTMENT OF COMPUTER SCIENCE**

**FACULTY OF COMPUTER SCIENCE & IFORMATION TECHNOLOGY**

**VIRTUAL UNIVERSITY OF PAKISTAN**

**LAHORE, PAKISTAN**

# A COGNITIVE AGENT COMPUTING-BASED MODEL FOR THE PRIMARY SCHOOL STUDENT MIGRATION PROBLEM USING A DESCRIPTIVE AGENT-BASED APPROACH

By

**MUHAMMAD TAUSIF**

Thesis submitted in partial fulfillment of requirements for the degree of

MASTER OF SCIENCE

IN

COMPUTER SCIENCE

FACULTY OF COMPUTER SCIENCE & IFORMATION TECHNOLOGY

VIRTUAL UNIVERSITY OF PAKISTAN

LAHORE, PAKISTAN

2023

# ***DECLARATION***

I hereby declare that the contents of the thesis **"**A COGNITIVE AGENT COMPUTING-BASED MODEL FOR THE PRIMARY SCHOOL STUDENT MIGRATION PROBLEM USING A DESCRIPTIVE AGENT-BASED APPROACH**"** are the creation of my own research and no part has been copied from any published source (except the references, standard mathematical or geometrical models/equations/formulae/protocols etc.). I further declare that this work has not been submitted for the award of any other diploma/degree. The University may take action if the information provided is found inaccurate at any stage. (In case of default, the scholar will be proceeded against as per HEC plagiarism policy).

**Muhammad Tausif**
**MS120400102**

**To**

**The Controller of Examinations,**
**Virtual University of Pakistan,**
**Lahore.**

We, the supervisory committee, certify that the contents and forms of thesis submitted by Name Muhammad Tausif, VUID MS120400203 have been found satisfactory and recommend that it be processed for evaluation by the External Examiner(s) for the award of degree.

**SUPERVISORY COMMITTEE:**

1. Dr. Muaz Ahmad (Supervisor) Signature

2. Syed Shah Muhammad (Member) Signature

# ***DEDICATION***

I dedicate this thesis to the most important people in my life who have supported me throughout my academic journey: my parents, who have always been my source of motivation and inspiration, and my family members, who have always encouraged me to pursue my dreams and ambitions.

This work is also dedicated to my friends who have been a constant source of encouragement and support during the course of my studies. Your support and guidance has helped me in overcoming the challenges I faced and has been a source of inspiration in completing my thesis.

Most importantly, I dedicate this thesis to my elder brother Abdul Qadeer, my mentors and supervisor, who have shared their knowledge and expertise with me and have helped me in shaping my ideas into a coherent and meaningful form. I am deeply grateful for the invaluable guidance and support that you have provided me with, and I would like to express my sincere appreciation for all that you have done for me.

## ACKNOWLEDGEMENTS

First of all, I would like to express my sincere gratitude to everyone who has supported me in completing thesis. This has been a long and challenging journey, but with the help of many people along the way, I was able to overcome all obstacles and successfully complete this important milestone.

I would like to extend my heartfelt thanks to my thesis supervisors Muaz Ahamad Niazi, and co-supervisor Syed Shah Muhammad Shah for their invaluable guidance, support, and encouragement throughout the entire process. Their expertise, knowledge, and dedication have been truly inspiring, and I am grateful for the opportunity to work with them.

I would also like to acknowledge the support and encouragement of my family and friends, who have always believed in me and encouraged me to pursue my dreams. Their love and support have been a constant source of motivation, and I could not have achieved this without them.

I would like to express my appreciation to my colleagues and students, who have provided me with constructive feedback, helpful insights, and a supportive community throughout my studies. Finally, I would like to thank all of the individuals who have contributed to my research, including those who participated in surveys, provided data, and offered their support and encouragement.

Thank you to everyone who has supported me on this journey. I am truly grateful for your help and encouragement, and I could not have completed this without you.

Sincerely,
Muhammad Tausif

# Abstract

Students' migration from public to private schools, due to lack of school performance of public schools, is one of the major issues faced by the Government of Punjab to provide compulsory and quality education at low cost. Due to complex adaptive nature of educational system, interdependencies with society, constant feedback loops conventional linear regression methods, for evaluation of effective performance, are ineffective or costly to solve the issue. Linear regression techniques present the static view of the system, which are not enough to understand the complex dynamic nature of educational paradigm. We have presented a Cognitive Agent Computing-Based Model for the School Student Migration Problem Using a Descriptive Agent-Based Modeling approach to understand the causes-effects relationship of student migration. We have presented the primary school students' migration model using descriptive modeling approach along with exploratory modeling. Our research, in the context of Software Engineering of Simulation & Modeling, and exploring the Complex Adaptive nature of school system, is two folds. Firstly, the cause-effect relationship of students' migration is being investigated using Cognitive Descriptive Agent-Based Modeling. Secondly, the formalization extent of Cognitive Agent-Based Computing framework is analyzed by performing its comparative analysis with exploratory modeling protocol 'Overview, Design, and Detail'.

# TABLE OF CONTENTS

# LIST OF FIGURES

# LIST OF ABBREVIATIONS

| | |
|---|---|
| CAS | Complex Adaptive System |
| ABM | Agent-Based Modeling |
| CNA | Complex Network Analysis |
| CN | Complex Network |
| PSSMM | Primary School Students' Migration Model |
| CABS | Cognitive Agent-Based Simulation |
| ODD | Overview, Design, Detail |
| ODD+D | Overview, Design, Detail + Decision |
| ODD+2D | Overview, Design, Detail + Decision, Data |
| DREAM | DescRiptivE Agent-based Modeling |
| FABS | Formal Agent-Based Simulation |

# Chapter 1

# Introduction

Students from public schools tend to migrate to private schools because of differences in quality of education between public sector schools and private sector schools (Alderman 1996). This primary school students' migration trend is causing segregation of bright, average and slow learners. It also results in segregation of students based on economic background (Bohlmark & Lindahl, 2007). This segregation strongly affects the scores and achievements of students. For example, poor students, studying in public schools, get relatively low marks as compared to richer students in private schools (Card & Krueger, 1990).

Experimental studies show that quality of education in private schools tends to be better as compared to the quality of education in public schools (Elder & Jepsen, 2014). However, the demand for a quality education requires a higher cost than is affordable for students from average and low-income families (Doherty, 2008).

The unequal educational quality between public and private schools leads to students being segregated based on their socioeconomic status. Wealthier students attend private schools for better education, while poorer students attend public schools with lower quality education. This migration reinforces the idea that a student's education is tied to their family's financial situation and exacerbates the gap between students of different socioeconomic statuses. (Elder & Jepsen, 2014).

In Complex Adaptive System (CAS), complexity and adaptation, due to the feedback loop, make these systems nonlinear (Mitchell, 2009). Due to this nonlinearity, in complex adaptive systems, conventional problem-solving techniques and Newtonian paradigm are unable to solve problems related to these complex systems (Newman, 2003). Student migration is a complex adaptive problem that needs to be solved using the post Newtonian paradigm. We use bottom-up approach to solve it by developing a Cognitive Agent-Based Computing Model of schools.

## 1.1 Students' Migration

Students, from public schools, tend to migrate to private schools because of differences in quality of education and availability of teaching staff. Even though, it is very difficult to measure quality of education that is the major cause of migration. However, measurement of the factors affecting the quality of education is possible. The quality of education in a school depends on student teacher ratio, and socioeconomic status index of the vicinity where the school is situated. Due to favorable student teacher ratio, the quality of education in private schools is better than the quality of education in public schools. However, maintaining a better student teacher ratio requires a higher cost. Due to this reason, students with better socioeconomic status would prefer to study in private school, while poor students because of their weaker social and economic background can only study in public school. The result is the migration of students with better socioeconomic status from public to private schools, which produce a segregation of students in public and private schools based on socioeconomic status.

## 1.2 Students' Migration as Complex Adaptive System

Social systems are purely adaptive, strongly interdependent, highly cohesive, and naturally complex. These are the macroscopic collection with partially connected microscopic structure. Nonetheless, there macroscopic interface is not a purely reflection of their individual entities due to continued adaption. The solution to such system is not possible with linear problem-solving techniques. These systems need a specific dynamic approach to be tackled effectively. Students' migration is also a social system which belongs to complex adaptive system. Hence, the solution to this problem also needs a dynamic problem-solving technique. Data manipulation techniques — spreadsheet, databases, graphs — present only static state of the system. The static state only helps in visualizing the situation at any point, or at the end of the result. In complex adaptive system, the static state does not reflect the underlying effect. In such system, the state is changing constantly due to permanent feedback loops and adaptive behavior based on these feedback loops. Hence, due to non-linearity of these systems, conventional data manipulation techniques are unable to solve the problem. Students' migration, in between schools, is a complex adaptive system, we will use Agent-Based modeling to investigate the causes of students' migration in schools.

## 1.3 Limitation of Conventional Techniques for Solving Migration Problem

In complex adaptive system, complexity arises due to feedback loops and continues adaption behavior. Conventional data manipulation techniques − Newtonian paradigm and linear problem-solving techniques − are unable to handle the complex situation in these systems. Data manipulation techniques can provide us static discipline to visualize the end results. However, we are not able to co-relate the causes to the effects in such a system where feedback loops are changing the behavior of the system continuously. In fact, in such a system, where larger numbers of agents are acting simultaneously, the prediction is not possible with graphs and charts. Data points, graphs, and charts show only a static picture of the domain. On the other hand, the system is changing dynamically. Hence, Newtonian paradigm and linear problem-solving techniques are unable to solve problems related to complex adaptive systems.

## 1.4 Cognitive Descriptive Agent-Based Modeling

The Agent-Based Modeling has been used for several decades for solving complex problems using simulation and modeling. However, informal presentation of models is a major problem with the presentation of the ABM. Mostly, these models are presented in textual forms which makes the difficult to comprehend and replication for verification and validation of models. We use cognitive descriptive agent-based modeling approach for formal presentation of ABM.

The cognitive descriptive agent-based modeling is multilevel semi-formal modeling approach. It is composed of a complex network, pseudo code-based specification, and model itself. The complex network projects an overall structure of the model and it makes comprehension and replication of the model relatively easy task. The pseudo code-based specification is the complete and clear source of model' documentation. Cognitive descriptive agent-based modeling makes the model description complete and clear for comprehension and replication. On the other hand, at exploratory modeling level, the presentation is based on a textual specification which contains ambiguity inherently. Using cognitive descriptive Agent-Based Modeling, we investigate the critical problem of students' migration from public schools to private schools.

## 1.5 Problem Statement

In the last two decades, it is observed that students from public school migrated to private school. Empirical data are evident of this fact for Pakistan, and specifically in Punjab province. Numbers of students in private schools are increasing day by day. Increase in number of private schools is also observed. This migration of students is alarming for government. Even though, several steps to retain the students in public schools have been taken place, however, the retention rate in public schools is not satisfactory. Due to complexity and adaptation in the school system, the problem cannot be easily traced out with conventional problem-solving techniques.

## 1.6 Main Contribution

Our principal contribution to the software engineering of simulation and modeling is to provide dual presentation – ODD and DREAM – of a model of complex adaptive system from domain of education, and to perform simulation to replicate the complex phenomenon of migration and segregation of primary school students. Specifically, we prove via semi-formal specification and complex network analysis of a complex adaptive system that formal description is an effective, and efficient way of modeling and simulation.

The sub goals of this thesis are:

- Developing a simulation model to investigate the relationship between causes of migration and segregation of primary school students in between schools because of heterogeneity in quality of education, teacher induction policy, and socioeconomic status of students.
- Replicating the complex adaptive system from educational domain that can assist the policy makers in decision making regarding improvement in quality of education and reduction in socioeconomic based segregation in schools.
- Comparing the effectiveness of descriptive cognitive agent-based computing framework with exploratory agent-based modelling framework for clear, complete, and concise presentation of the agent-based models.

## 1.7 Thesis Organization

The remainder of the thesis is organized as follows

- Chapter 2: **Background and Literature Review**. This chapter focuses on defining the basic concepts that are base for the rest of the thesis. It also contains the detailed literature review of the students' migration, quality of education, complex adaptive system, agent-based modeling, complex networks, and the modeling framework for complex adaptive systems.

- Chapter 3: **Methodology and Model Design**. This chapter is devoted to Methodology and Model Design. We present the detailed description of how we would start by developing the model, and presentation of the model. Basically, this chapter is focused on a mathematical model of the system, descriptive cognitive and exploratory descriptions of the model. First, formal equations of the model are precisely presented. Secondly, descriptive modeling, DREAM of PSSMM, is presented, and thirdly, in this chapter exploratory modeling, ODD of PSSMM, is presented.

- Chapter 5: **Results and Discussion**. This chapter contains the comparison of ODD protocol and DREAM; it also includes the results of the simulation, discussion on these results, and conclusion.

- Chapter 6: **Summary**. This is the last chapter; it includes the summary of the thesis. It also contains the direction to the future in the context of work presented in this thesis.

# Chapter 2

# Background and Literature Review

Students tend to migrate from the public-sector schools to the private sector schools since students with better socioeconomic status have a greater interest to study in private schools where quality of education is comparatively batter (Alderman 1996). This migration is causing segregation of bright, average and slow learners. It also results in separation of students based on economic background (Shabbir, 2017). Due to complex adaptive nature, and heterogenous inter and intra dependencies of education system, conventional methods are unable to solve the prescribed problem (Levin, 2017). In this chapter, we elaborate complex adaptive systems, modeling frameworks, students' migration and thus make the foundation to solve the complex adaptive problem of primary school students' migration.

## 2.1 Complex Systems Background

Complexity science is the study of complex systems, which are characterized by a large number of interacting components, non-linear relationships, and a high degree of uncertainty. The study of complexity has its roots in physics, mathematics, and computer science, but has since evolved into an interdisciplinary field that encompasses fields as diverse as sociology, economics, biology, and ecology. Complex systems exhibit a wide range of fascinating and sometimes unexpected behaviors, such as self-organization, emergent behavior, and chaotic dynamics. These behaviors arise from the interactions between the components of the system and the non-linear relationships that exist between them. Understanding these behaviors and the underlying mechanisms that drive them is one of the central goals of complexity science. Examples of complex systems abound in the natural world, including ecosystems, weather patterns, and the human brain. In recent years, the study of complex systems has also been applied to a wide range of social and economic systems, including financial markets, transportation networks, and political systems (Holland, 2006).

### 2.1.1 Complex Adaptive System

Complex Adaptive System (CAS) is a special case of the complex systems. A complex system that arises due to the nonlinear interaction of smaller components can be defined as CAS. Holland notes "examining the literature, it appears that CAS may be more of a notion rather than a formal classification, even though these complex systems differ in detail, the question of coherence under change is the central enigma for each." (Holland, 2006). In CAS smalls and numerous give raise the big picture. These small, numerous, and simple entities' interactions emerge the complex behavior if seen from the macro perspective.

**Dictionary Definition of Complex Adaptive System**

The term "Complex Adaptive System" consists of three words, "Complex", "Adaptive", and "System". Let's define, using "scholarpedia", each of these words in reverse order to understand the whole term "Complex Adaptive System".

***System:*** Set of components having a specific relation with each other at a given instance of time with in defined boundaries is called system.

***Adaptive:*** Ability of components in a system to change its state to maintain the stability and existence in a response to some actions by other components or environment.

***Complex:*** A system whose components are coupled to each other such that sum of actions of each component is not equal to the whole action of the system. Such a system must be irreducible to smaller components and its components are strongly dependent on each other.

Hence "Complex Adaptive System" can be defined as

"A system composed of large set of strongly interconnected, and irreducible components that are dependent on each other such that they cannot exist independently or work independently and change their state upon a response from other components or external entity from environment to maintain their stability and existence". Life is a simple example of CAS. It has a dual nature. It is fragile if seen from the top down, but also very complex if observed from the macro perspective (Niazi M. A., 2011).

### 2.1.2 Complex Adaptive System Modeling

Modelling is an abstraction of a system. Epstein defines a model for development of simplified representations of something. There are two types of Modelling, first is the "Implicit modelling" and the second is the "Explicit modelling". In implicit modelling the assumptions are hidden, internal consistency is untested, logical consequences and relations to data are unknown. Implicit modelling is the basic type of modelling (Epstein J. M., 2008).

**Agent-Based Modelling**

Weimer states "Agent-based models offer some interesting opportunities to explore the notion of communication to organize complex systems. Using such models, we can begin to explore 'strategic' communication in which agents must decide what to say to others and how to react to what others say to them." In the different field of science, the use of the computer for problem-solving has been proved. Computer systems are used to figure out the problem in physics, social sciences, biology, chemistry, and economics. Advances in computing in term of processing power, improved and effective software engineering techniques, faster communication network, and enhanced computer graphics established a foundation for artificial laboratory. A recent development in the problem-solving domain is the induction of ABM. Intrinsically speaking, ABM models the CAS that is difficult to understand with the help of differential mathematics. In CAS where it becomes hard to infer the complex phenomena with differential equations ABM provides an effective approach (Weimer, 2016).

**Complex Networks Analysis (CNA)**

A network consists of a set of items called as nodes or vertices and connection between them referred as edge. Graph theory is the origin the complex network (CN). CN is a special type of network, modeled by graph theory. Specifically, it is somewhat advanced and dense graph, containing more information than traditional theoretical graph in the mathematics literature. There are different many mathematical and computational tools used for developing and analyzing the CN. In CAS, CN are used to analyze different adaptive systems like Social Network Analysis to Biological Networks and in Citation Networks (Newman, 2003).

## 2.2 Problems with School Administration to Handle Migration

Computational models are used increasingly to assist in developing, implementing and evaluating public policy. Policy models can have an important place in the policy process because they could allow policy makers to experiment in a virtual world, and have many advantages compared with randomized control trials and policy pilots. The main benefit of designing and using a model is that it provides an understanding of the policy domain, rather than the numbers it generates. Models are designed at an appropriate level of abstraction so that appropriate data for calibration and validation may be available. Policy modelling will continue to grow due to its importance as a component of public policy making processes. However, if its' potential is to be fully realized, there will be a need to have a melding culture of computational modelling and policy making (Gilbert, 2018).

The typical software developed by people only handles record-based Management Information System (MIS). Such type of MIS is not capable of modelling/simulation aspects to analyze the complex behavior. Traditional software is usually designed for data insertion, manipulation, and retrieval which are essentially only record-based systems as a replacement of manual file-based work (Levin, 2017).

Student Migration is a Complex Adaptive System (CAS). Student in a school, adapt migration behavior according to the school performance in term of educational quality. To get better educational quality, student tends to study in a better school. This simple pattern of school selection choice emerges the complex migration behavior of students from school to school. Due to the complexity and adaptation, it is very hard to correlate the causes of migration to the overall emergent behavior of migrating and vice versa (Burns, 2011).

There is a disconnect between the software developed for schools and what is typically needed by the school management. While the management needs tools to help them, understand the dynamics of complex problems such as the migration of students by being able to model problems for their day-to-day affairs. In other words, management tries to understand the cause-effect relationship – mapping of the causes to the effects that result in these effects – for better performance and output of the school (Beckner, 2009).

## 2.3 Literature Review

### 2.3.1 Complex Adaptive Systems' Modeling framework

Even though the Overview, Design concepts and Details protocol is widely used for describing Individual- and Agent-Based Models documentation for providing a consistent, logical and readable structure and dynamics. There are still several limitations to ODD that hindered its widespread adoption. For example: length, limitation on documenting complex models, and guidance on how to use ODD etc. In the second update, steps for complex model structure, guidance, rationale, and evaluation are documented (Grimm V. R., 2020).

Turning verbal theories into formal models is an essential need of a science. Smaldino elaborated taxonomies of models and provided ten lessons for translating a verbal theory into a formal model. He also discussed the specific challenges involved in collaborations between modelers and non-modelers (Smaldino, 2020).

Laatabi (2018) proposed 'ODD+2D', which is an extension to 'ODD+D', which is an extension to ODD – Overview, Design, and Detail. It uses DAMap – Data to Agenet Map, diagram which is inspired by UML. It intends to describe role of data inside ABM. It improves the ODD+D by integrating data. It is generic, structured and detailed as ODD, and it gives new ways to understand and consider data for integration into agent models. It reuses its ancestor's architecture and adds four new blocks inside Input Data part for data integration (Laatabi, 2018).

Currently, there are three types of model descriptions used in modeling community: natural languages, formal language, and graphics; which are used for either of eight purposes: communication of the model, in-depth comprehension of the model, model assessment, model development, model replication, model comparison, theory bonding, and code generation. However, no single model description type alone can fulfil all purposes simultaneously. Hence, a minimum standard consisting of a structured natural language description plus the provision of source code is needed. Such description frame is particularly important for academic purposes, favoring in-depth model comprehension and model assessment (Müller B. B., 2014).

Representing human decisions is of fundamental importance in agent-based models. However, the rationale for choosing a human decision model is often not sufficiently empirically or theoretically substantiated in the model documentation. Furthermore, it is difficult to compare models because the model descriptions are often incomplete, not transparent and difficult to understand. An extension 'ODD+D' to the 'ODD' protocol establishes a standard for describing ABMs that includes human decision-making (Müller, et al., 2013).

The Cognitive Agent-Based Computing: A Unified framework for modeling, provides a common platform for researchers from multi-domain to model CASs under one umbrella. It is divided horizontally into four levels of abstraction, according to the need of researches. These levels are: i) Complex Network modelling level, ii) Exploratory agent-based modelling level, iii) Descriptive ABM level, iv) Virtual Overlay Multi-Agent System (Niazi M. A., 2011).

The major objectives of ODD were to make model descriptions complete and understandable. After few years, the existing uses of the ODD protocol evaluated and identified, and it is observed that parts of ODD needing clarification and improvement. The definition of ODD is revised to clarify aspects of the original version and thereby facilitate future standardization of ABM descriptions, along with several other parts are reshuffled, modified, and improved. The ODD improves the rigorous formulation of models and helps make the theoretical foundations of large models more visible. Although the protocol was designed for ABMs, it can help with documenting any large, complex model, alleviating some general objections against such models (Grimm, et al., 2010).

Grime along with 28 developers presented the ODD protocol for text-based documentation of Agent-Based Model (ABM). ODD was presented as a standard protocol for documentation. The primary objectives of ODD are to make model descriptions clear, concise, more understandable, and complete, thereby making ABMs less subject to criticism for being irreproducible. The ODD protocol has three blocks Overview, Design concepts, Details. These three blocks are subdivided into seven subsections. Overview and Detail blocks have six sections as purpose, state variable and scale, Process overview and scheduling, initialization, input, and sub models. Each section holds a specific meaning and context to maintain the conceptual flow of the reader (Grimm, et al., 2006).

### 2.3.2 Agent-Based Modeling in Education

An agent-based model is used to explorer the factors influencing educational choices at the tertiary level and the impact of three alternative scenarios aimed to increase the level of enrollment in higher education. It takes into account both economic and social motivations, approaching the economics of education with an innovative methodology and a multidimensional perspective. The ABM approach in fact allows considering social influence deriving from interaction with a set of neighbor agents as it would happen in real life, and to account for multiple dimensions of one phenomenon, making it a useful tool to reach proximity with reality (Leoni, 2022).

The critical issues for building (Agent-Based Modeling) ABM include integrating human behavior and modeling human behavior in AMB, transparency and useability of ABM, verification and validation models, Big data and high performance ABM, and presenting Spatially explicit ABMs. We can solve these issues with help of advancement of Artificial Intelligence, big data, qualitative data, data science, and from artificial neurons (An, 2021).

With the help of simulation, using ABM, Reardon (2017) explored three types of important patterns after applying SES based affirmative plan to maintain a reasonable diversity in colleges. First, race-based affirmative action plan is much effective than SES-based affirmative action plan or race recruiting polices in producing racial diversity. Second, diversity decreases in colleges not having affirmative plan due to the colleges having it. Third, academically overmatched minority students cause an overall decrease in the academic achievement in colleges (Reardon S. F., 2017).

Reardon et al explored how dynamic processes related to socioeconomic inequality operate to sort students into, and create stratification among, colleges. They used simulation of students' application decisions and focus on horizontal segregation – with in groups – of students rather than vertical stratification. Firstly, they described the mechanism behind the stratification of students with in groups. Secondly, they ran a series of scenarios that investigate the relative influence of each of a set of mechanisms that have been hypothesized to link resources and college destinations. Using simulation model, a tractable model of the application, admission, and enrollment process was implemented that can be extended to simulate the study aspect beyond sorting of college students (Reardon 2016).

An agent-based model of the transition to school choice was used by Maroulis (2016) as platform for examining the school choice treatment effects' sensitivity from lottery-based studies. Result showed that lower treatment effects were observed in districts with higher participation rates, even when there were no differences in student preferences between districts and distributions of school quality. This is because capacity constraints increasingly limited the number of students who can attend the highest quality schools, causing the magnitude of the treatment effect to fall (Maroulis S. , 2016).

Models have been used to judge the comparative performance of schools, known as 'differential school effectiveness'. An Agent-Based Model (ABM) as an exploratory model, along with Multilevel Model (MLM) as variable based models are used to compare their performance in term of productivity and explanatory power. Although, MLM provides more fit than ABM, but later also provide casual mechanism for explanation that are absent in MLM (Salgado, 2014).

An Agent-Based Model is developed by Millington to investigate the implication of location-based school-allocation in London, UK. Parents' aspiration to send their pupil to best performing school is the main attribute of parent agent, which also constraint the effect of location and movement in the model. Model shows how simple micro level rules emerges the macro level pattern which are close and consistent replica of real-world pattern (Millington 2014).

To address school differential effectiveness, two models, MLM and ABM, are presented and compared. Hierarchical nature of school and educational process is considered in MLM, while in ABM, micro level pupil attainments are formalized to generate macro level phenomenon. MLM is found more accurate than ABM in prediction. MLM is data driven model, and fast as compared to ABM. On the other hand, ABM is also data driven and theory based also which allows to formalize and falsify in silico plausible mechanism (Marchione, 2011).

Tools that can characterize the individual entities like school and student, quantity the relationship between these individuals helps in investigating, how micro level actions results and aggregates into macro level emergent behaviors in schools. The Complex System approach would certainly help to integrate insights from different types of research and provide infrastructure for better policy making. It is very important to consider not only the working of a system, but also on how and why the system works (Maroulis S. G., 2010).

### 2.3.3 Primary School Students Migration and Segregation

The predictive role of socioeconomic status on academic achievement in three transition systems in Turkey is investigated. The investigation results are, with regard to socioeconomic status, students from private middle schools are the most advantaged, while students from regional boarding middle schools are the most disadvantaged group in all three transition systems (Suna, 2020).

This study was aimed to identify and describe possible relationships between trends in the inequality of educational opportunities related to the differing socioeconomic status (SES) of the students. Thirteen educational systems were assessed as having sufficiently comparable data. The research revealed some tentative patterns that may be worthy of deeper investigation. In the second decade (2003–2015), larger reductions in the achievement gap between low and high-SES students tended to accompany steeper increases in the corresponding education systems (Broer, 2019).

Equal distribution of 26070, 8 years old, students form public and private schools are surveyed to assess the proficiency of students in math, English, and reading. Binary regression logistic analysis method reveals that the main difference in the test outcomes is due to the difference in intake of students from different families having different socioeconomic status. Most of the out-school students are girls and low-income families' children (Siddiqui, 2017).

Ordinarily least regression methods and logistic regression methods are used to study the household school choice, and relationship between students' test score and SES, mother educational class test, family size, study hours, and teacher's qualification. Ordinary least regression method reveals positive relationship between test score, and SES, mother's education, family size, and study hours (Shabbir, 2017).

Educational quality in low-income countries, stakeholder's perspective and school benefits are researched by Jarrad (2016), in rural areas of Sindh, Pakistan. Unsatisfactory educational achievements were indicated in under developed countries, study explored. Furthermore, the employment of local teachers, teachers' training, and development of private school are suggested to achieve the goal of education for all (Jerrard, 2016).

Comparative analysis between NGO operated schools, public, and private schools showed that NGO operated schools are leading in term of quality of education, followed by private school, but public schools are lacking basic indicators of quality of education. The main indicators included are quality, access, and affordability of education. Education accessibility, and affordability based on parents' income may also cause disparity in the society (Saqib, 2015).

A co-relational study is conducted in 25 randomly selected school from district Gujrat, Punjab. More than average class size, as recommend by research, is found in majority of school for grade five students. Academic achievements were found below the average. There is no specific difference for effect of class size and student achievement for gender. Negative and strong correlation is found between class size and the academic achievement of students (Laraib, Saira, & Mobeen, 2015).

Private schools are providing better educational quality than public school. It is because students attending private sector schools, like catholic school require extensive involvement of parents, which result in the differentiation of educational outcomes of students between public and private sector school. The grades of students are not a complete measure of performance of a school as socioeconomic status had a greater impact on over all students' outcomes (Elder & Jepsen, 2014).

Due to poor performance of public schools in Pakistan, the number of private schools has increased in past few decades. This also caused the increase in the enrolment rate in private sector school and decline in enrollment in public sector school. At time of this research more than 30% students are studying in private school with a favorable increase in enrollment (Ravish & Gordon, 2014).

Teacher migration is defined as transferring of a teacher in the educational sector for a better educational environment for teaching, and home to school distance while teacher attrition is defined as teacher removal from the educational sector. Teacher migration is one of the causes of student migration. Well reputed, effective schools with the relatively better performance have a greater opportunity to retain a teacher (Loeb et al., 2012).

It was noted that teachers tended to migrate from school to school and even tended to leave the teaching profession. Causes of migration and attrition of teachers from the teaching profession included social, economic, and spatial position of the teacher with respect to school. Most of the teacher who had left the teaching profession started working in management staff or got job in universities or colleges (Ingersoll and May, 2012).

There are several changes occurred in past two decades in the educational system of Pakistan. Overall net primary enrolment rate, for five to nine years old children, has been increased from 42% to 57% from 1999 to 2009. So, about fifteen percent changes occurred over a decade. However, primary schools' enrolment decreased from 75% to 70% from 2001 to 2009. This modifications in the sector of education of Pakistan generate various challenges, and opportunities for improvement in policy (Amjad, 2012).

In Pakistan, public-sector school system is providing largest service. Despite lack of quality of education, it covers primary to higher secondary education in schools free of cost. The reason for low quality of education in public sector is due to lack of physical infrastructure, absence and shortage of teachers, old teaching methods, and non-availability of learning materials. Due to these unhealthy environments with respect to quality of education in public sector school's parents have started to shift their children from public to private schools (Barber, 2010).

The study is conducted in Pakistan for period of 1999-2008. Number of private schools has increased by 69% as compared to 8% increase of public schools. 71% primary students, in 2008, were enrolled in public primary schools, but the overall enrolment decreased by 2.6%. On the other hand, with 11% increase in enrollment, the private primary schools had a low share, 29%, in total enrolment. Private schools are widespread in rural and urban areas. Due to availability of low paid teachers, private schools with lower fee cater lower income groups also. Infect, private school pay very low average salary 1,084 as compared to public school 5,620 per month. Even though, the number of teachers in private primary schools increased from 75,924 to 88,195. (I-SAPS, 2010)

# Chapter 3

# Methodology and Model Design

We present a Cognitive Agent Computing-Based Model of student migration using Descriptive Agent-Based Modeling approach along with the textual description of primary school student migration model. The developed model simulates the CAS of school in an agent-based computational environment to investigate students' migration. The model presents the visual description of the structure of ABM and the emergent behavior of the CAS. This descriptive approach helps the administration of a school to understand the causes of the emergent behavior of migrating students. In short, cognitive agent-based computing using a descriptive approach provides the cause-effect relationship of students' migration CAS. The Netlogo is used as a simulation modeling tool for implementation of ABM and the R is used as a data processing tool to manipulate the data exported from experiments carried upon using behavior space in Netlogo. The R presents the results in the form of expressive graphs to verify the output of the model.

A visual description of the overall research methodology is given in **Error! Reference source not found.** It contains: exploratory modeling, descriptive modeling, and the comparative analysis of both techniques. We will begin with a concrete question of student migration with a need to explore the cause-effect relationship of students' migration from school to school. The research methodology starts with the exploratory modeling of ABM, going through descriptive modeling of ABM and ends with the comparative analysis of both approaches. After comparative analysis, verification of the hypothesis using graphs and visual output of the model is presented.

This chapter also provides a brief mathematical foundation, descriptive and exploratory presentation of the primary school student migration model. The general concept of migration, and segregation derived from the literature defined. After that agent-based model of PSSMM, along with complex network analysis and pseudo code based would be presented. At last, we present the ODD based specification of the model.

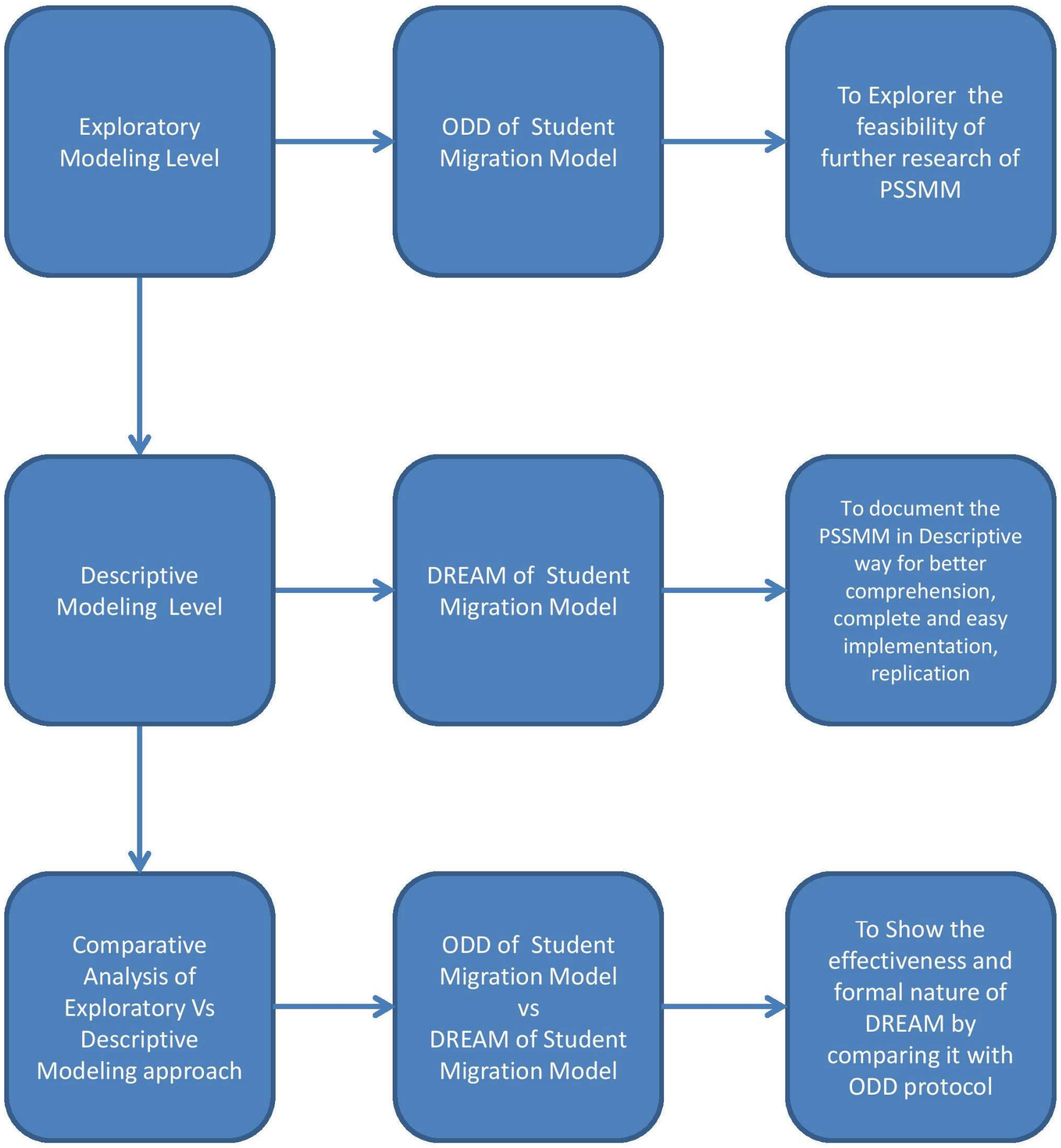

**Figure 3.1: A visual description of the overall research methodology**

## 3.1 Creating ABM: Primary School Student Migration Model (PSSMM)

Primary School Student Migration Model consists of two agents: students and schools. These agents would be simulated on a two-dimensional grid, replicating the real-world scenario of educational institutions. The students get enrolled into one of the schools where they can afford to get the education. The schools' agents would enroll students and would induct new teachers to maintain a specific class size. The rational behavior of schools and students would lead to emergent behaviors of students' migration from lower quality schools to better quality schools.

## 3.2 Textual Based Specification of ABM: ODD Description of PSSMM

We will develop ODD of Students' Migration Simulation Model to understand the feasibility of the research for the CAS of students' migration. The intent of the research is two folds. Firstly, it is to demonstrate that the ABM can be used to investigate the student migration causes, effectively and efficiently. Secondly, the exploratory agent based modeling, using the ODD protocol, helps to determine the extent of formalization of the DREAM, by presenting a comparative analysis of between both approaches. Students' Migration Simulation Model implementation based on ODD is the exploratory agent-based modeling level. As discussed above, ODD is, a text based, weak formalized approach for presenting ABM of CAS. It cannot encompass the complete picture of a CAS under study correctly and completely. Due to this reason, the descriptive approach (DREAM) is used to describe the model in a complete and clear way.

## 3.3 DescRiptivE Agent-Based Modelling (DREAM) Specification of PSSMM

DREAM is a semi-formal method for describing ABM. It presents the model using complex network, and pseudo-code based specification. The complex network of the model assists the readers for easy comprehension of the model. It also helps the reader for better comparison of the model with other models without digging down the code. The pseudo code based specification helps the modeler in clear and complete presentation of the model. The clear and complete presentation of the model makes it possible for the modeler to replicate the model. The replication of the simulation model is the basic need in the scientific community for verification and validation of the model. In Figure 3.2 the sequential representation of the DREAM shows the flow of research methodology in depth.

After the overview, going through complex network construction and analysis, and the model ends with pseudo-code based specification of each entity and processes of the model. More specifically, we will use DREAM to make a cognitive descriptive simulation model of Students' Migration. The results and discussion of the simulation are presented in a separate chapter. We will use NetLogo as the IDE for ABM construction. The data would be extracted from the behavior space of NetLogo. The R would be used for drawing the graphs and charts of the data extracted from NetLogo.

Using the ABM of Students' Migration Simulation Model, and by performing different simulation experiments, we will discover the emergent behavior of the CAS. Simulation experiment with different inputs probably emerges different emergence behaviors. These emergence behaviors help in understanding the cause-effect relationship of the migration of students. After that, the emergent behaviors will be validated by the empirical data extracted from the literature. The validation of the model indicates that the causes are correctly mapped to the effect in the model. This mapping of emergent behavior to the base causes with the help of simulation provides a cause-effect relationship. Hence, major causes can be determined with simulation and modeling, which are hard to correlate in CAS with traditional approaches.

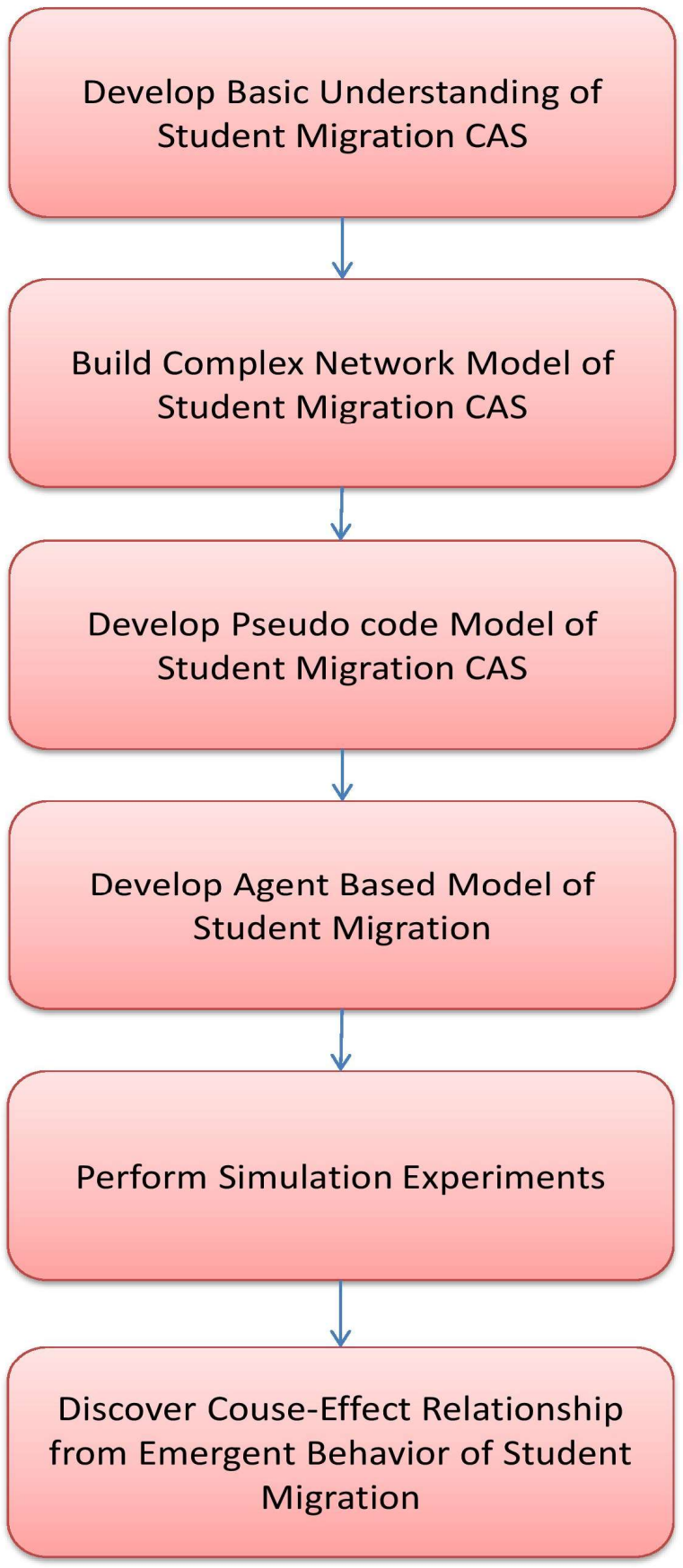

**Figure 3.2: Overall layout of DREAM of PSSMM**

## 3.4 Comparison of ODD and DREAM

In the conclusion, we will present the comparative analysis of exploratory and descriptive modeling approaches. This comparison is to show the effectiveness of each approach. In this regard, comparative analysis, documentation, implementation, execution, and emergence behavior of DREAM and ODD of Students' Migration Simulation Model will be carried out. That analysis demonstrates the formalization extent of each approach. Figure 3.3 show the pictorial representation of the DREAM

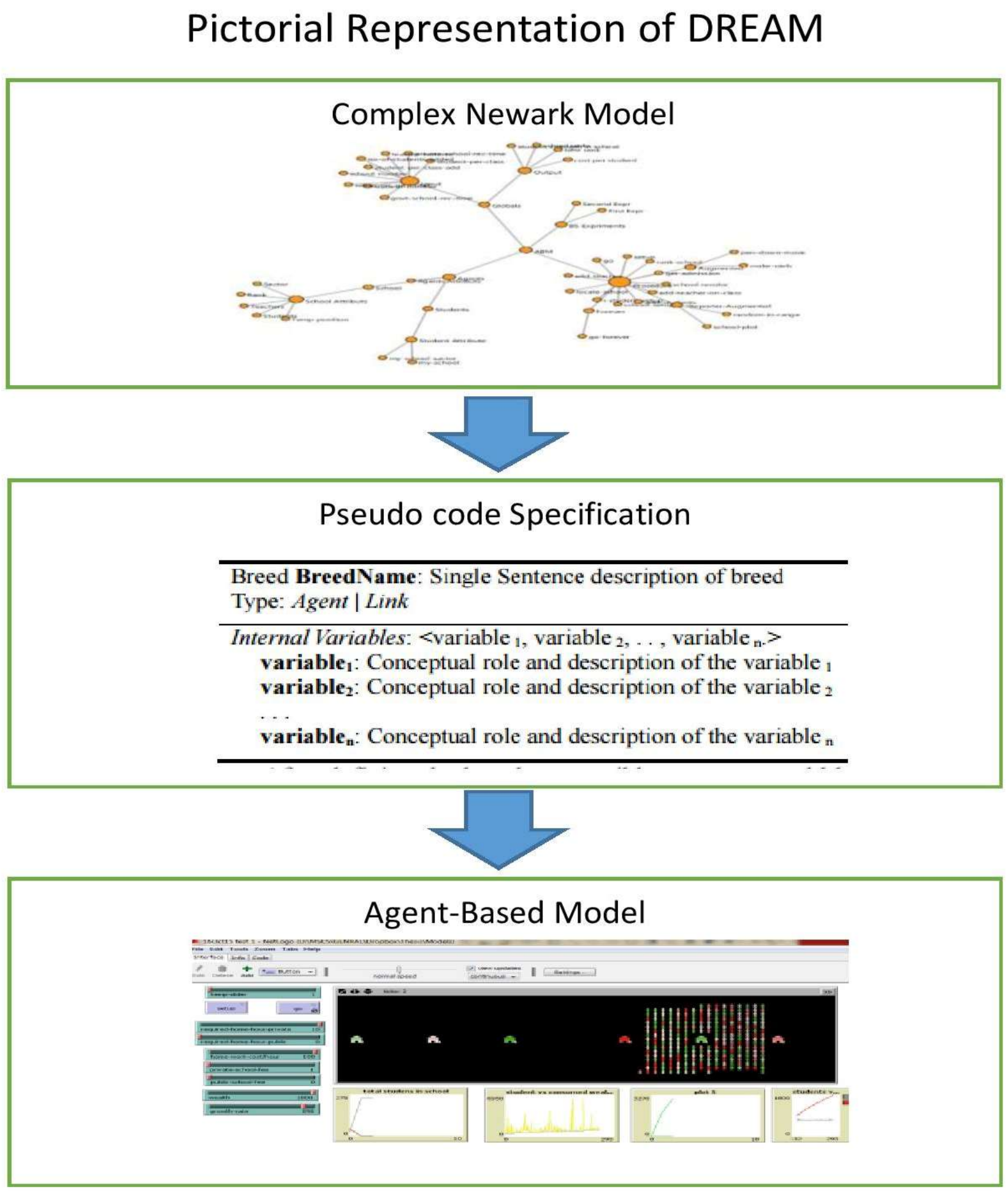

**Figure 3.3: DREAM pictorial view**

## 3.5 Verification of PSSMM: Result and Discussion

ABM can be used as an effective Artificial Intelligence Laboratory to understand the complex adaptive behavior of student migration. Simulation of student migration CAS provides an opportunity to investigate the complex adaptive behavior which is not possible to discover with differential mathematics.

Simulation output as emergent behavior of Students' Migration Simulation Model is validated against empirical data extracted from the literature and surveys. It proves that the students' migration CAS is correctly implemented in the form of ABM. Different emergent behaviors relate to different inputs. In simulation, this correlation represents a relationship between a cause – which are the input – and its effects – which are the emergent behavior. In the real-world case study, it is hard to make clear that which a cause results in the specific effect in a CAS. Hence, verification and validation of the Students' Migration Simulation Model against experimental data and extraction of correct cause-effect relationship are the criteria of success.

## 3.6 Students' Migration and Segregation in Schools

In this section, we would briefly describe what migration and segregation are, and how these are defined in the literature. Later, in this chapter, we would technically describe the mathematical model of the migration and segregation which would be used to develop our model for simulation.

### 3.6.1 Students' Migration in Schools

Migration can be defined a permanent movement – or at least more than six months staying at the destination – of humans from one country, or locality to another in search of a better environment for survival. The migration presented by [1] deals with human as migrant moving from one place to another. (Migration, 0000). The total number of enrollments of new students in a school at any given time instance is called ***in-migration*** of that school at that time. The total number of students gets unrolled (left the school) from a school and gets enrolled in any other school that is supposed to be the best school for those students is defined as ***out-migration***. This migration is out-migration of the school from where student migrated and in-migration for another school to where student migrated.

The net students' migration in primary schools, now, can be defined as the movement of students in between schools to get better educational quality. In this movement, students get unrolled from one school with lower educational quality and get enrolled to school with better educational quality. For any school, the student migration is the difference of in-migration and out-migration. This number may be positive or negative depending upon the value of in-migration and out-migration. Net student migration would be positive for any school if students get enrolled in that school are greater than the students get unrolled. In reverse, the student migration would be negative if the students who left the school are greater than the incoming students.

### 3.6.2 Students' Segregation in School

According to (Echenique & Fryer Jr, 2007) "At an abstract level, segregation is the degree to which two or more groups are separated from each other. However, practical definitions can be quite distinct from one another, conceptually and empirically." Massey and Denton (Massey & Denton, 1988) group indices into five classes: i) evenness, ii) exposure, iii) concentration, iv) centralization, and v) clustering. Out of five indices, first two indices "Evenness", and "Exposure" are used commonly in the majority, measuring the segregation. Evenness means differential distribution of the two groups while exposure measures the approximate the amount of interaction between group members (Echenique & Fryer Jr, 2007). Specifically, segregation means the act of separation of people based on gender or race into separate groups based on similarity with dominance of someone's on others.

For SES segregation in schools, segregation means separation of students in different school, specifically into public and private school, based on SES. In the current context, separation does not mean the complete existence of each student group in a separate school, rather separation is the dense and low concentration of students' groups in different schools. For example, if 20% of students belong from families having 80% income, then each school in the respective territory should have 20% enrollment of such student. In that case, segregation is considered as zero. However, if, in this territory some schools have more than 20% of such students belonging to families having 80% income, and some other schools have less than 20% concentration of such students then the schools have segregated number of students with respect to the income of their families.

## 3.7 School Performance Factors

School's performance depends on different factors. We denote these factors as School Performance Factors (SPF). These factors include school fee, home to school distance, class strength, student's socioeconomic status etc. (Livaditis, et al., 2003) (Furstenberg Jr & Hughes, 1995) (Tahir & Naqvi, 2006) (Farooq, Chaudhry, Shafiq, & Berhanu, 2011) (Hanushek, 1997) (Stiefel, Iatarola, Fruchter, & Berne, 1998) (Livaditis, et al., 2003)(Alderman, elt, 2001) (Alderman, elt, 1996) . Due to complex systems, the school's performance factors would reside inside and outside of school.

One of the most known measures of the school's performance is the quality of education delivered by any school in term of a student's achievement. However, the quality of education is not a well-defined term and its meaning varies from culture to culture (Michael, 1998). In the current context, quality of education is the achievement of students in the respective medium of education. Factors that enhance the students' achievement in school increase the quality of education of that school.

### 3.7.1 General Mathematical Model of School's Performance Dependencies

Now, we would define the index of quality of education of a school. Let $\mathcal{H}$ represents the SPF. It is directly proportional to school fee represented by F, and study hours at home represented by H. Contrarily, it is inversely proportional to the distance from home to school as D, and class strength as C. We introduce φ to denote the other unknown factors in a school complex adaptive system that are not considered or not investigated due to domain limitation.

$$\mathcal{H} = \alpha F + \beta H + \frac{\gamma}{D} + \frac{\delta}{C} + \varphi \quad (3.1)$$

Where $\alpha, \beta, \gamma, \delta$ are proportionality constants, home working hours, home to school distance, class strength respectively. The constant φ is SPF constant for indicators that are not included in the model. The SPF index notation is partially derived from student's performance as presented by Tahir (Tahir & Naqvi, 2006), and Bohlmark (Bohlmark & Lindahl, 2007).

### Socioeconomic Status (SES) Relation with School Fee and Study Hours

The SES is most often representation of parental education, income, occupation, and other facilities, used due to inheritance or demography by individuals separately or collectively. Students with better SES with respect to their peers in school relatively show better performance (Farooq et.al 2011). To understand the relation of SES and SPF, we would study the correlation of school fee F and home working hours H with SES. All other factors are denoted by $\Phi_f$ and $\Phi_h$ constants in Equation ( 3.1 ) for fee and home working hours respectively.

$$\mathcal{H}_f = \alpha F + \Phi_f \qquad (3.2)$$

$$\mathcal{H}_h = \beta H + \Phi_h \qquad (3.3)$$

Where $\mathcal{H}_f$ represents the SPF for considering school fee, where $\mathcal{H}_h$ represents the SPF for considering school study hours.

### Students Migration and Class size

A prominent fact of student's migration from one school to another is class size. Students tend to migrate to schools having relatively smaller classes (Stiefel, Iatarola, Fruchter, & Berne, 1998). Migration of students is inversely proportional to the class size of the school from where student migrated. Let $\mathcal{M}$ denotes in-migration of students from a school whose class size is C

$$\mathcal{M} = \lambda \times \frac{1}{C}$$

Where $\lambda$ is proportionality constant.

To investigate the effect of class size on migration of students from school to school all other indicators of school performance other than class size are kept constant as $\Phi_c$. Equation ( 3.1 ) becomes

$$\mathcal{H}_c = \frac{\delta}{C} + \Phi_c \qquad (3.4)$$

### 3.7.2 Students Segregation Index ($I_s$)

Research on segregation and its measurement has been conducted by various researchers from different prospective (Frankel & Volij 2011) (Frankel, etl, 2007) (James & Taeuber, 1982). Racial segregation, socioeconomic segregation, and white and black school segregation are most commonly types of segregation that remained under debate for decades. As cited by (Frankel, etl, 2007) there are more than 20 segregation indices used by different researchers to measure the segregation of concern (James & Taeuber, 1982) (Massey & Denton, 1988). "Atkinson indices", "Mutual Information index" proposed by Theil, "index of isolation", and "Dissimilarity index" are most commonly used segregation indices used to measure school segregation.

**Mathematical Model of Students' Segregation Due to SES**

The segregation index is inspired from the index of dissimilarity, a commonly used index for segregation measurement (Jahn, Schmid, & Schrag, 1947). The notations in the mathematical model are derived from model "Measuring School Segregation" proposed (Frankel, etl, 2007).

We define schools, students, and segregation of the students with respect to SES as follows.

- $N(X)$ represents the set of schools, and which has two subsets $N(X) = Public \cup Private$
  - $N_r(X) = Private$ is the sequence the private schools own and governs by private authorities. In this model, these schools are given id as *even* numbers.

    $$Private = \langle N_2, N_4, N_6, ...\rangle$$

  - $N_b(X) = Public$ is the sequence of public schools governs or funded by government. Further public schools are given odd ID.

    $$Public = \langle N_1, N_3, N_5, ...\rangle$$

- $G(X)$ represents the set of student groups with respect to SES. We consider only two groups in this scope, poor and rich students as $G(X) = Poor \cup Rich$
  - $G_p(X) = Poor$ is the set of poor students
  - $G_r(X) = Rich$ is the set of rich students
- $T_g^n$ is a non-negative number that represents the count of the total members of group $g$ who attend a school $n$, for each $g \in G(X)$, and each school $n \in N(X)$ in the sector.

The other useful derived notations are.

| Notation | Description |
|---|---|
| $T_g = \sum_{n \in N} T_g^n$ | Number of students from any *SES* group $g$. |
| $T_g = \sum_{g \in G} T_g^n$ | Total number of students in school $n$ |
| $T = \sum_{g \in G} T_g$ | Total number of students. |
| $P_g = \frac{\boldsymbol{Tg}}{\boldsymbol{T}}$ | Proportion of students of *SES* group $g$ |
| $P^n = \frac{\boldsymbol{T^n}}{\boldsymbol{T}}$ | Proportion of students in school $n$. |
| $P_g^{\ n} = \frac{\boldsymbol{Tg}}{\boldsymbol{T^n}}$ | Proportion of students in school $n$, of group $g$. $X = \frac{T_g^n}{X}$ |

To define the poor and rich student empirically, let $w_i$ be the wealth of any student $i \in g \in G(X)$, then $W_m$ be the pivot partitioning *poor* and *rich* students

$$Wm = \frac{\sum_{i=1}^{m} w_i}{|G(X)|} \qquad (3.5)$$

In set notation *Poor* and *Rich* students are defined as

$$Poor = G_p(X) = \{x \in G(X) | w_x < W_m\}$$

$$Rich = G_r(X) = \{x \in G(X) | w_x \geq W_m\}$$

We specify list $L_g^{\ n}$ as a sequence of sorted school $\langle (L_g^{\ n})_{g \in G} \rangle_{n \in N}$ containing the presence of different groups in the school. For example, it's an instance ⟨ (150, 100), (90, 160) ...⟩ denotes proportion of poor and rich students in any school. The pair (150, 100) contains *150 poor* and *100 rich* students. The $\boldsymbol{L_g^{\ n}}$ is generated from the $G_p(X)$ and $G_r(X)$ in form $\boldsymbol{G_p(X)}$ **x** $\boldsymbol{G_r(X)}$.

$$L_g^{\ n} = \langle (T_p^{\ 1}, T_r^{\ 1}), (T_p^{\ 2}, T_r^{\ 2}), (T_p^{\ 3}, T_r^{\ 3}), ..., (T_p^{\ n}, T_r^{\ n}) \rangle$$

If $T_p^n$, $T_r^n$ represents the number of $Poor$, and $Rich$ students respectively in any school $n$; then, $T^n$ denotes the total number of students in a school

$$T^n = T_p{}^n + T_r{}^n$$

Let $I^n$ represents the Index of segregation for school $n$, which would be either public or private

$$I_p^n = \frac{T_p^n - T_r^n}{T^n}$$

$$I_r^n = \frac{T_r^n - T_p^n}{T^n}$$

Mutual Segregation Index $M^n$ is the absolute value of $I_p{}^n$ or $I_r{}^n$

$$M^n = \left|\frac{T_p^n - T_r^n}{T^n}\right| = \left|\frac{T_r^n - T_p^n}{T^n}\right| \tag{3.6}$$

Now, Let $I_G{}^N$ represents the segregation of $N$ schools in the system

$$\varphi I_G^N = \frac{\sum_{n=1}^{N} M^n}{N} \tag{3.7}$$

Due to the fuzzy distinction between poor and rich students, we use the average wealth of the students in each to determine the segregation index $I_s$. Let $W_g{}^n$ represents the average wealth of students belong to group $g$ in school $n$. Replacing the group strength $T_g{}^n$ of students in a school in Equation 5.9↑ by average wealth $W_g{}^n$

$$I_s = \frac{\sum_{n=1}^{N} \left|\frac{W_p^n - W_r^n}{2}\right|}{N} \tag{3.8}$$

### 3.7.3 Student Migration Index ($M_{s(t)}$)

Migration defined by Merriam-Webster is "to move from one country, place, or locality to another". Mathematical model of Migration corresponds to the human migration form one place to another place. In the case of students' migration in schools, migration can be defined as the movement of students from one school to another to study in school where STR is low. Here, the movement is the process of unrolling of a student from one school and then enrolling in the other school. In other words, students' migration is the combined process of leaving of schools by students and enrollment to another school to get a better and affordable education. In students' migration **Student** is the migrant who migrates from one school to another school. **Schools** are places from where a student migrates and to where a student migrates.

**Mathematical Model of Migration: Migration Index (I%)**

Net migration of students in a school is the total number of increase or decrease of students after one educational year. It can be defined as the ratio of students' retention by a school. Migration index (%) is used to represent the Net Migration. Migration Index can be positive or negative depending upon the in and out migration. Higher value of Migration Index indicates a larger number of students were migrated from school. In opposite case, lesser value of the Migration Index shows that lesser students were migrated from that school to other schools.

Mathematically Net Migration of a school *S*, represented by Migration index (%) $M_{s(t)}$ at any time *t* is defined as

$$M_{s(t)} = \left(\frac{I_{s(t-1)} - O_{s(t)}}{I_{s(t-1)}}\right) \times 100 \qquad (3.9)$$

Where $I_{s(t-1)}$ is the total number of students in school *S*, before the time interval *t*, and $O_{s(t)}$ is the total number of students at time $t + 1$ in school *S*. Migration index '$M_{s(t)}$' is normalized to % to standardize the migration index and to make it more expressive.

## 3.8 Agents and Environment Design

In this section, we present agents and environment design in detail. There are two types of agents in the model: the school, and the student. School (agents) hires the teachers, and provides education related facilities to the students. Students (agents) pay fee, and study in schools to get the education. The quality of education is the measure of school performance. It depends on many factors. In this model, we consider only two performances effecting measures − class size, and study hours at home − for educational quality in the schools. Table 3.3-3.5 contains all variables for agents of schools and students.

The school and its related phenomenon like migration are complex adaptive systems. School inducts teachers to the school following a specific teacher induction policy. A school following dynamic teacher induction policy hire teacher dynamically when the need arises, while a school following static teacher induction policy hires teachers after a long and specific time. Due to the difference in teacher induction policy deployed by different schools, different class size is maintained in the respective schools. A school having a dynamic teacher induction policy can maintain a smaller class as compared to school having static teacher induction policy. Students tend to migrate to schools having a dynamic teacher induction policy because of smaller class sizes maintained by such schools. We would discuss in detail how teacher induction policy affects the class size in a school, which in term is the cause of students' migration.

Table 3.1 describes the Agents in the PSSMM. There are two types of agents: Schools and Students. The School agents are responsible for managing school resources and providing teaching facilities to the students. On the other hand, the Student agents study in schools and pay the required fees. They also have the ability to migrate between schools in search of better education opportunities.

### 3.8.1 Specification of Task Environment

According to Russell "A task environment specification includes the performance measure, the external environment, the actuators, and the sensors. In designing an agent, the first step must always be to specify the task environment as fully as possible." (Russell, Norvig, Canny, Malik,

& Edwards, 2003) Our task environment, migration of students in between schools, include two agents, **schools**, and **students.** The **Performance**, **E**nvironment, **A**ctuators, and **S**ensors **(PEAS)** of our task environment are fully described in Table 3.2, 3.3. For Students, their performance objective is to maximize grades while minimizing educational costs. They operate within an environment consisting of schools and teachers. The students can take actions such as enrolling in schools, studying, migrating to different schools, and getting graded. They gather information about schools' classes through their sensors. On the other hand, Schools aim to maximize their income by minimizing class size. They interact with students and teachers within their environment. The schools utilize actuators to hire teachers, collect fees from students, enroll students, and grade them. Their sensors allow them to monitor the status of classes, teachers, and students.

The Table 3.4 describes variables related to schools and their corresponding values. The "School_ID" variable represents the unique identification number of a school. The "Teachers" variable represents the teachers working in a school and can include information like names, qualifications, or experience. The "Students" variable indicates the number of students enrolled in a school. The "RWH" variable represents the required home working hours for school agents, which can include the number of hours or other relevant information. Lastly, the "Sector" variable denotes the sector or type of school, with "P" representing public and "R" representing private. These variables provide important details about schools, including identification, personnel, enrollment, home working requirements, and sector classification.

Table 3.5 describes the variables related to student agents capture essential aspects of their educational journey and personal circumstances. The "School" variable uniquely identifies the school attended by the student. "Grades" represents the academic performance, "Wealth" denotes the current financial resources of the student, and "Class" signifies the grade level they are studying. Additionally, "Home" provides information about the student's living place. Together, these variables encompass crucial dimensions such as school affiliation, academic achievements, financial status, grade level, and living situation of the student agents.

**Table 3.1: Agents in PSSMM**

***Agents:*** Agents in the model

| Agent | Brief Description |
|---|---|
| School | School agents manage the school resources and provide teaching facilities to students. |
| Student | Student agent studies in school and pay the fee. Student agent migrates in between schools to get better education. |

**Table 3.2: PEAS description of the Agents in PSSMM**

| Agent | Performance | Environment | Actuators | Sensor |
|---|---|---|---|---|
| Students | Maximize the grades, Minimize educational cost. | School, Teachers | Get enroll, Study, migrate, get graded | View schools' class |
| Schools | Maximize income by minimizing class size | Student, Teacher | Hire teacher, get fee, enroll student, grade students | View class, teacher and student |

**Table 3.3: Attribute and Action of Agents**

| | School | Student |
|---|---|---|
| **Variable** | ID, Name, Sector to which school belongs | ID, Name, Grade, School, Parent |
| **Methods** | Hire teacher, enroll student, Grade Student, pay teacher, expel student, expel teacher, get fee | Get enroll in a school, migration from the school, view top school, get graded |

**Table 3.4: Variables of school agents**

| Variable Name | Brief Description | Value |
|---|---|---|
| School_ID | ID of school | $S_{id} \in \mathbb{N}$ |
| Teachers | Teachers in school | $T_i \in \mathbb{W}$ |
| Students | Students in school | $St \in \mathbb{N}$ |
| RWH | Required Home working hours | $R_i \in \mathbb{W}$ |
| Sector | Sector of school | $S_c \in \{P\|R\}$ |

**Table 3.5: Variable of Student Agent**

| Variable Name | Brief Description | Value |
|---|---|---|
| School | School of the student | $St_{id} \in \mathbb{N}$ |
| Grades | Grades of the student | $G_i \in \mathbb{Z}$ |
| Wealth | Current amount of the wealth | $W_i \in \mathbb{Z}$ |
| Class | Class in which student is studying | $C_i \in \mathbb{N}$ |
| Home | Living place | $H_i \in \mathbb{N}$ |

### 3.8.2 School

School agents are educational institutions in the model. They provide educational facilities to the students. There are six schools created in this simulation model. Out of these six schools, three schools are from the public sector and three schools are from the private sector. All schools in the model follow a specific teacher induction policy, and assign a specific homework assignment to the students. Properties of the school include the number of teachers, teacher induction policy, number of students, and homework study hours for students. Initially, teachers in the school are hired from a normal distribution, then every school hire teacher, according to a specific teacher induction policy. Figure 3.4 shows the complete life cycle of a school. The school agent has different variables: ID, Name, Sector which are presented in Table 3.1 and 3.3

**Public school**

In this model, as in a real-world situation, public schools are managed by government authorities. Teacher induction follows local government policy, according to the budget of the districts. Public schools hire teacher following teacher induction policy approved by the government. In public school class strength is relatively high due to the late and long hiring process of teachers. We only consider the teacher induction, class size, and homework assignment in this context.

**Private school**

Private schools, which are managed by non-government organizations or individuals, operate under the laws imposed by the local government. However, they have complete independence when it comes to managing their finances, teacher induction policy, student enrollment criteria, and other managerial procedures. The quality of education offered by private schools is widely considered to be superior to that provided by public schools, even when taking into account the basic facilities available.

One of the reasons for this is the smaller class sizes in private schools. To maintain this smaller student-to-teacher ratio and ensure the quality of education, private schools adopt a dynamic teacher induction policy, allowing them to hire teachers as needed. Additionally, private schools place a greater emphasis on homework assignments and formal evaluation, assigning more homework to students than what is typically given in public schools.

This enhanced focus on education and smaller class sizes contribute to the overall higher quality of education in private schools compared to public schools. Furthermore, the independence that private schools have in managing their finances, teacher induction policies, and other procedures allows them to prioritize the needs of their students and maintain the highest standards in education.

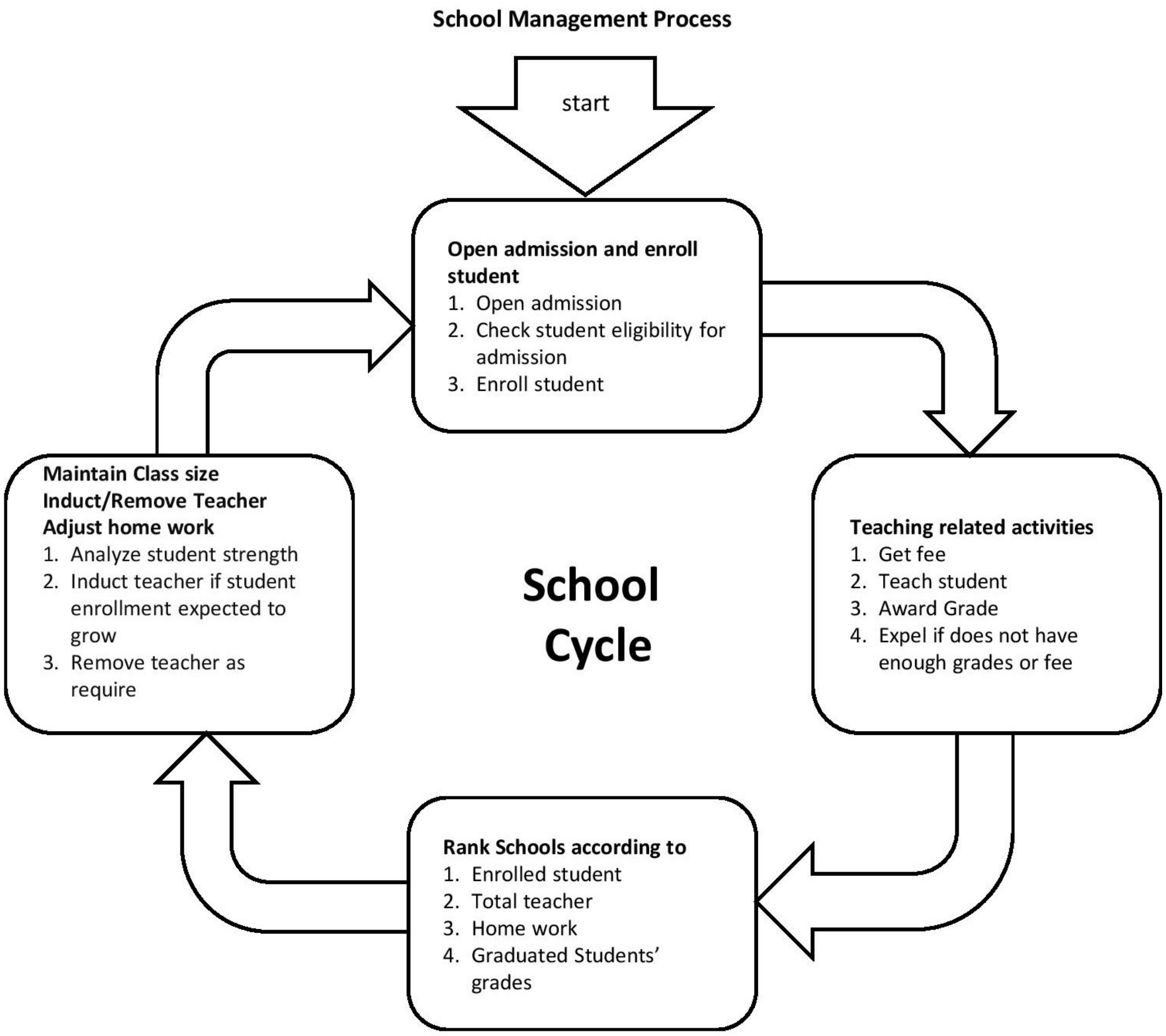

**Figure 3.4: School Management Yearly Cycle**

### 3.8.3 Methods of School Agent

Following is the detailed description of the functions of a school in the model which is also presented in Figure 3.5 as flow chart. The schools' methods are summarized in Table 3.3.

**Teachers' induction**

The school hires teacher according to the teacher induction policy of sector to which it belongs. Initially, there are 7-10 teachers in each school. These teachers are hired randomly. However, after first educational life cycle, which is one-year, new teachers are being hired according to specific teacher induction policy of the school. A school needs to maintain a specific class strength to cope with the educational quality challenges. To do this, schools must induct teacher on a regular basis as the need arises.

Teacher induction policy is of two types, dynamic teacher induction policy, and static teacher induction policy. Intuitively, dynamic teacher induction policy can be defined as induction of a teacher immediately to the school when the need arises. Dynamic teacher induction is not an absolute term rather than a relative term. Schools, those are hiring teachers after a shorter time as compared to other schools hiring teachers after a longer time, are considered following dynamic teacher induction policy as compared to later schools following static teacher induction policy.

**Remove Teacher**

Schools remove teachers according to the predefined teacher removal policy after each educational cycle of one year. Teacher removal policy is same as the teacher induction policy in the sense that both takes place at the same time, however, both effects the school in reverse order. In teacher removal, specifically, teachers are being removed based on the school preferences and policy.

**Enroll student**

Schools enroll students at the start of the educational session. While student enrollment, student's previous grades, socioeconomic status is considered. A Student having better socioeconomic status have a greater chance to be admitted into a school. Student previous grades are also a strong influencing factor in student's admission and enrollment.

### Get fee from students

The school will get the fee from the student. After educational life cycle, each school collects a fee from the student according to the fee policy. Student pay fee to the school according to their socioeconomic status in the society.

### Teach student

The school provides teaching facilities to students using teaching staff. Every school has a specific number of teaching staff. The staff at the initialization stage is taken from normal random distribution. However, with the passage of time new teacher are being inducted and older teachers are being removed according to teacher induction and removal policy respectively. A school inducting teaching staff more dynamically tends to be able to provide quality education.

### Award student grades

Study output of students is the result of the study of the students at home and at school. Schools award students with grades according to their performance resulted due to study effort. There are two measures of student's performance. The first measure of performance is the study at school, and second is study at home. A student in a school having smaller class size gets more grade, similarly, a student having batter socioeconomic status performs better at home.

### Expel student

School expels students because of low grades or wealth. A student grows wealth after each year, and must pay the fee to the school along with other expenses like residential, and food expenses. Students either heaving wealth lesser than the school fee, or lower grades would be expelled from their respective schools.

### Rank school

A school will be ranked based on student teacher ratio in the school, and grades obtained by the students of the respective school. A school with lesser student teacher ratio is classified as the best school in the educational sector, this lesser student teacher also results in better grade award to students. Both these factors collectively define the rank of each school.

Flow Chart of School

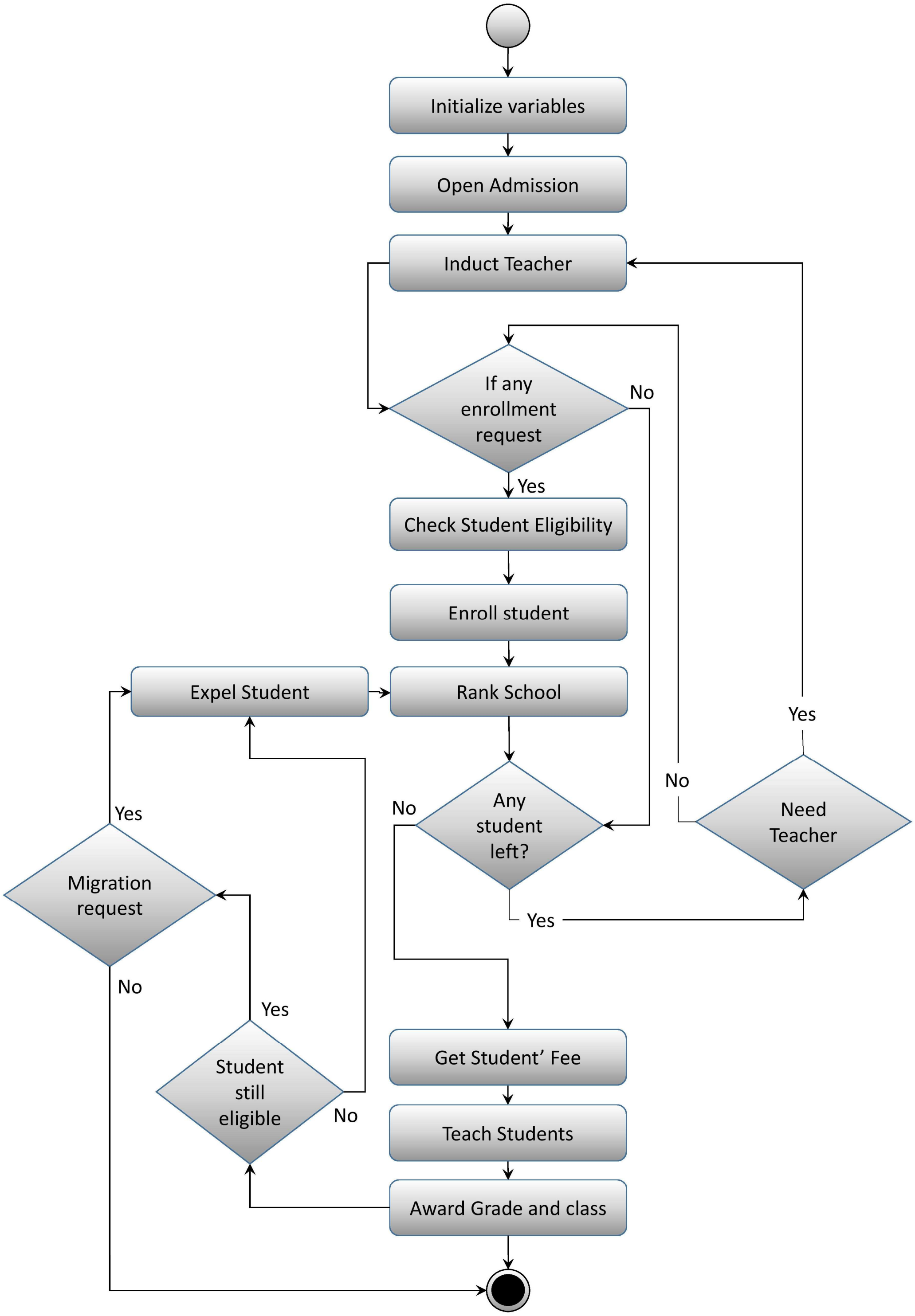

**Figure 3.5: Flow Chart of School Agent for one Cycle**

### 3.8.4 Student

Students are the main agents of the model. Each student in the model gets enrolled in a school which is best suited with respect to expected grades and is in affordable range with respect to the fee. After getting enrolled in the school, students get the education in schools and pay a specific fee according to the chartered fee of that school. The fee of the student is indirectly paid by the student's parent. A student can see schools in its approachable range. Based upon unsatisfactory results, a student will migrate to the other school with better performance. To accomplish this migration, a student after a yearly cycle analyzes the school performance then unroll from current school if the performance is not satisfactory. After leaving the school, student analyzes all other schools in range, and get enrolled in the best school with respect to grade awarding scheme and within affordable fee range. To uncover the causes of migration is the prime concern of this model. Table 3.1, and 3.3 present the student, and Figure 3.6 represent the student complete life cycle in school.

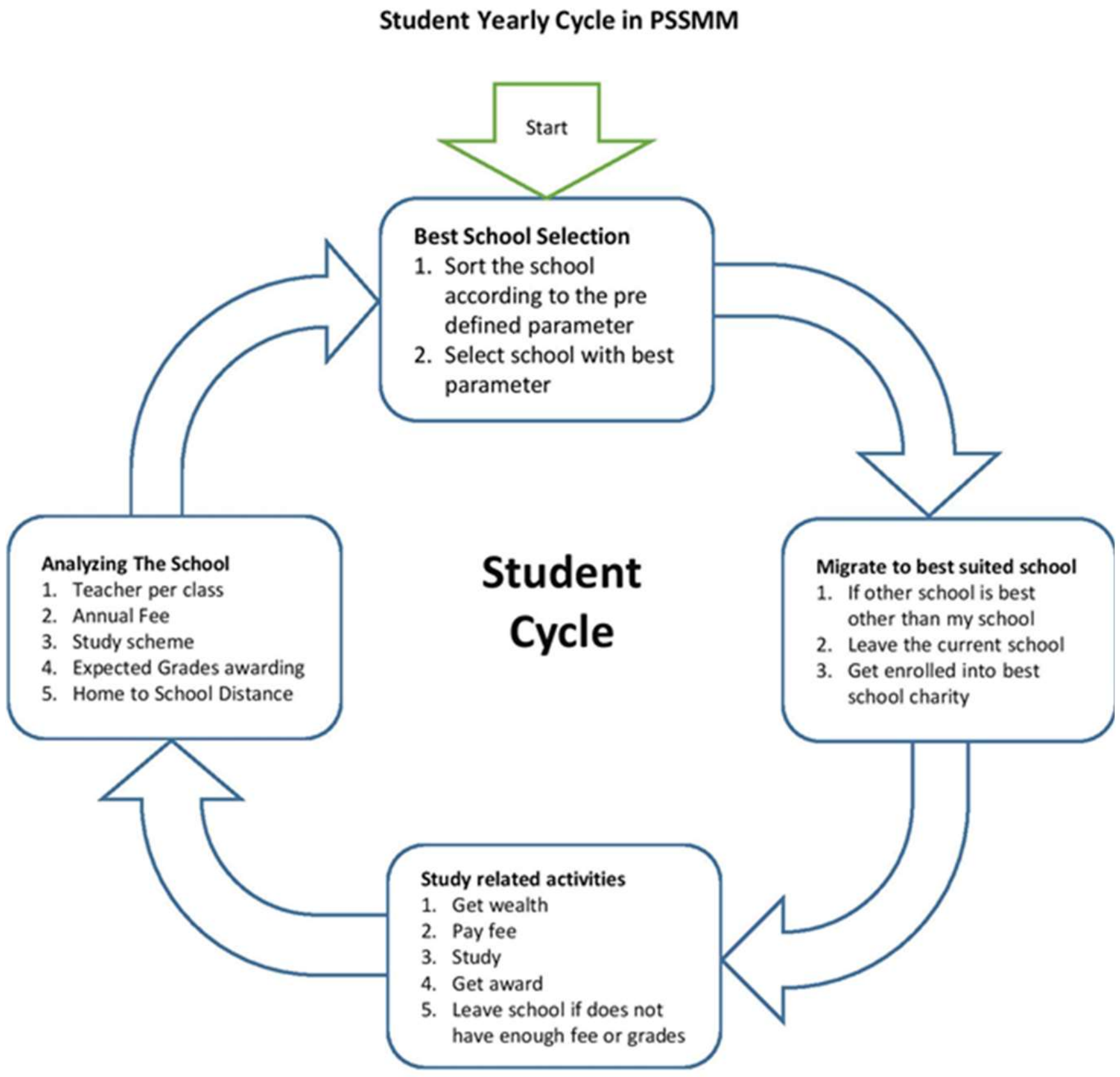

**Figure 3.6: Student Year Cycle**

### 3.8.5 Methods of Student Agent

The student agent has following methods to execute his function as presented in Table 3.3.

**Get Wealth**

The student will get wealth from the parents. Each student has a specific rate of wealth growth that is assigned from random distribution at the start of the simulation. Each year a student gets wealth according to this rate. Student total wealth is the sum of total wealth of previous year plus the growth rate, subtracted from school fee of the present year and the uncertainty in expenses.

**Analyze School**

At first, the student analyzes all school, and get the best-ranked school in the educational sector. Student views the list of school with respect to performance. The best-ranked school will be selected. The selection of the school will be determined based on several different factors. The most prominent factors are the school class strength and grade awarding scheme. A school with least students in a class is the best school. It is since a school with less student per class is more tends to perform better. Contrarily, a school with huge class cannot be predicted as best school. A big class needs more concentration to be taught, and it also needs many resources to be handled.

**Get Enrolled**

Student gets enrolled into best-ranked school. A student with enough wealth and grade will enroll to a school. While analyzing the schools, a student will get a list of all affordable schools. After that, the list would be sorted according to the performance of the schools. A student will get enrolled in the school which is in the top of the sorted list. If the list is empty student will not get enrolled and remains out of A student enters in the model at age of five year and may remain in the model for maximum age of nineteen years. So, from these fourteen years of life, student may be at school or at home when declared as out of the school. In a school, a student can study at most for ten years, and solely at home a student can stay at most four years.

**Pay Fee**

The student will pay a fee of the school. Every school charge a specific fee from the students each year. This fee includes tuition fee, stationary charges etc. As in public school, a minimal amount with respect to tuition is being charged, but in this model, all other living charges are also considered. The total wealth of the student is reduced with respect to the fee of the student. On the other hand, wealth will be increased according to the growth rate of the students. Students get admission into the schools where the fee is in an affordable range.

**Study**

Students get the education in a school and get home tuition. The study in school is inversely proportional to the class strength in the school. A school having larger classes tends to provide less education in school. This effect has been discussed before. Students also study at home. Quality, and quantity of study at home are determined by the socioeconomic status of the student. As study at home is directly proportional to socioeconomic status, a student having better socioeconomic status has a greater opportunity to get better home tuition. Contrarily, a student with weaker socioeconomic status may have to earn money for the family instead of arranging tuition at home.

**Get Grades**

The student will get grades according to the study hours. Study hours includes study at home and study at school. At school, study means to study in class. At home, study means study by home tuition or the homework assigned. The school will award grades to students after each year. A student must have to retain specific grades to remain in a school.

**Leave School**

A student may leave school if there is either lack of wealth, or deficiency in grades. If a student becomes deficient in money, then it is mandatory for that student to leave the school. However, a student is still able to switch to a school where a fee is in the affordable range. On the other hand, if a school could not find a school whose fee is not in affordable range, then a student cannot enroll in any school. In such a case, a student is declared as out of school children. Such a student will earn money according to growth rate, and after each year the list of affordable school would be searched. If there would be enough money with respect to lowest school fee, then this student would be able to enroll in that school. To get enrolled in any school at any time trigger, a student must have to fulfill all other requirements along with these requirements.

### Migrate to Another School

The student will migrate to other school either if there is another better school, or student is struck off due to lack of fee or deficiency in grades, as discussed before in the last paragraph as presented in Figure 3.7. So, migration can take place in two cases. In the first case, a student switch to a school which is with the higher fee and better ranking scheme. Secondly, a student gets stuck due to a deficiency in grades or fee, then re-enrollment takes place in the affordable school.

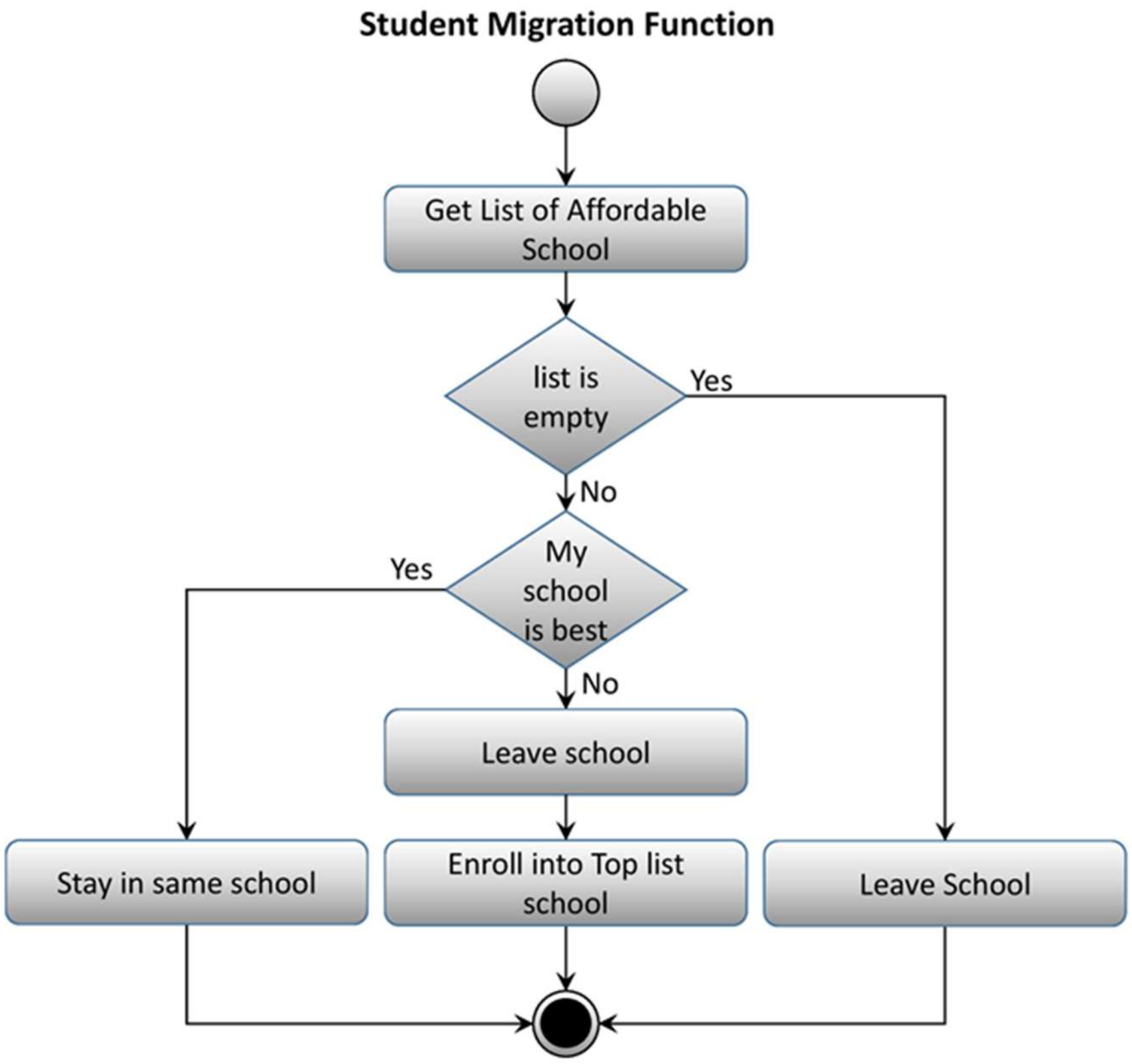

**Figure 3.7: Students Migration Function**

### 3.8.6 State Diagram of Agents in Model

State diagram of the school and student agent in the model are as follows

#### State Diagram of School

In the model, school agent, as shown in Figure 3.8, has seven states in total. School starts from initialization of variables. After that, a school enters to teacher hiring state according to the teacher induction policy. First time hiring of the teacher is a selection from a random distribution pool. However, after the first iteration, each school is allowed to hire teachers in accordance with specific induction policy. After enrollment of teaching staff, every school enters in students' enrollment state by opening admission. In the enrollment state, a school enrolls students and switched to the ranking state. In that state a school would be ranked according to students' enrollment. This dynamic ranking causes to upgrade or degrade a school depending upon the teacher induction policy and student enrollment. All enrolled students will pay a fee to their respective schools. All schools would provide teaching facilities to all respective enrolled students, and then award students with grades according to their performance. A school has a right to expel a student based on their poor study performance or based on their low socioeconomic status. However, after the expulsion of a student, the respective school must be re-ranked.

#### State Diagram of Student

The state diagram of student agent has eight states as shown in Figure 3.9. It is just reverse of the school agent with minor changes. A student starts its life from initialization state. In the first iteration, a student is allocated with a reasonable amount of wealth from a random distribution of wealth. However, after the first iteration, a student gets wealth according to the growth rate which is again randomly assigned. After getting wealth in compliance with growth rate, every student gets the sorted list of schools with respect to performance ranking. The student selects the top school for study. In that state, the student gets enrolled into a school. After enrollment, a student pays a fee to the school and start studying. Each student, after yearly cycle, gets grades according to the performance and then may remain in the school or migrate to the school. This migration is based on the performance of the school. The third option is a student leave a school due to a shortage of resources or lowering of grades.

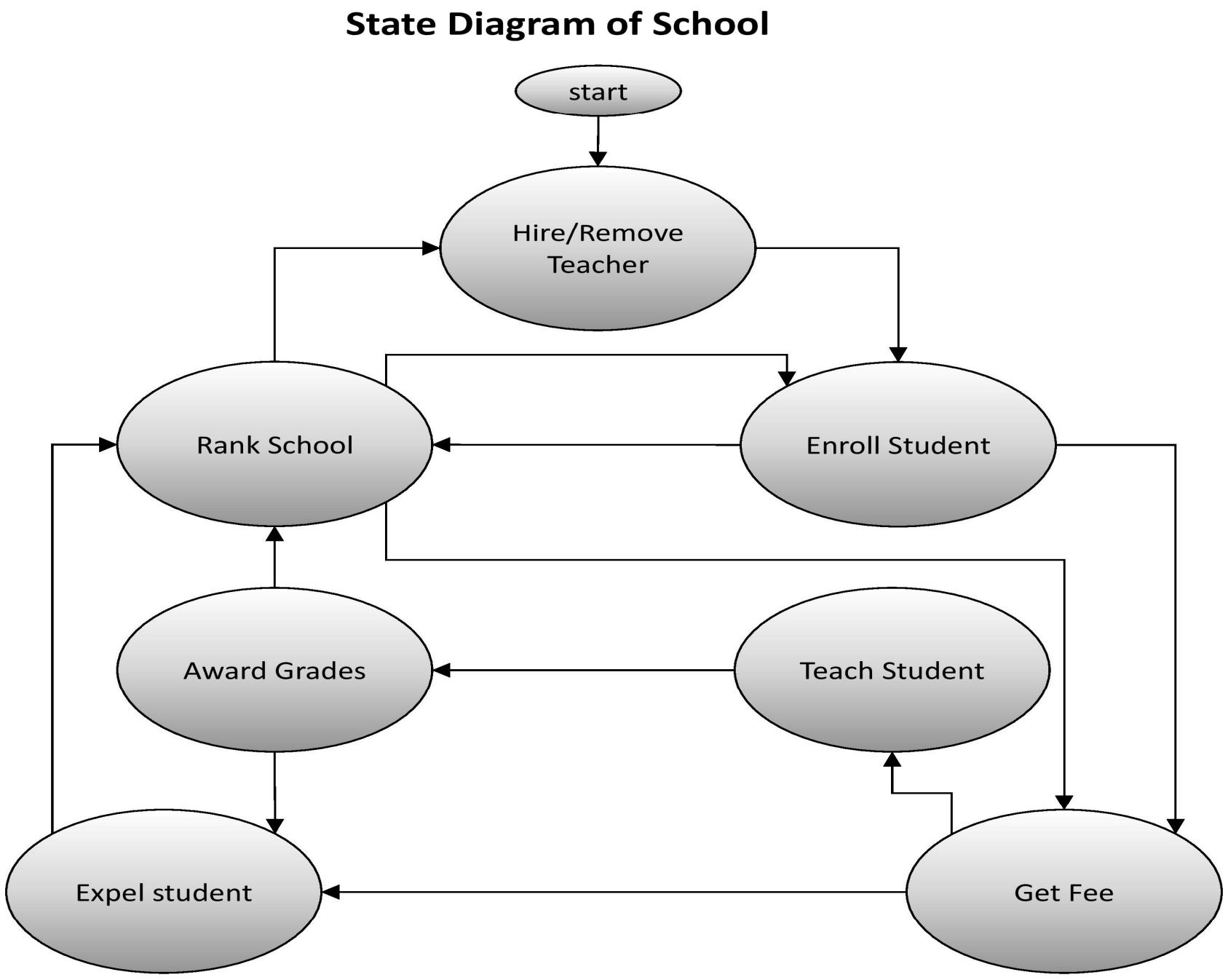

**Figure 3.8: State Diagram of School Agent**

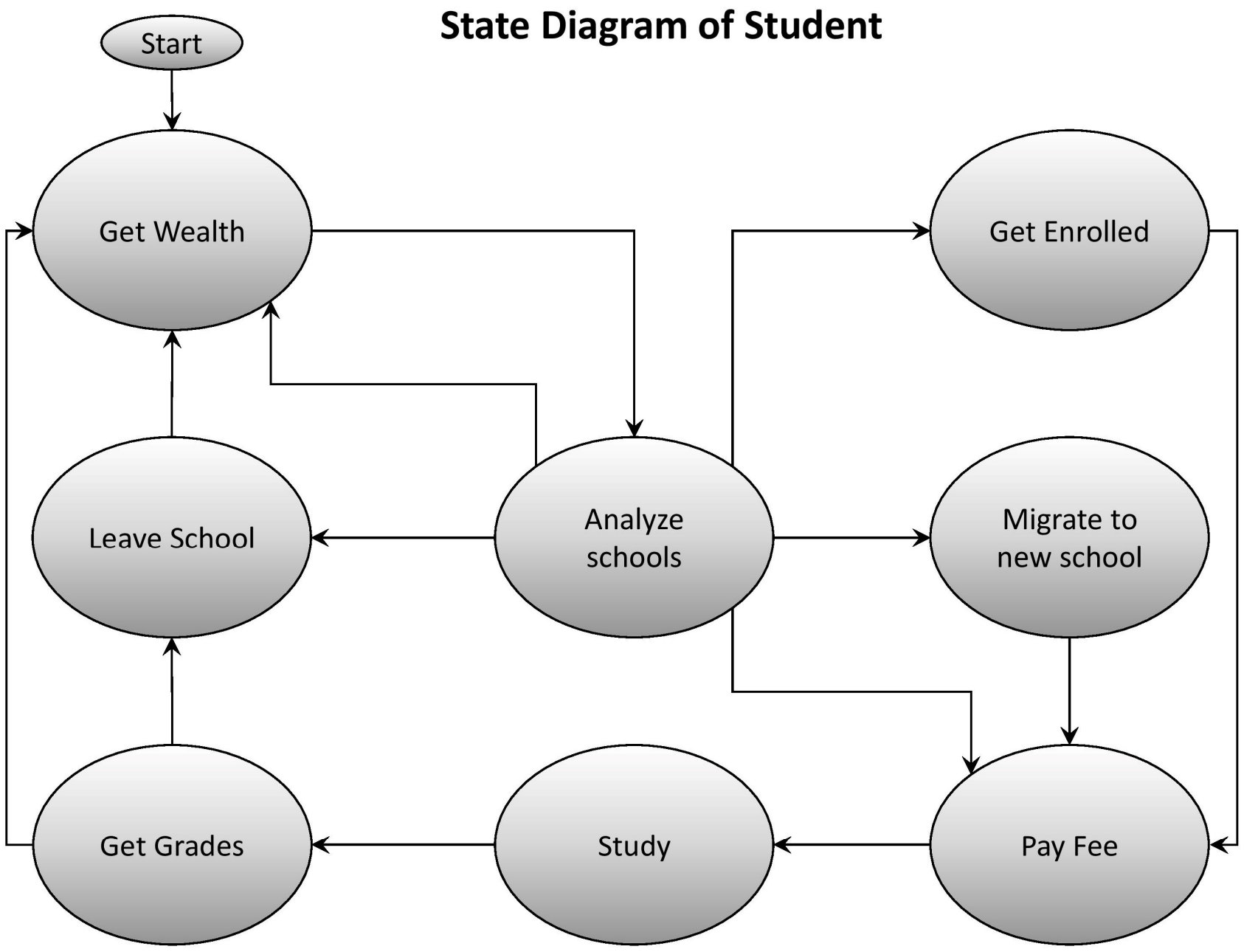

**Figure 3.9: State Diagram of Student Agent**

## 3.9 Complex Network Model

The Complex Network model is an abstract representation used for analysis, comprehension, replication, and comparison. It can be presented in three forms: random network model, gem layout, and tree layout. The random network model is the preliminary stage in network design. The gem layout is a detailed and structured representation that makes it easier for the reader to understand the model's structure. The tree layout is an extension of DREAM, designed to facilitate easier comparison of models.

### 3.9.1 Random Network Model

The Random network model is a useful tool to gauge the density of a complex network. Although the logical complexity of the network cannot be fully reflected in the number of nodes and edges, the quantity often corresponds directly to the complexity. The random network model provides an overview of the complexity of the model and serves as a foundation for developing more intricate layouts, such as the gem and tree layouts. Figure 3.10 illustrates the Random network model.

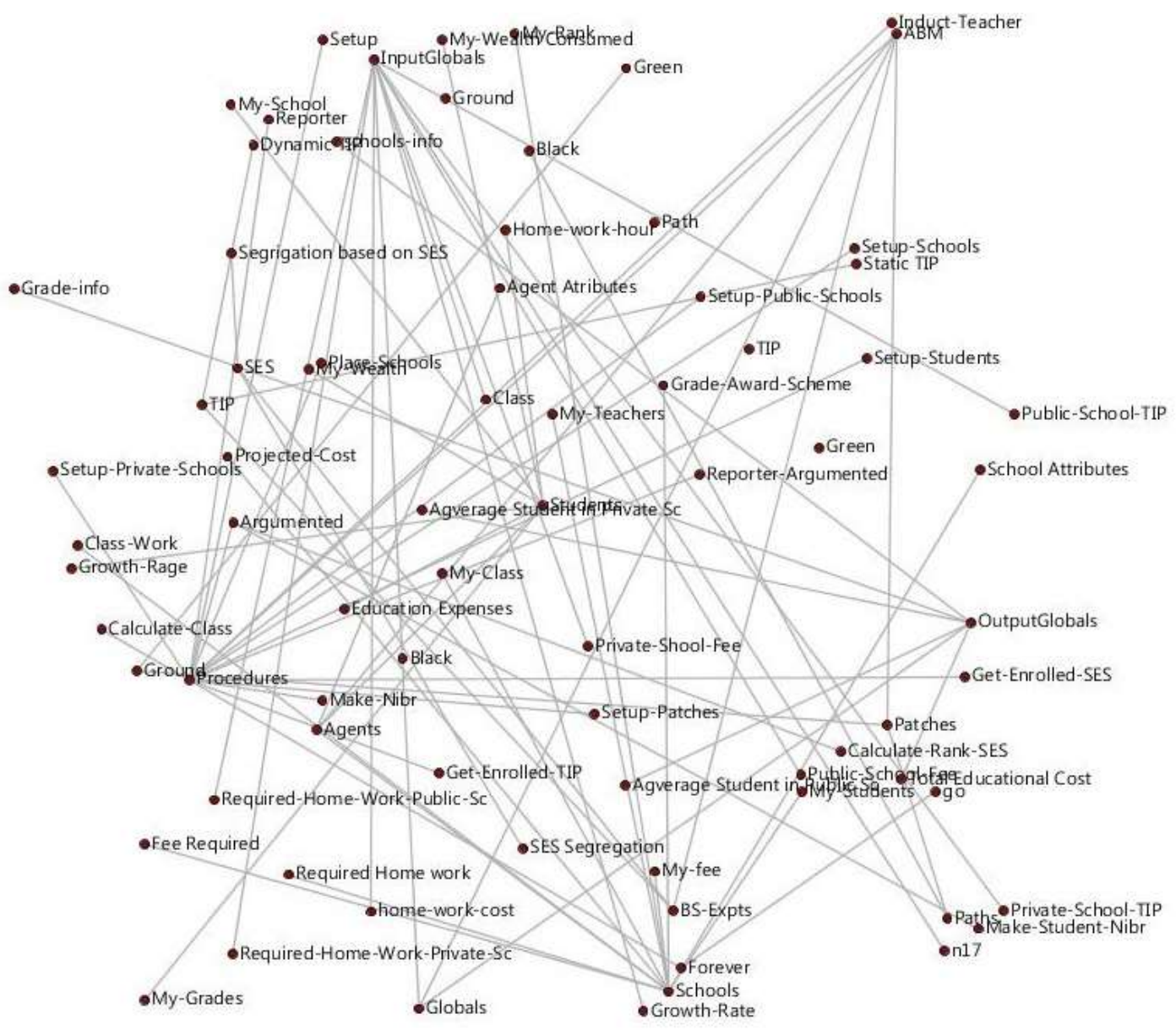

**Figure 3.10: Random Layout**

### 3.9.2 Gem Layout

Gem layout of the complex network model is a key component in showcasing the complexity of the model. This layout presents the logical structure of the model in groups, with each logical item or entity attached to its parent node according to its scope of work. This organization makes it easier for the reader to understand the structure of the model, and is therefore an important factor in replication of the model. The clear representation of the logical structure also enables comparison between different models from different scientific domains.

The gem layout of the complex network is designed to provide a detailed and structured representation of the model. It offers a clear view of the relationships between the different elements, making it easier for the reader to grasp the model's structure without having to analyze the code. The gem layout is therefore a crucial component in comprehending, replicating, and comparing complex network models.

The complex network starts with the Agent-Based Model (ABM) and expands in all possible directions based on its attributes. The ABM node is followed by nodes for procedures, breeds, global variables, and patches. The procedures node includes categories of agents' actions, such as simple procedures, forever procedures, and argumented procedures. The breeds node encompasses the agents in the model, such as students and schools, and provides an in-depth pictorial description of the agents' states.

The global node contains input and output global variable nodes, which are further explained. The patch node encompasses the basic environmental nodes that belong to the environment configuration. This provides a convenient approach for comparing the model. The network offers a comprehensive view of the model, starting from the ABM and going into the details of agents' actions, their states, and the environment in which they operate. The network is designed to give a clear picture of the model and its components, which makes it easier to comprehend, replicate, and compare with other models.

Figure 3.11 is a gem-layout which contains the description of school model.

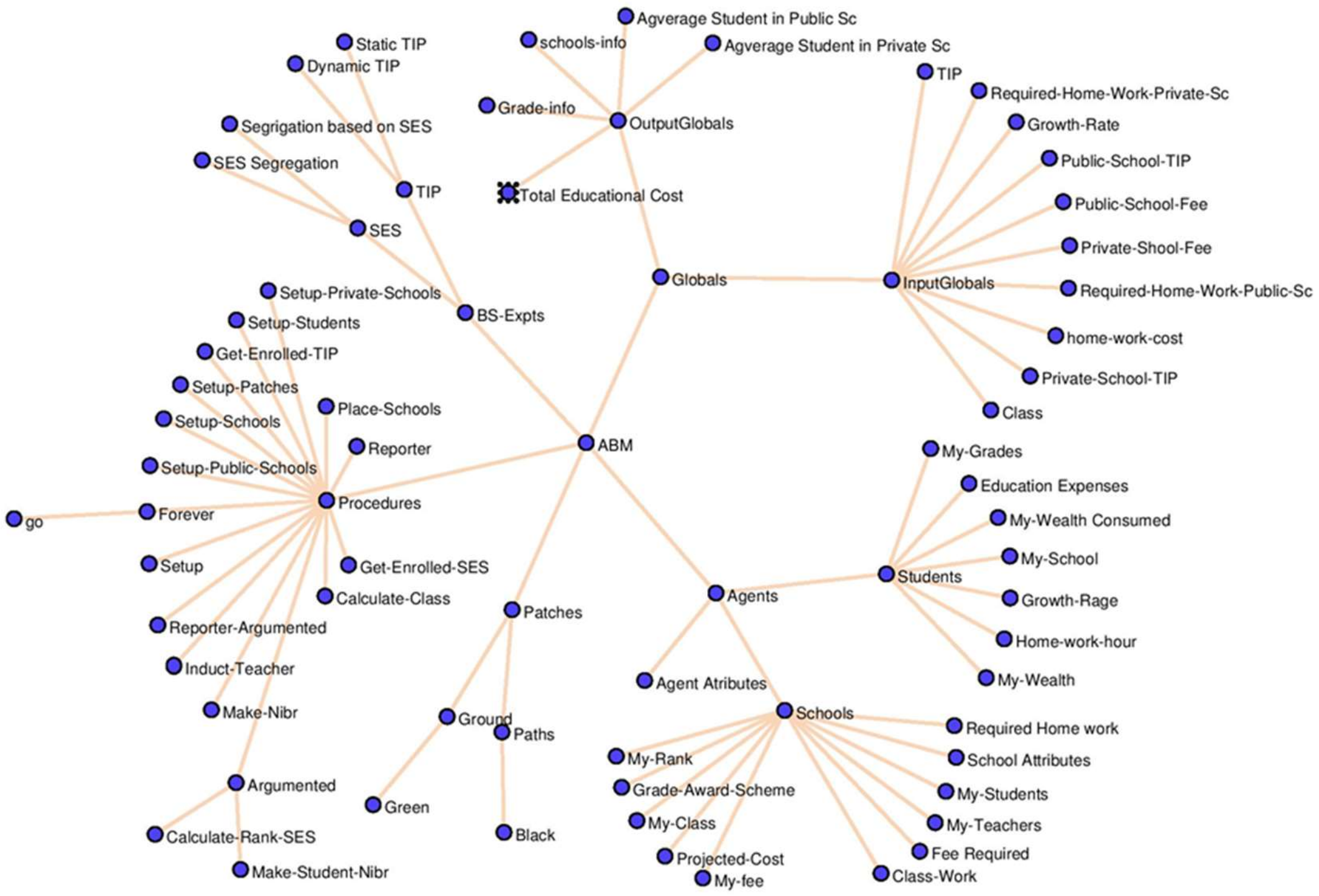

**Figure 3.11: Gem Layout**

### 3.9.3 Tree Layout

The Tree layout of the complex network model is an innovative and valuable addition to the DREAM framework. This layout provides a unique and beneficial way of visualizing the complex network structure, offering a vertical presentation that allows for more effective comparison of different models from different scientific domains. By presenting the network structure in a vertical manner, the Tree layout makes it easier to understand the hierarchical organization of the model, which is crucial for replication and comprehension. This visualization also makes it possible to compare models without requiring a deep knowledge of their coding details.

The hierarchical presentation of tree layout starts from the tree main node as “ABM”. The procedures, agents, globals, patches, and BS-Expts are the sub node of the main node. The procdures node contains forever, argumented, reporter, reporter-argumented nodes. Agents’ node contains agents, and their details. Patches node contains environmental specific configuration. And the last node in this hierarchy contains the experiments' details performed in the simulation.

In conclusion, the Tree layout of the complex network model offers numerous advantages over other layouts, such as the Random network model and the Gem layout. This layout provides a clearer and more comprehensive view of the structure of the model, making it easier to replicate, comprehend, and compare different models. Whether you are a researcher, a student, or simply someone interested in understanding complex models, the Tree layout of the complex network model is a valuable tool that can help you gain a deeper understanding of the models you are studying. Figure 3.12 is a detailed Tree layout of the model.

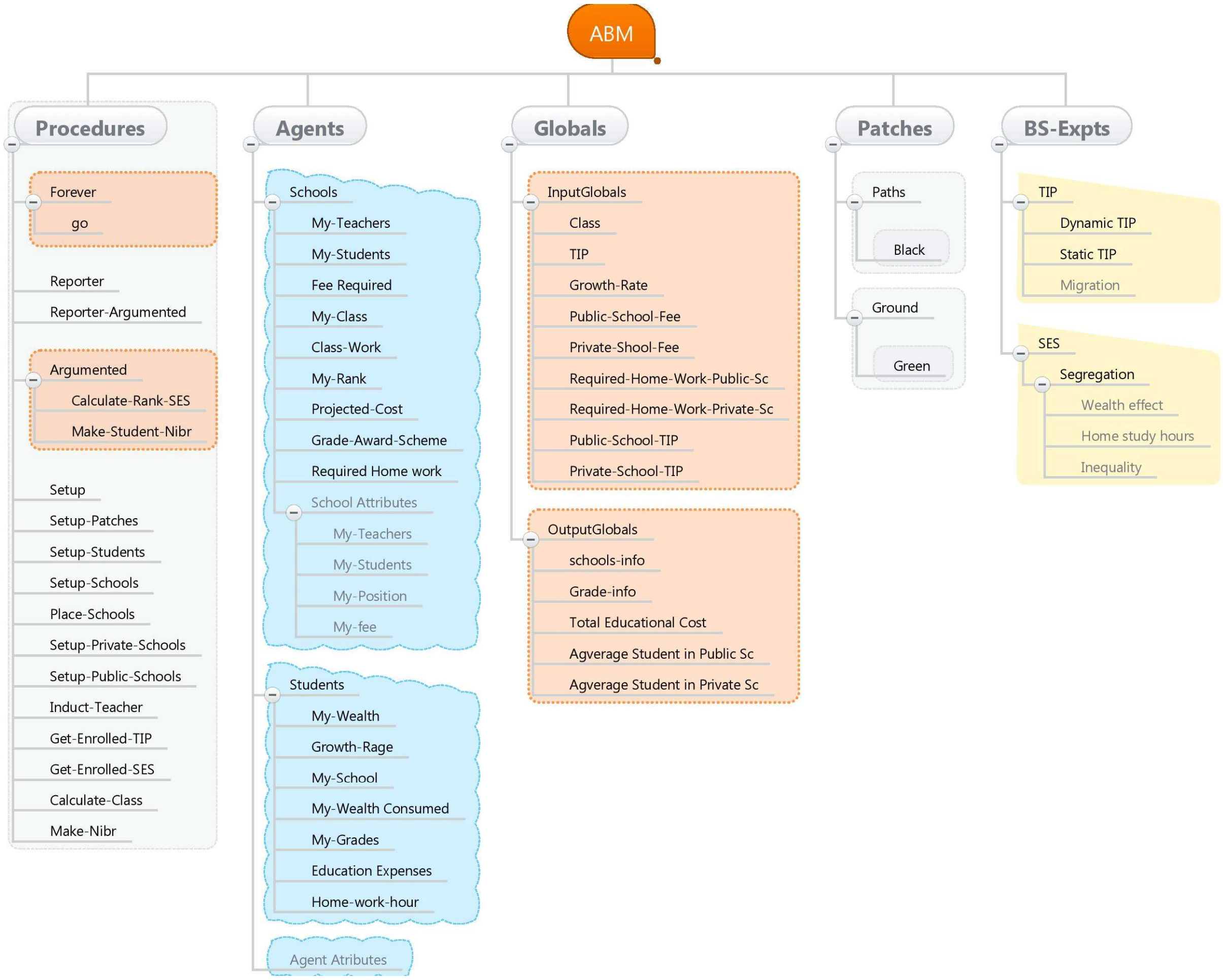

**Figure 3.12: Tree Layout of Network**

## 3.10 Pseudo Code Model

Pseudo code-based specifications of the model are as follows

### 3.10.1 Agents Specification

**Student**

Student agent pseudo code-based specification.

**Breed Student:** This agent represents a student in the model

*Internal Variable:* <growth-rate, wealth, grades, school, home-work, expenditure>

| | | |
|---|---|---|
| | | **growth-rate:** Growth rate of the student. |
| | | **wealth:** Total wealth of the student. |
| | | **grades:** Total grades of the student. |
| | | **school:** School of the student. |
| | | **home-work:** Homework doing capability of the student. |
| | | **expenditure:** Total expenditure of the student. |

**School**

School agent pseudo code-based specification.

**Breed School:** This agent represents a school in the model

*Internal Variable:* <teachers, students, rank, sector, position, fee, income, req-home-work, class-work, projected-cost, class, grade-award-scheme, TIP >

| | | |
|---|---|---|
| | | **teachers**: Total number of teacher in a school |
| | | **students**: Total number of students in a school |
| | | **rank**: It is the evaluated variable. It is calculated by dividing students by teachers |
| | | **sector:** The sector of school. |
| | | **position**: The Position of the school. |
| | | **fee:** Fee of the school. |
| | | **income:** Total income of the school. |
| | | **req-home-work:** Required homework demanded by school from students. |
| | | **class-work-hours:** Class work hours in school. |
| | | **class:** Class size in the school. |
| | | **grade-award-scheme:** Grade awarding scheme for the students of the school. |
| | | **TIP:** Teacher induction policy of the school, either dynamic or static. |

### 3.10.2 Global Variables

**Input variables**

Specification of Input variables

| **Input Global:** <class-size, TIP, growth-rate, public-school-fee, private-school-fee, required-home-hours-public, required-home-hours-private, home-work-cost, public-school-rec0-time, private-school-rec-time> | |
|---|---|
| *Switch:* | |
| | **TIP:** Teacher Induction Policy |
| *Slider:* | |
| | **class-size:** Class size for the schools. |
| | **growth-rate:** Random number for growth rate of students. |
| | **public-school-fee:** Public school fee. |
| | **private-school-fee:** Private school fee. |
| | **required-home-hours-public:** Required working hours for study at home by public school. |
| | **required-home-hours-private:** Required working hours for study at home by private school. |

**Output Global**

Specification of Input variables

| | |
|---|---|
| **Output Global**: <growth-grades, wealth-grades, students-count, teachers-count, avg-cost > | |
| *`Plots:* | |
| | **growth-grades**: Plot for grades of students' vs growth of students. |
| | **wealth-grades:** Plot for wealth of students' vs grades of students. |
| | **students-count:** Number of students in public and private school. |
| | **teachers-count:** Number of teacher in public and private schools. |
| | **avg-cost:** Average cost per students. |
| *Simulation visualization output Grid:* | |
| | **min-x:** 0 |
| | **max-x:** 60 |
| | **min-y:** 0 |
| | **max-y:** 40 |
| | **Origin:** Bottom left, corner. |

### 3.10.3 Procedures

Pseudo-code specification of procedures.

**Simple Procedure**

1. Pseudo-code specification of **setup** procedure.

| Procedure **setup**: Setting up the simulation |
|---|
| *Input*: Uses global variable from the user interface for setup |
| *Output*: Two type of agents are created and set upped for simulation |
| *Execution*: Called at the start of the simulation |
| *Context*: Observer |
| *Dependency*: setup-patches, setup-schools, setup-students |
| **begin:** |
| 1. Clear all (Reset all patches, all agents, ticks, and clear drawing)<br>2. Setup patches with **setup-patches**<br>3. Create school's agent according to number given by the slider<br>4. Setup the school with **setup-schools** procedure<br>5. Create students according to the number given by slider in user interface<br>6. Setup students with **setup-students** procedure<br>7. Place all student around their schools.<br>8. Reset timer to zero by reset-ticks |
| **End** |

2. Pseudo-code specification of **setup-patches** procedure

| | |
|---|---|
| Procedure **setup-patches**: Setup all patches of the simulation grid. | |
| | *Input*: Number of schools. |
| | *Output*: Set all patches green or black according to school number. |
| | *Execution*: Called by ***setup*** procedure. |
| | *Context*: Patches |
| **begin:** | |
| | 1. Set patches color black with coordinate (x, y) multiple of by school number.<br>2. Set patches color green with coordinate (x, y) not multiple of school number. |
| **End** | |

3. Pseudo-code specification of **place-schools** procedure.

| | |
|---|---|
| Procedure **place-schools**: Place school on the grid. | |
| | *Input*: School agents |
| | *Output*: Place schools on the grid. |
| | *Execution*: Called by **setup-schools** procedure. |
| | *Context*: Turtles |
| **begin:** | |
| | 1. Move to school to empty green patch with coordinate multiple of its ID. |
| **End** | |

4. Pseudo-code specification of **setup-schools** procedure.

| |
|---|
| Procedure **setup-schools**: Setup the students |
| *Input*: Schools. |
| *Output*: Set all schools. |
| *Execution*: Called by setup procedures |
| *Context*: Turtles |
| *Dependency*: place-school, setup-private-school, setup-public-school |
| **begin:** |
| 1. Set the shape of school to building.<br>2. Set size double.<br>3. Allocate teachers from random distribution.<br>4. Place schools by calling **place-school**<br>5. Setup public schools by calling **setup-private-school**<br>6. Setup private schools by calling **setup-public-school**<br>7. Calculate rank and class size of schools. |
| **End** |

5. Pseudo-code specification of **setup-public-school** procedure.

| Procedure **setup-public-school**: Setup the public schools |
|---|
| *Input*: Public school agents, slider input for public schools. |
| *Output*: Setup all public schools. |
| *Execution*: Called by **school-setup** procedure. |
| *Context*: Turtles. |
| **begin:** |
| 1. Set color of the public schools to red.<br>2. Set fee of public schools according to slider input.<br>3. Set class work of public schools according to slider input.<br>4. Set income zero, and project-cost of school according to input parameter. |
| **End** |

6. Pseudo-code specification of **setup-private-school** procedure.

| Procedure **setup-private-school**: Setup the students |
|---|
| *Input*: Slider input for private schools, and school agents. |
| *Output*: Set all privates schools. |
| *Execution*: Called by **school-setup** procedures |
| *Context*: Turtles. |
| **begin:** |
| 1. Set color of the private schools to yellow.<br>2. Set fee of private schools according to slider input.<br>3. Set class work of private schools according to slider input.<br>4. Set income zero, and project-cost of school according to input parameter. |
| **End** |

7. Pseudo-code specification of **setup-students** procedure.

| Procedure **setup-students**: Setup the students |
|---|
| *Input*: Students agents, slider input for student agents. |
| *Output*: Set all student agents. |
| *Execution*: Called by **setup** procedures |
| *Context*: Turtles. |
| **begin:** |
| 1. Set size of student agents to 0.9.<br>2. Set shape of student agents to person.<br>3. Place student agent to empty green patch.<br>4. Set grades zero and wealth from random distribution.<br>5. Set growth-rate from random distribution.<br>6. Enroll me in the best affordable school.<br>7. Increase the number of students in my school. |
| **End** |

8. Pseudo-code specification of **calculate-class** procedure.

| Procedure **calculate-class**: Calculate the class size of schools. |
|---|
| *Input*: School agents |
| *Output*: Return the class size of each school. |
| *Execution*: Called by **get-enrolled-TIP, and get-enroll-SES** procedure. |
| *Context*: Observer |
| **begin:** |
| 1. Calculate the class size of each school.<br>2. Assign the respective class size value to each school. |
| **End** |

9. Pseudo-code specification of **get-enrolled-ses** procedure.

Procedure **get-enrolled-ses**: Setup the students

*Input*: Students agents, school agents, and sliders.

*Output*: Enroll students to best affordable schools.

*Execution*: Called by **go** procedure.

*Context*: Observer

*Dependency*: calculate-class, calculate-rank-ses

**begin:**

1. Calculate class size and rank of schools.
2. Increase the wealth of each student by its growth rate.
3. Pay fee of the students to the schools.
4. Decrease the wealth according to the fee paid by students.
5. Increase the income of the school with correspondence of fee received.
6. Award grades to the students according their schools.
7. Unroll, and enroll the students according to their wealth and the school preferences.
8. Place the student around the new school if migration takes place.
9. Initialize the respective cardinals of students, and schools after unrolling and enrolling processes.
10. Recalculate the class size, and rank of schools by calling **calculate-class** and **calculate-rank-ses** respectively.

**End**

10. Pseudo-code specification of **calculate-rank-ses** procedure.

Procedure **calculate-rank-ses**: Calculate the rank of schools.

*Input*: School agents.

*Output*: Calculate the rank of each school.

*Execution*: Called by **get-enroll-SES,** and **get-enroll-TIP** procedures

*Context*: Observer

**begin:**

1. Calculate the rank of each school.
2. Assign the respective rank value to each school.

**End**

11. Pseudo-code specification of **get-enrolled-tip** procedure.

Procedure **get-enrolled-tip**: Enroll student according to class size of school.

*Input*: Students agents, school agents.

*Output*: Unroll, and enroll students to schools according to class size.

*Execution*: Called by **go,** and **setup** procedures

*Context*: Observer

*Dependency*: calculate-class

**begin:**

1. Calculate the class size of schools by calling **calculate-class.**
2. Award grades to all students according the school grade awarding policy.
3. Unroll, and enroll students to the schools according the class size of schools.
4. Induct new teachers for each school according to teacher induction policy.
5. Recalculate the class size of each school and place the student around the new school

**End**

12. Pseudo-code specification of **induct-teacher** procedure.

Procedure **induct-teacher**: Setup the students

*Input*: School agents.

*Output*: Induct new teacher according to teacher induction policy.

*Execution*: Called by **get-enroll-TIP,** and **setup** procedures

*Context*: Observer

**begin:**

1. Check the required time for teacher induction.
2. Check the teacher induction policy for school.
3. Induct teacher if policy and time constraints are fulfilled.

**End**

13. Pseudo-code specification of **make-nieb** procedure.

Procedure **make-nieb**: Place the students around their respective school.

*Input*: Students', and schools' agents

*Output*: Student place around their schools.

*Execution*: Called by **setup**, **go**, and **get-enrolled-TIP, SES** procedures

*Context*: Observer

*Dependency*: make-neib

**begin:**

1. Call **make-nieb** procedure for each student to place around respective school.

**End**

**Forever procedure**

Pseudo code-based specification of **go** procedure.

Procedure **go**: Simulate the model

*Input*: According to the slider and switches in user interface

*Output*: Simulate the model, migrate the student, add teacher, rank school.

*Execution*: Called by setup procedure, called to start simulation

*Context*: Observer

*Dependency*: get-admission

**begin:**

1. Stop simulation if the times becomes more then 250 ticks
2. Call **get-admission** procedure
3. Migrate the student to the best ranked school
4. Add teacher according to the policy of the school
5. Add teacher on class limits if the slider is turned on
6. Add student if the switch is turned on
7. Update the plot and view according to the respective variables

**End**

**Argumented Procedure**

1. **Make-Student-Nibr** procedure pseudo code-based specification.

Procedure **Make-Student-Nibr**: Place a student around respective school.

*Input*: Coordinate of school, and ID of student

*Output*: Place student around respective school.

*Execution*: Called by **setup, get-enrolled-TIP,** and **get-enrolled-SES** procedures

*Context*: Observer

**begin:**

1. Search empty green patch around the respective school.
2. Place the student at the empty patch around the school.

**End**

2. **Calculate-Rank-SES** procedure pseudo-code based specification.

Procedure **Calculate-Rank-SES**: Calculate the rank of schools

*Input*: Schools agent.

*Output*: Calculated ranks are assigned to schools.

*Execution*: Called by **setup, get-enrolled-TIP,** and **get-enrolled-SES** procedures.

*Context*: Observer

**begin:**

1. Calculate the rank of school.
2. Assign rank to respective school.

**End**

### 3.10.4 Pseudo-Code Experiments Specifications

**Pseudo-Code Specification of Very Class Size Experiment**

Experiment **very-class-size:** Experiment with effect of varying class size of public and private schools.

*Input:*<schools, student, public-school-class, private-school-class, TIP>

*Setup procedures:*<setup>

*Go procedures:* <go>

*Repetition:* 10

**Inputs**:

- **Schools**: 6
- **Public-schools**: 3
- **Private-schools**:3
- **Students**: 250
- **Public-school-class-size**: [1 → 10 → 100]
- **Private-school-class-size**: [1 → 10 → 100]
- **Public-school-home-study-hours: 3**
- **Private-school-home-study-hours: 3**
- **TIP?** True

*Stop condition*: Ticks=100

*Final commands:* None

**Pseudo-Code Specification of Very-Home-Study-Hours Experiment**

Experiment **Very-home-study-hours:** Experiment with effect of varying home study hours of public and private schools.

*Input:*<schools, student, public-school-home-study-hours, private-school-home-study-hours, TIP>

*Setup procedures:*<setup>

*Go procedures:* <go>

*Repetition:* 10

**Inputs**:

- **Schools**: 6
- **Public-schools**: 3
- **Private-schools**:3
- **Students**: 250
- **Public-school-class-size**: 30
- **Private-school-class-size**: 30
- **Public-school-home-study-hours:** [1 → 1 → 10]
- **Private-school-home-study-hours:** [1 → 1 → 10]
- **TIP:** False

*Stop condition*: Ticks=100

*Final commands:* None

## 3.11 Overview, Design, Detail (ODD) of PSSMM

Simulation models — Agent-Based Models or Individual-Based Models — have been widely used since few decades to describe the complex systems in ecology, social science, economy, and natural sciences. However, even having wide application, there was no formal protocol for presenting Agent-Based Models and Individual based model. Researcher from different scientific domain used textual representation for presentation of these models. This makes the comprehension, replication, and verification of the model very.

The need of a standard protocol for presentation of simulation models was felt to enhance the comprehension and ease the replication. In this regard, a large group of researchers from ecology and other simulation modeling domain comes together to propose a standard protocol to represent ABM and IBM. The main intent of standard protocol was to present the simulation model in precise and complete way. Nevertheless, it was the first effective attempt; the need of formal presentation of simulation model is still felt.

ODD stands for Overview, Design, and Detail. It is presented by Grimme et al(2006). It comprises of three sections "Overview", "Design", and "Detail". The 'Overview' section contains the basic definition of the model. The design section presents a generic list of abstract ideas about the model that is emerged from it. The last section of the ODD is 'Details'. It contains details in three sub sections as 'input', 'initialization', and 'sub models'.

### 3.11.1 Overview

**Purpose**

The purpose of the Primary School Student Migration model (PSSMM) is to understand, how class size in schools, and socioeconomic status of students effect the migration and segregation in schools respectively. Intuitively, the purpose of the model is two folds. Firstly, PSSMM helps to investigates the effect of class size in schools on migration of students in between primary schools. Due to variation in class size, migration of students takes place. Secondly, it shows how students get segregated based on their socioeconomic status.

### Entities, State Variable and Scales

This section contains entities, state variables, and scales of PSSMM.

***Entities:***

## a. Agents

There are two agents in the model, schools and students. The school agents are places where student agents get education. Student agents study in schools. They can migration from one school to other school. Their migration is based on the performance and reputation of schools.

i. *Schools*

There are total six schools in the model. Out of these six schools, three schools are public, and three schools are privates. Public school differs from private schools by teacher induction policy, class size, and required study hours at home from students. Internal variables of school agent are location of a school on the lattice, class size, a specific teacher induction policy, and required study hours at home mandatory for students. Teacher induction policy determines the class size in a school. Figure 6.1 shows the school agent state diagram.

ii. *Students*

There are total 1000 students in the model. Internal variables of the students are wealth growth rate, grades in the school, class, total wealth, and age. There is no hardline distinction between students. Main differentiating quality of the students is assigned from random distribution of growth rate. A student with higher growth rate can afford a school with higher fee and required study hours. Student can be divided into two categories: poor students, and rich students.

***Scales: Environment***

All these agents are placed on a two-dimensional lattice of 2500 patches with dimension as $250 \times 250$. Schools are static with respect to location, and students can move from school by migration. All schools are placed with equal distance from each other and students are placed around the schools where they got enrolled. A student, after migration, changes its location.

## Process and scheduling

### *Teacher agent processes*

The Primary School Student Migration Model proceeds in annual time steps. In a year, a school induct teacher according to teacher induction policy, open admission, enroll student according to student enrollment eligibility criteria, charge school fee according to fee plan, providing teaching related services according to pre-defined and announced policy, award grade according to the performance of the student, retains students who are fulfilling the retaining criteria, and rank school according to the teacher, student ratio, and grade awarded to the successful students.

### *Student agent processes*

Student agent share most of the processes of the school in the senses that these processes are done by school by directly involving or interacting with students. These processes are enrollment, pay fee, get grades, leave school. In addition to shared processed with schools, students have processes like study at home after school timing each day, and migrate to a school. Migration process is, infect, a combination of three processes. In the first step, a student analyzes all school, and gets the top suitable school. After knowing that the best school is other than my school, student decides to migrate. In the second step, student leaves the school current school. Then, the third step, if the student fulfills the requirement of admission, then he is being get enrolled in the best school according to his choice.

After each year, when annual exams held, students are awarded with grades based on their performance. A school having more students with better grades have better rank. After each year migration takes place based on the performance of the schools. School charges the fee, and poor student forced to stick in the public schools where no fee is being charges. The economic segregation often leads to a chaotic situation when students with worst economic status leave school until the next year, and this process may continue until student exceeds the educational age limit if he could not get enough wealth to pay educational expenses. This segregation projects an uneducated and illiterate society portion based on the standard deviation of people living below economic threshold level.

### 3.11.2 Design Concept

**Emergence:** The PSSMM is built to explorer the migration and segregation of students in schools. In this model, there are two emerging phenomena. Firstly, students' migration from schools having static TIP to schools having dynamic TIP. Secondly, the segregation of students based on their socioeconomic status. Students having better socioeconomic status will get better education and students belong to poor families would tends to be less educated

**Adaption:** Student having enough resources to study in private school will adapt the behavior of migration from public schools to private schools. This adaptation behavior helps students to survive in the practical life after getting degree from the school.

**Fitness:** All students and schools are assigned a fitness value during initialization. These values are drawn from pre-defined random distribution and not subject to change during the remain course of running the model. The fitness values of the students' agents are determined by two variables: students' grades and student wealth, and the fitness of schools are the number of enrolled students.

**Sensing:** Student agent sense its grades, school, growth rate, total wealth, and ranks of all other schools. The school agent can sense its number of teachers, students, rank, class strength, and wealth of students etc. "Information gathered by students about schools is certain and complete. School only sense the total number of teachers, and students and their grades".

**Prediction:** Student agents predict the expected grades and wealth usage in a school and then get enrolled. The school agent predicts about class size by induction of teacher in the school.

**Heterogeneity:** Students and School agents are heterogeneous with respect to each other and within context. The heterogeneity of students comes from the fact that all have a random growth rate that determines their success in the model. The school agents are heterogeneous with respect to the teacher induction policy.

**Stochasticity:** Random number teacher, from a given range, are allocated at the start of the model. In addition, students are initialized with growth rate according to the random distribution.

### 3.11.3 Detail

#### Initialization

The environment of the model is initialized by creating a grid of 250 × 250 cells. There are total six schools and one thousand students created during initialization. All schools are placed at equal distant from each other's. As there are only six schools at the start, schools are placed in three rows and two columns with equal distance. Each school is initialized with a random number of teacher ranges from 5 to 10. Student agent are enrolled in schools randomly during initialization. However, specific parameters — age, growth rage — of each student are taken a pre-defined random distribution. Each student gets color of the school in which enrollment taken placed.

#### Input data

Following initialization, the conditions of environment remain constant with respect to space and time in PSSMM. There is no model spatial, imposed or temporal heterogeneity. So, in this model further data is not required when the model simulation started once. There is no exogenous input data required for the model during simulation.

#### Sub models

***Enroll students***

Each school enrolls students at the start of the educational session. Before opening admission, a school declares its educational policy and resources for education. The educational policy is based on teacher induction policy, grading scheme, total number of teachers, students in that school. Based on these preferences, a student decides to get enrolled in to the school. Students' admission in a school is based on their previous grades, and socioeconomic status. A school preferred to enroll students with better SES and grades.

Total number of students in any school at year t is given by.

$$St_y = St_{y-1} + \sum St_{enrolled}$$

### *Pay Fee*

The student will pay a fee to the school. Every school charges a specific fee from the students. As in public school, a minimal amount with respect to tuition is being charged, but in this model, all other living charges are also considered. The total wealth of the student is reduced with respect to the fee of the student. On the other hand, wealth will be increased according to the growth rate of the students. Students get admission into the schools where the fee is in an affordable range. After fee payment student have following wealth in order to sustain.

$$W_s(t) = \delta + W_s(t-1) + Gr_s - F_s$$

Where $W_s(t)$ is the total wealth of student $s$ at any time $t$, $\delta$ is the proportionality constant, $W_s(t-1)$ is the total wealth of student $s$ at any time *t-1*, $Gr_s$ is the specific growth rate of student, and $F_s$ is the fee of the school in which student is registered.

### *Teach Student*

A school provides teaching facility to its students. The whole process is composed of three steps.

1. Get fee
2. Conduct classes
3. Take exams and award grades

After enrollment, a school charges fee from the students to get managed its expenses and run its business. Total wealth $W_t$ of any school is the sum of fee charged from the students excluding the expenses in term of teacher fee and non-salary budget.

$$W_t = W_{t-1} + \sum Fee_t - \sum E_t$$

Where $\sum Fee_t$ is the total fee collected from thnbe students, $\sum E_t$ is the total expense of the school, and $W_{t-1}$ is the total wealth of the school before that year.

***Get Grades***

The student will get grades according to the study hours. Study hours includes study at home and study at school. At school, study means to study in class. At home, study means study by home tuition or the homework assigned. The school will award grades to students after each year. A student must have to retain specific grades to remain in a school. Let $Gr_s(t-1)$ be the grades of an student at time *(t-1)*, $R_s$ be the Study hours at school, and $H_s$ be the study hours at home, then current grades $Gr_s(t)$ of student can be calculated as, by taking $\kappa$ as proportionality constant.

$$Gr_s(t) = \kappa + Gr_s(t-1) + R_s + H_s$$

***Rank School***

In this process school is ranked according to the strength of teachers and students in the school. Ranking of the school is also based on the teaching activities of the school, SES of students, and the grades awarded by the schools.

Mathematically, the rank of the school is calculated as

$$R_{a(t)} = R_{a(t-1)} + \frac{\sum T_{a(t)}}{\sum St_{a(t)}} + G_{Avg(t)}$$

where $\alpha$ the random factor to neutralize the factors that are not included in the model to make it simple. It varies between range 0 to 1. The $R_{a(t-1)}$ is the rank of school before new ranking, $\sum T_{a(t)}$ is the total teacher at any school *(a)* at any time $t$, $\sum St_{a(t)}$ is the total students at any school *(a)* at any time $t$, and

$$G_{Avg(t)} = \frac{G_{Avg(t-1)} - \sum G_{Avg(t)}}{\sum (St_t - St_{t-1})}$$

In PSSMM, we rank school with respect to students' strength in school, grades of respective students, and the teachers present at the time of ranking. In other words, we consider the two major factors of school performance that determines its ranking are class strength and grades awarded to the students.

### *Induct Teacher*

During initialization, teachers are inducted scholastically. After that, every school induct teacher according to the teacher induction policy.

### *Teacher Induction Policy*

Teacher induction is the main process to affect the performance of a school. For teacher induction, every school has specific teacher induction policy. Based on this teacher induction policy, school induct teacher to teach student effectively. Teacher induction policy is a relative term. According to the time span of induction, teacher induction policy can be divided into two types: dynamic teacher induction policy, and static teacher induction policy.

- *Dynamic Teacher Induction Policy*: Dynamic teacher induction policy, intuitively, is defined as immediate induction of teachers to the school according to needs. However, in the real circumstances, it is not possible to induct a teacher into the school due to fluctuation in demand and supply of human resources and process of induction. Due to this fact, dynamic teacher induction is considered as a relative term instead of considering it as absolute. In this sense, a school inducting a teacher in lesser time span as compared to another school who is inducting a teacher after long time span is considered as following dynamic teacher induction policy.
- *Static teacher induction policy:* Static induction policy is opposite of dynamic induction policy. As it has been discussed before that teacher induction policy is a relative term, so, we can say that, in static teacher induction policy, teachers are being inducted into the school after relatively longer time. In addition, this time does not, correctly, relate to the needs of on time induction of teachers for school.

Mathematically, absolute meaning of dynamic and static policy can be formalized as under

$$TIP_t(dynamic \mid static) = \begin{cases} TIP_t(dynmaic) & lim_{t\to 0} \\ TIP_t(static) & lim_{t\to\infty} \end{cases}$$

### *Migration to Another School*

A student may have to leave school, if there is either lack of wealth, or deficiency in grades. If a student becomes deficient in money, then it is mandatory for that student to leave the school. However, a student is still able to switch to a school where a fee is in the affordable range. Nevertheless, if a school could not find a school whose fee is not in affordable range, then a student cannot enroll in any school. In such a case, a student is declared as out of school children. Such a student will earn money according to growth rate, and will analyze the list of affordable school each year. If there would be enough money with respect to lowest school fee, then this student would be able to enroll in that school. To get enrolled in any school at any time trigger, a student must have to fulfill all other requirements along with these requirements.

The student will migrate to other school either if there is another better school, or student is struck off due to lack of fee or deficiency in grades, as discussed before in the above paragraph. So, migration can take place in two cases. In the first case, a student switch to a school which is with the higher fee and better ranking scheme. Secondly, a student gets structed from the school due to a deficiency in grades or fee, then re-enrollment takes place in the affordable school. In the third case, when the re-enrollment does not take place, student is considered as out of school. The formal representation of the whole process is as follows.

$$M_s(S_a) = \begin{cases} M_s(S_a) = 0 & if(W_n > F_n \; and \; R_n \geq R_b) \\ M_s(S_b) = 1 & if(W_n > F_b \; and \; R_n < R_b) \\ M_s = -1 & if(W_n < F_n) \end{cases}$$

- $M_s(S_a) = 0$ means that student would stay in its old-school $a$, and would not migrate.
- $M_s(S_b) = 1$ means a student get another better school. Student would migrate to school $b$ from school $a$.
- $M_s = -1$ means that student could not find any school to whom admission is possible. Student is now out of school.

# Chapter 4

# Results and Discussion

In this chapter, results will be presented. Results contain verification of the following hypothesizes using descriptive agent-based modeling of primary school student migration phenomenon:

- Descriptive Agent based Modeling (formal specification and complex network analysis) of agent-based model makes the model easier for comprehension, comparison, and replication.
- Students' migration from public schools to private schools can be mitigated by reducing class size in public schools by hiring teaching staff dynamically and instantly as per need.
- Socioeconomic status of students and the difference in educational quality in public schools and private schools are collectively causing poor-rich students' segregation in both sectors' schools, which can be reduced by enhancing the quality of education in public schools.

## 4.1 Models Comparison

In this section, we present a comparison of both these presentations. We proved that using the traditional textual protocol completeness at such a level is difficult to achieve or even impossible in some cases. We conclude that a verifiable and scalable model is only possible using descriptive agent-based model instead of ODD protocol. We started by presenting the pros and cons of the ODD description of the PSSMM. This presentation contains the detail description of the advantage of describing the model in ODD followed by short comings brief description of ODD. In the third section, the fitness of the DREAM was compared with already presented ODD. In the last, the comparison is concluded, and future work is presented.

### 4.1.1 Reflection of ODD on PSSMM Specification

Even though similar models as PSSMM have been presented using ODD protocol. However, for the sack of one to one comparison, and verification of the descriptive description of PSSMM, the need of standalone and specific presentation using ODD protocol was mandatory. Following section contains the shortcoming of using the ODD standard protocol presentation of PSSMM.

**Ambiguity:**

The PSSMM model specification following ODD is ambiguous. The main reason for ambiguity is the textual description. Textual presentations are inherently ambiguous as naturally developed languages have much flexibility for presentation. It is not possible for the user to communicate exactly what they want while using textual format. User from different region and background can extract different meaning from the same text. This is not specific to the ODD presentation, but it is a general fact.

**Redundancy:**

ODD protocol is highly redundant. Several arguments have stepped in to prove the need of redundancy in the protocol. Either due to the need of repetition of information, or due to the lack of formalization, the redundancy makes it difficult for presentation and comprehension. This redundancy makes the reader very chaotic, and it becomes hard to communicate the intents of modeler. For example, the ‘overview’ and ‘detail’ sections are highly cohesive and interdependent due to redundancy. Additionally, it is not easy for a user to define the exact domain information to be present in the section scheduling and sub models.

**Absence of Meta Model**

ODD has been just a checklist of the components needed for the so-called complete presentation of the CAS model. It does not contain a meta model that is necessary for the modeler to communicate with the reader. In PSSMM, the ODD presentation does not assist the reader in this manner. The reader must have to make a virtual meta model to understand the model. The replication of the model is beyond the sense if the comprehension becomes a difficult task. The need of meta model is further elaborated in the reflection on a replication section in detail.

**Strong Cohesion Between Data and Meta Model**

CAS presentation needs to separate the data from the model structure when there is a heavy use of field data. We cannot find any help in this regard from ODD. Infect, ODD does not contain the separate section for the data. In PSSMM, we deal with a dense data collected from the students' interaction with teachers, and schools. To handle the situation, that is mandatory for comprehension of the model, the absence of the separate section creates a difficult task for correlating the results. In short, the ODD mix the presentation with respect to 'why' and 'how'.

### 4.1.2 Reflection of DREAM on PSSMM Specification

There has been a lot of critics on ABM that these are presented informally. It is a fact that most ABM are presented in textual form. Even, the most used protocol ODD, for documentation of ABM is also textual in nature. Due to this reason, the comprehension and replication become a severe problem for the researchers. The problem becomes more critical for researchers from different domains having alike meta model for different ABMs. To cope with this problem, a formal approach was needed. The first effort in this regard was DREAM: Descriptive Agent-Based Modeling framework. The intention behind the presentation of DREAM was to cope with the challenges raised because of limitation of ODD. In the remaining section, we will elaborate the main advantages of DREAM.

### 4.1.3 Complex Network

Complex Network of ABM helps the model documentation in the following way.

**Structural Abstraction**

Descriptive presentation starts from translating the mental model into Complex Networks (CN) Model. This CN is purely an abstract representation of the ABM. It is built from the mind map to reflect the overall structure of the model. Going through different stages, a complete complex network contains all structural information for comprehension and replication of the model.

**Comprehension**

Complex network is a pictorial representation of the model. A picture is worth of thousand words. Hence, the complex network presents the comprehensive interface of the structure of the model. It is composed of connected graphs of tree shape started from ABM. It is the initial draft of descriptive model. The graph was presented in a circular layout with ABM at the center, however, in the extended form of DREAM the complex network is presented in the form of a tree with a tree as root node. This representation helps researchers from different domains to grasp the overall structure of ABM without digging down the code.

**Comparison**

The comparison of two different models is a matter of difficulty. This becomes even more cumbersome when researchers from different domains have to compare similar models from different domains. The pictorial form of ABM — complex network — enhance the understanding of the model which ease the comparison of the models. This CN is specifically useful for comparison of model for researchers from different domains.

**Replication**

The Complex Network of the Agent-Based Model helps in replication of the model. Most often, in ABM, the replication of the models has been possible by regeneration of code from the text based template of the model. However, this replication does not become the exact copy because of ambiguity in the text. On the other hand, naive user must dig down the code to understand and reproduce it. In a complex network of the ABM, this task becomes so easy. The modeler convoys the overall structure of the model. This theoretical abstraction of the model enhances the correct replica of the model.

### 4.1.4 Pseudo Code-Based Specification

In descriptive cognitive modeling, the pseudo code-based specification helps in presenting the model documentation in a complete and clear way. Pseudo code-based specification is semi-formal presentation of the agent-based model. It helps the modeler to present the model's description in much clear and structured way as compared to textual presentation.

### Complete and Clear Presentation

This pseudo code-based specification makes the model complete. Pseudo code-based specification contains the complete set of descriptors of every component of CAS model. It starts from the specification of Agents, going through a detailed mathematical description of global variables, procedures, end with specification of experiments. It covers all aspects of model's description using semi-formal presentation.

### Structured Presentation

The Dream provides a way to present the model in a structured way. This structure includes complex network, pseudo code-based specifications, and agent-based model. The pseudo code-based specification is the comprehensive and structured representation. Reader from different domains can understand the model by understanding the predefined structure of the model. It increases the level of comprehension and readability.

### 4.1.5 Agent-Based Model

This is the last major step of descriptive cognitive model development. It contains the construction of ABM. In this phase the ABM is developed, imposed and the simulation experiments are carried out. Based on a complex network of the ABM, and derived pseudo code-based specification, the ABM would contain all necessary detail for model simulation experiments and their replication.

### Simulation Experiments

This section contains the complete detail of experiments being done in the model. It is the core of the model. It reflects the output of the model. We would present it at the end of the model. It contains data and figure after manipulation of the data. It may contain the output global variables. It helps in determining the emergent behavior of the complex adaptive systems.

Table 4.1 shows the comparative analysis of DREAM and ODD.

**Table 4.1: Comparative analysis of ODD and DREAM**

| No | Feature | ODD | DREAM | Remarks |
|---|---|---|---|---|
| 1 | Formalization | No | Yes | DREAM is a formal specification while ODD is textual based specification. |
| 2 | Completeness | Lesser | Complete | Due to textual nature ODD, does not ensure Completeness of the presentation. On the other hand, due to pseudo code-based specification DREAM describe the model completely. |
| 3 | Non-Ambiguous | Lesser | Good | DREAM description is non-ambiguous as compared to ODD. |
| 4 | Clearness | Lesser | Better | DREAM presents the specification clearest. |
| 5 | Community | Large | Lesser | ODD is more popular in the community than DREAM. |

## 4.2 Complex Network Analyses of the Network sub-model

In complex network analyses, different centrality measures of the complex network of agent-based model will be presented. Complex network analysis is performed on the random network, then centrality measures are calculated, and corresponding graphs are drawn. The Figure 4.1 shows a simple sub-network of the ABM.

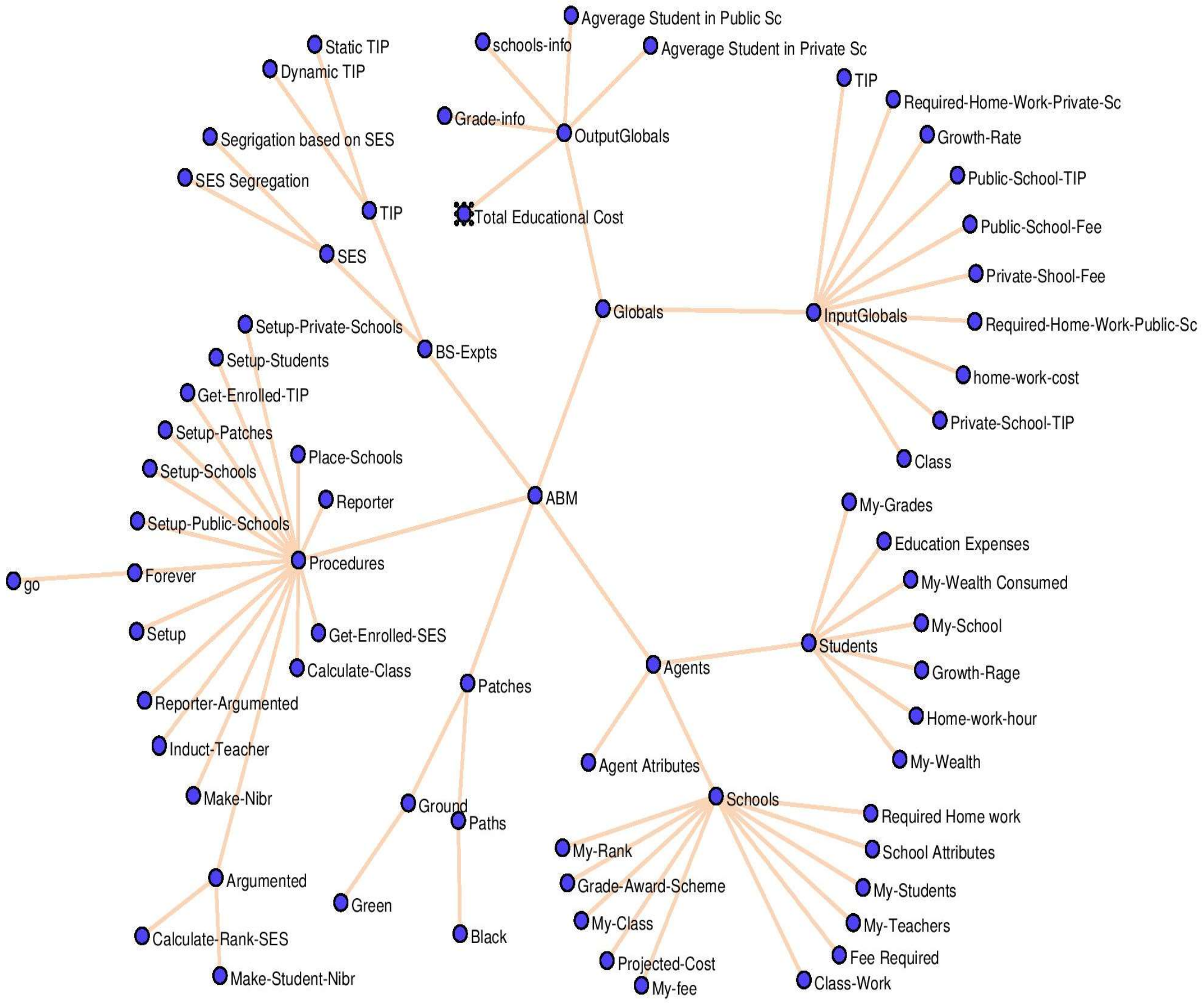

**Figure 4.1: ABM Sub Network**

### 4.2.1 Centrality Measures: Betweenness Centrality

Betweenness of a node is defined as the frequency of a node that exists on the shortest paths in between other nodes. It is a measure of the ability of any node to monitor the communication over the network, due to its occurrence in the path. In Figure 4.2 nodes are colored and re-sized according to the betweenness centrality measure of the nodes in the sub model.

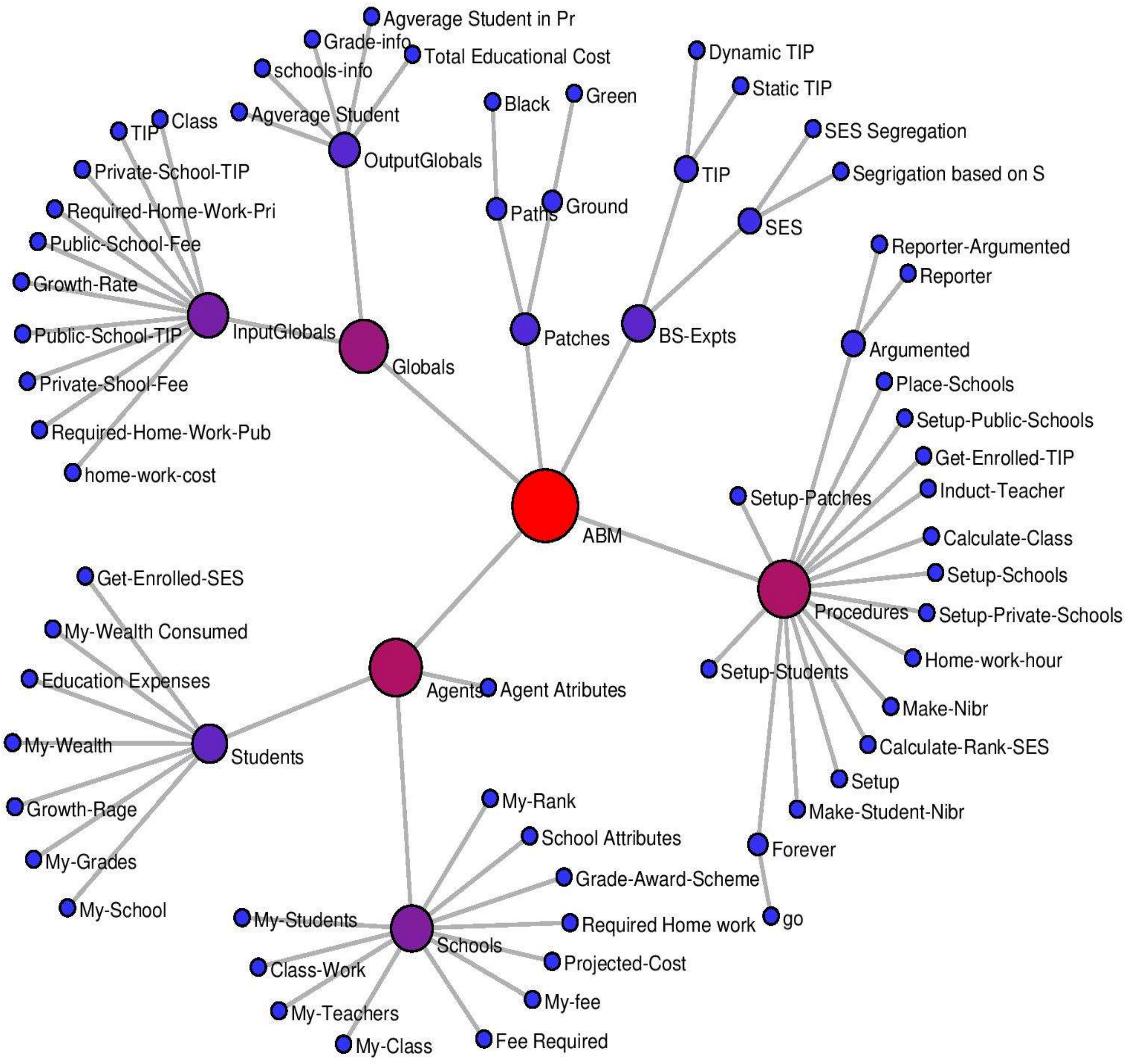

**Figure 4.2: Betweenness Centrality of Complex Network of Agent Based Model**

Bigger sized and darker red colored nodes show higher degree measure. Contrarily, smaller size nodes with blue color are the representation of a lesser value of betweenness centrality of that node. It can be observed that node ‘ABM’ in the model is darkest and have a bigger size in Figure 4.2, this shows that AMB is situated in the network in such a way that communication over the network is most visible to it. In Figure 4.3 the ABM has the highest peak, which also means that it is the most central node in the network. Other nodes with higher centrality are global variables, procedures and student agent itself. The higher centrality of these nodes is a clue of how dense these nodes are. This helps in demonstrating a comparison between two different ABM from the different domain without knowing the code of the respective model. Quantitative display of betweenness centrality of the nodes in the sub model is presented in Figure 4.3. Nodes with highest betweenness centrality are ABM, Globals, Agent attributes, and procedures.

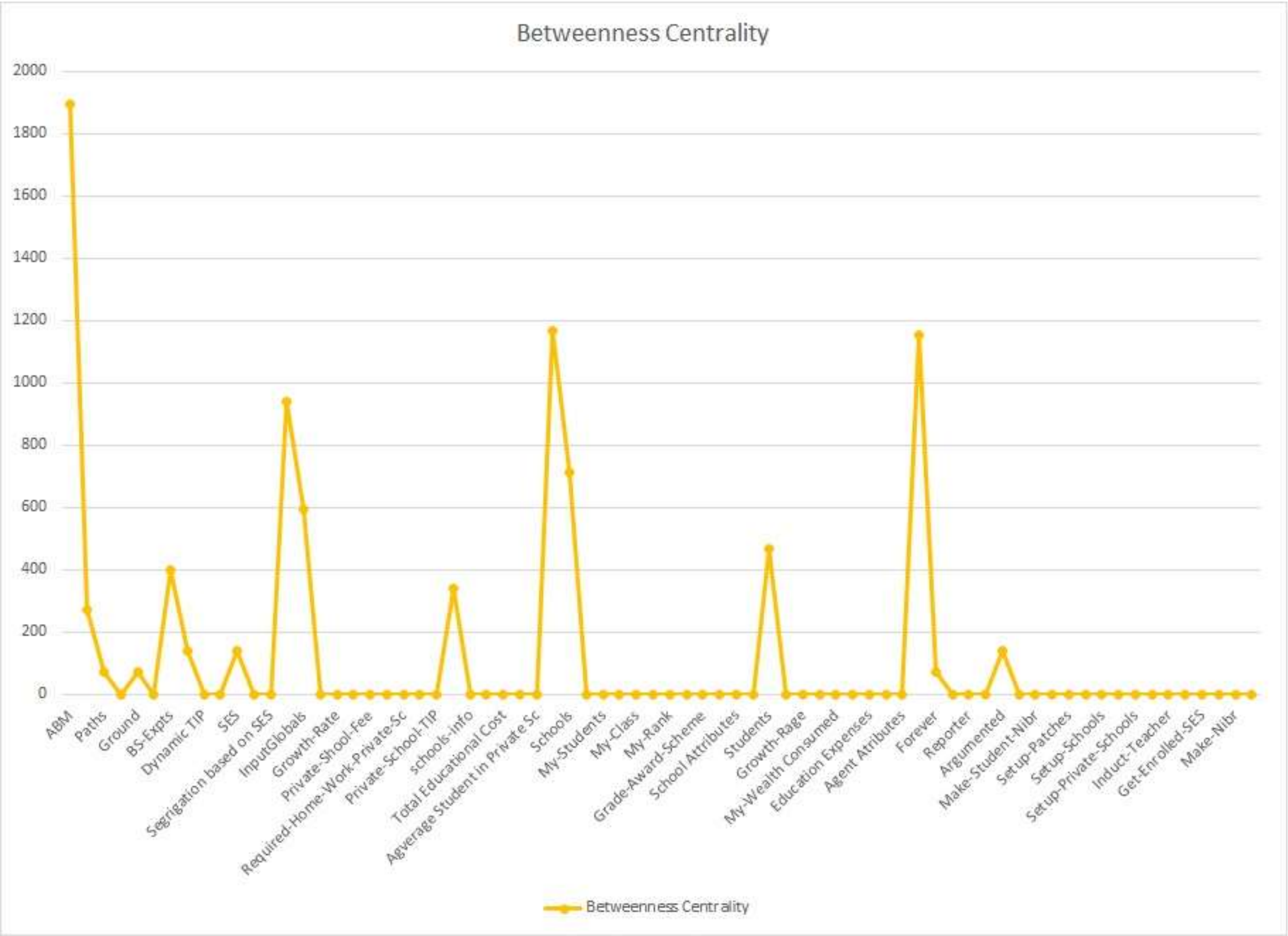

**Figure 4.3: Betweenness Centrality Measures Histogram**

### 4.2.2 Centrality Measures: Degree Centrality

Degree centrality is a measure of a node's connections within a network. It shows the number of connections a node has with other nodes. In a directed graph, degree centrality is classified as in-degree (number of nodes directed toward the node) and out-degree (number of nodes the node directs to). In an undirected graph, degree centrality is simply the number of nodes connected to the node. In Figure 4.4 sub-network regarding degree of nodes is presented. As the ABM network in PSSMM is undirected, we only consider the total number of connections for a node. The procedures, input Globals, and school has the most connections, so have the highest degree centrality.

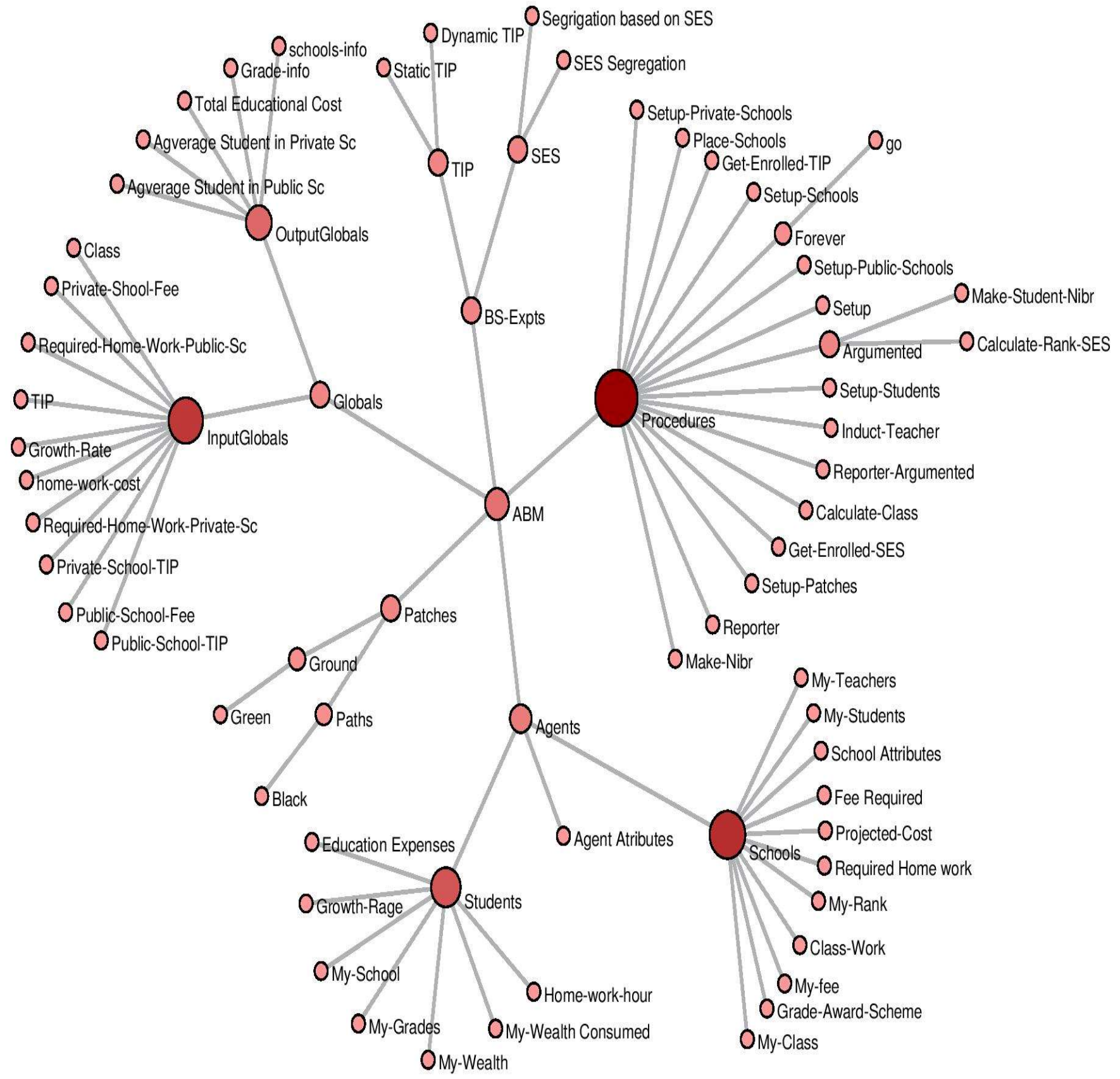

**Figure 4.4 : Degree Centrality of network sub model**

The network is colored and resized with respect to degree centrality. Nodes with darker red color and bigger size show a higher value of the degree centrality measure. It means that darker and bigger nodes are more connected in the network. The degree centrality analysis of the network depicts that model code concentration and logic strength areas. The Procedure, schools, and global variables are most connected to the network, and hence, show the most important role in the model. Degree centrality of the network sub model is shown in Figure 4.5. Nodes input global, school agent, student agent, and procedure have a much higher degree centrality. These peaks are observed due to the higher connectivity of the nodes in the model. The quantitative representation of the degree centrality provides a more detailed and categorical view of the nodes' characteristics of ABM.

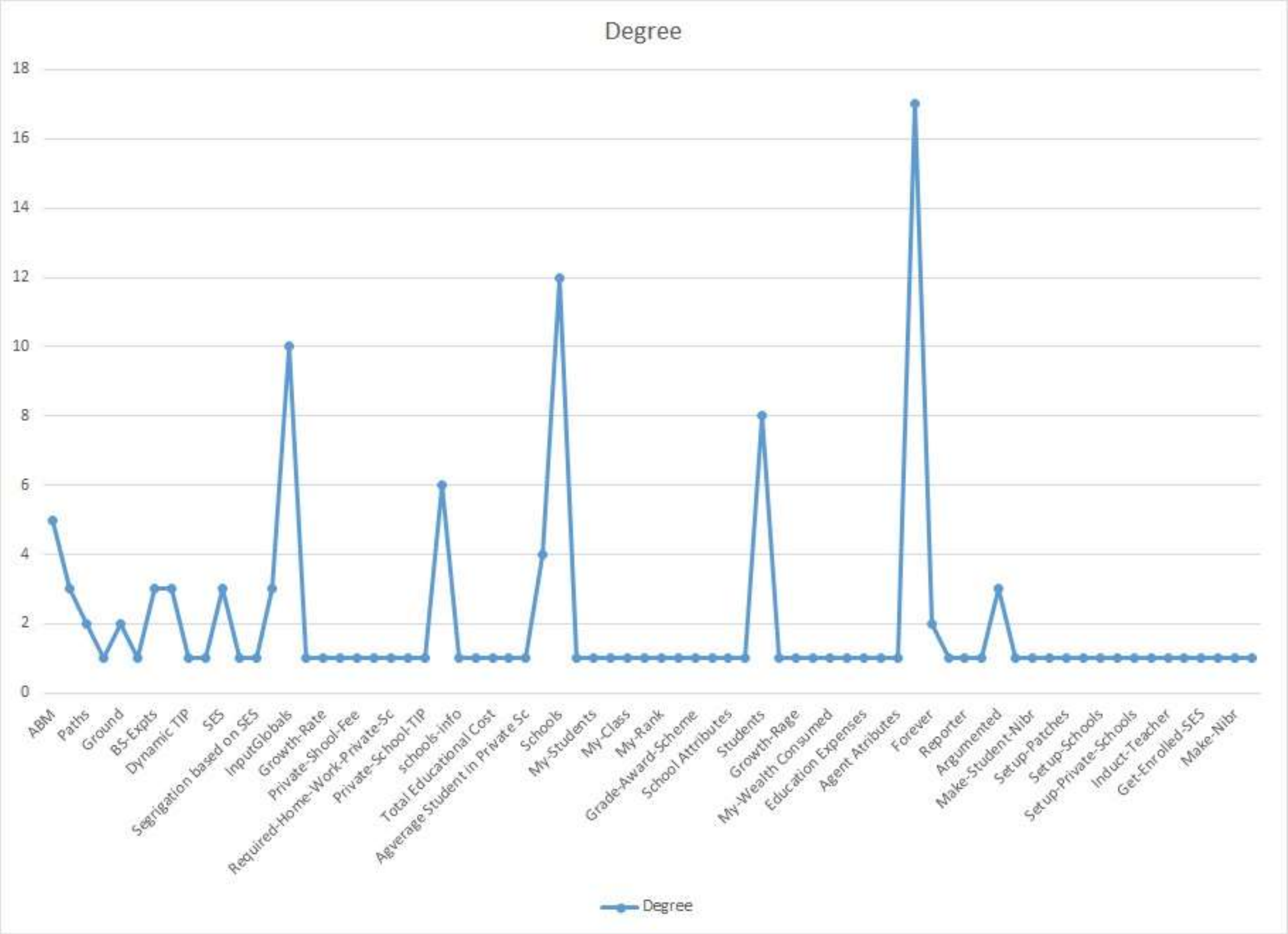

**Figure 4.5: Degree Centrality Histogram**

### 4.2.3 Sub-Tree Network of model

Tree network is an extension to DREAM to make it more expressive, easy to comprehend without digging down the code, and replication of the model across cross disciplines for research. In Tree network of the agent-based model, base node of the tree is ABM, from where different branches drop like procedures, agents, and patches. The branch Agents, and procedure have the most depth and hence contain the most functionality of the model. In Figure 4.6 the sub-tree network model is presented.

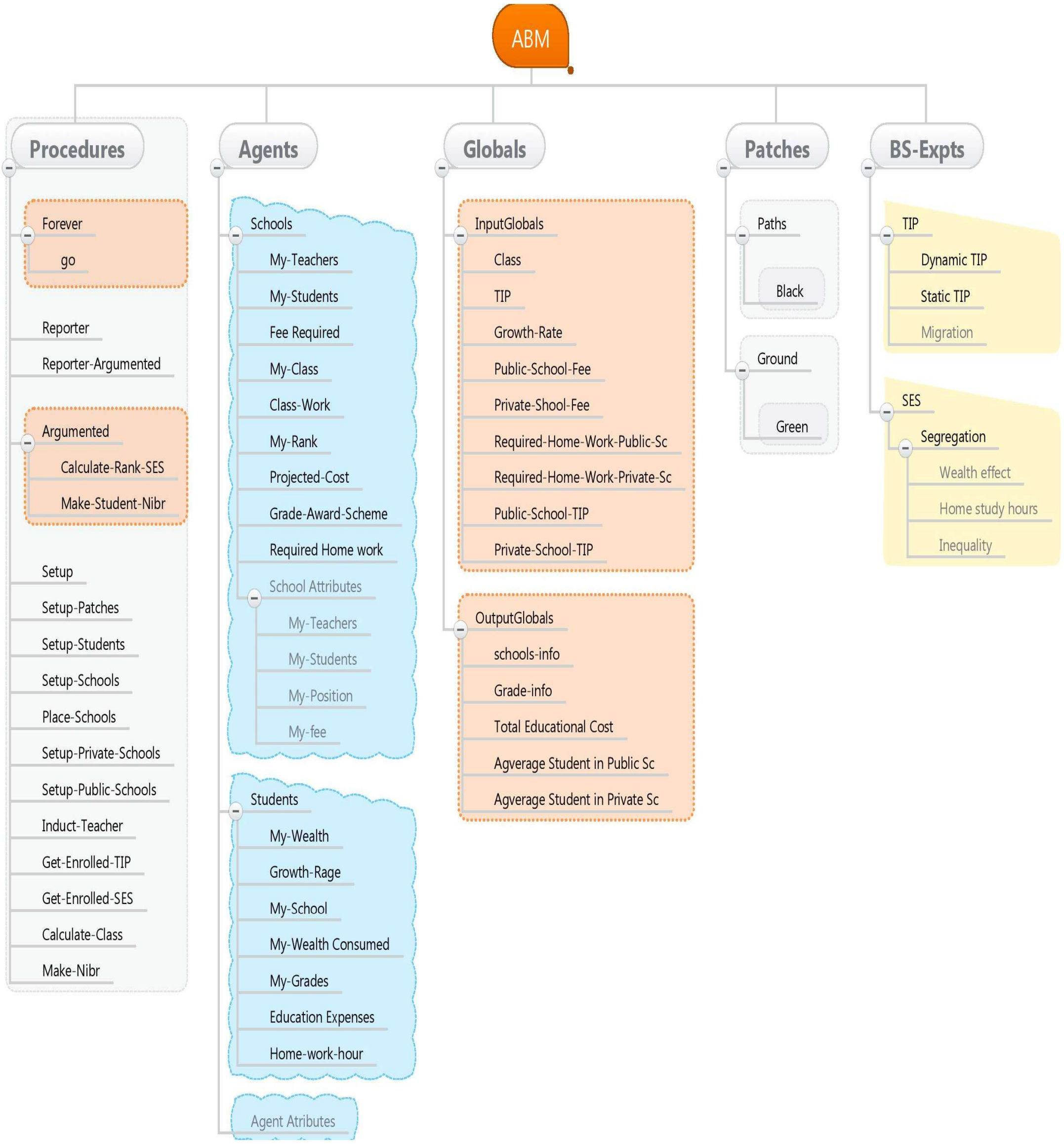

**Figure 4.6: Tree Layout of Network**

## 4.3 Quantitative Results

We did two experiments to prove two main hypotheses. Firstly, students migrate from schools with larger classes to schools with smaller classes. Secondly, students are segregated in between, public and private schools based on socioeconomic status due to the difference in educational quality in public and private schools. The first experiment was related to the migration of the students from public to private school due to class size, and the second experiment proves the segregation of students in public and private schools based on SES. In this section, we present the detailed analyses of quantitative values for each of these experiments.

In the first experiment, we vary the class size of public and private schools to test the hypothesis that students would migrate from schools with larger classes to smaller classes. Descriptive statistics for the first experiment are shown in Table 4.2↓. We can note that total simulation step for this experiment was 0.1 million. Total simulation runs were 100, and each simulation step would carry for 100 steps, so total 10,000 steps. Each run is executed 10 times to minimize the stochastic effects. Hence, a total number of run-numb steps become equal to 1000,000. A total number of schools in the vicinity are six, a total number of students in the vicinity are 250 and each school is allocated with the teacher from a random distribution of 7-10. Three schools are from the public-school sector and three schools are from the private school sector. Each school from respective sector hire a teacher, according to the teacher induction policy specifics detailed. Each school in the sector hires a teacher from one-time interval to 96. Each time interval supposes to be equal to one month. All other variables are assumed to be constant.

In the second experiment, required **study-work** at home are considered only to vary, all other variables are kept constant. Total "Run-numb" are 0.1 million. The simulation step for each run are 100, and we are varying two variables from 1 to 10 with a difference of one, so total runs in one experiment are 100. So, total running steps in one experiment become equal to 10,000. Now, each experiment is repeated ten times, so, for the whole experiment, the total number of running steps becomes equal to 100,000. The Table 4.3↓ shows the exact descriptive detail of the second experiment.

**Table 4.2 : Experiment one descriptive statistics**

| | N | Minimum | Maximum | Mean |
|---|---|---|---|---|
| Run-numb | 100,000 | 1 | 100 | 50.5 |
| Schools | | 1 | 6 | 3.5 |
| Students | | 250 | 250 | 250 |
| Steps | | 0 | 100 | 50 |
| Public School Class | | 1 | 96 | 48.5 |
| Public School Class | | 1 | 96 | 48.5 |
| Migration Index % | | -61.348 | 32.560 | -14.394 |

**Table 4.3 : Experiment two descriptive statistics**

| | N | Minimum | Maximum | Mean |
|---|---|---|---|---|
| Run-numb | 100,000 | 1 | 100 | 50.5 |
| Schools | | 1 | 6 | 3.5 |
| Students | | 250 | 250 | 250 |
| Steps | | 0 | 100 | 50 |
| Public School RSH | | 1 | 10 | 5.5 |
| Public School RSH | | 1 | 10 | 5.5 |
| Segregation Index % | | 3.154380 | 3.925 | 3.539 |

## 4.4 Migration of Students from Public to Private Schools

In this section, we present the graphs. These graphs will demonstrate the verification of the hypothesis that students migrate to schools having smaller size classes.

### 4.4.1 Visual Output of the Migration of Students

In this section, we described the visual results of simulation in detail. Figure 4.7 (a) is the screenshot of the visual simulation output when the simulation started, while Figure 4.7 (b) shows the visual simulation output when the simulation ended. We colored schools corresponding to their TIP policy and sector of the school. Yellow colored are private schools, while red colored schools are public. This colorization is based on the fact that Private schools have dynamic TIP, and Public schools have static TIP. A student after joining a school changes its color to the color of the school. Students' adoption of color helps in the visualization of migration and segregation apparently.

Figure 4.7 (a) shows a representation of the simulation that was conducted to study the migration of students from public to private schools. At the beginning of the simulation, each school is depicted with a random distribution of teachers. In the first iteration, students are enrolled in the schools based on the number of teachers in each school. The results of the simulation show that over time, the number of students in the red school, which represents a private school, increases while the number of students in the public-school decreases.

This increase in the number of students in the private school is due to the dynamic Teacher Induction Policy (TIP) that is maintained by the private schools. The dynamic TIP results in smaller class sizes, which is a key factor in attracting students to enroll in the private school. As the simulation progresses, the migration of students from the public school to the private school becomes more apparent, leading to a significant decrease in the number of students in the public school and a corresponding increase in the number of students in the private school. This trend is a clear indication of the impact that the dynamic TIP has on student migration and the effectiveness of this policy in attracting students to the private school.

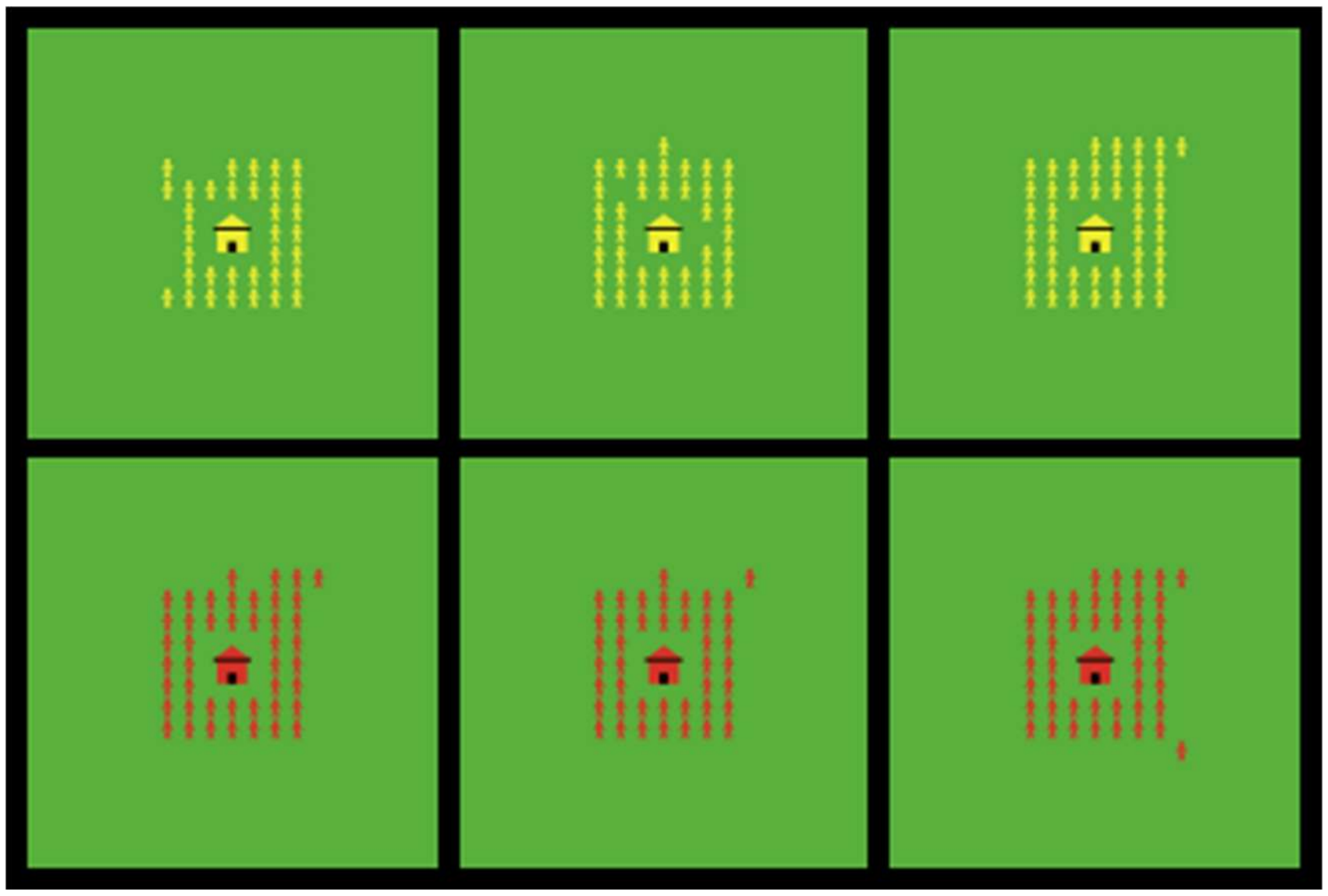

(a) Start view

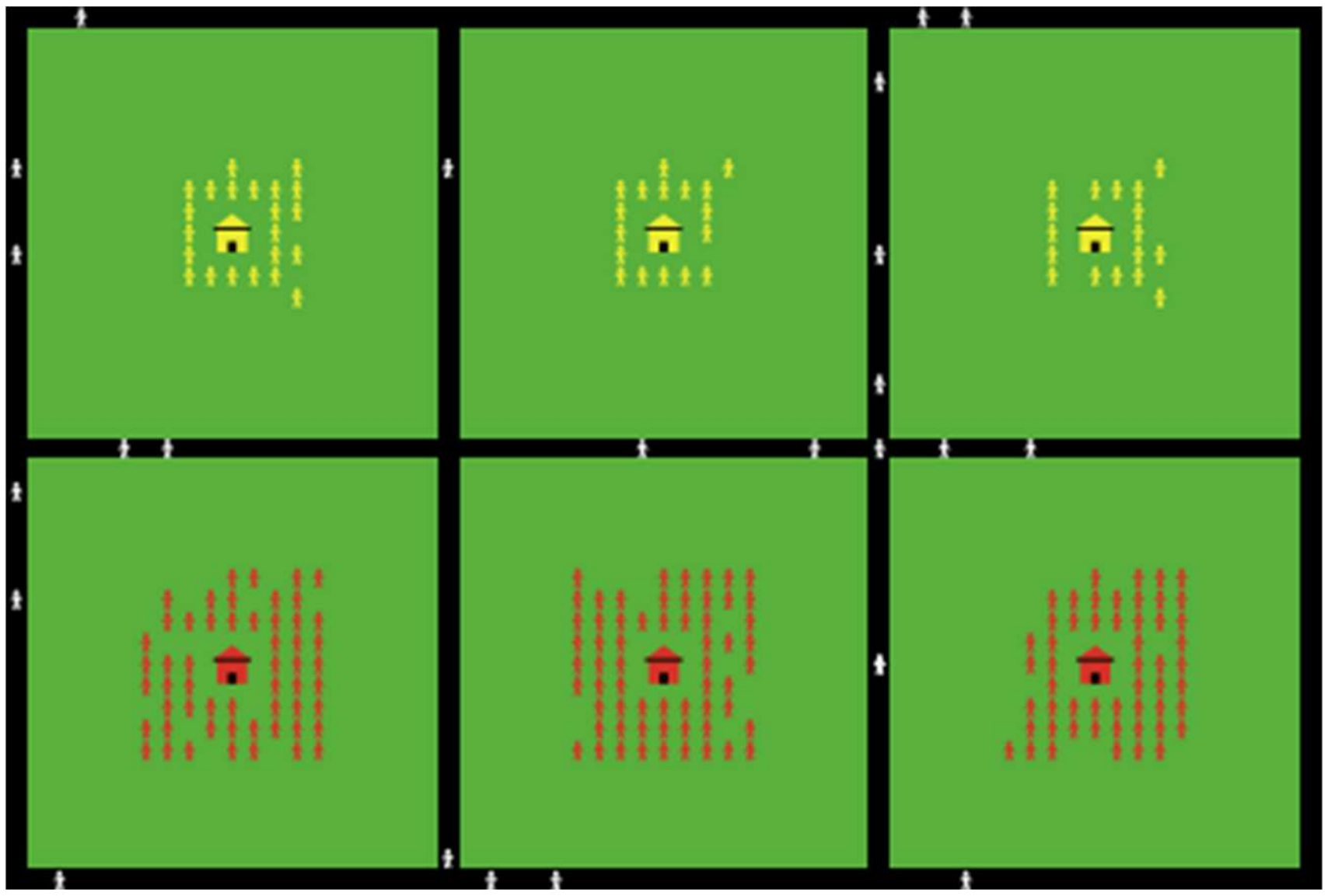

(b) End View

**Figure 4.7: (a) Start view of simulation, (b) End view of simulation**

### 4.4.2 Effect of Class Size in Schools on Students' Migration

Results in this section would prove the following hypothesis

- Migration of students from public to private schools can be reduced by reducing class size in a school.
- Migration of students from public to private schools can be reduced by hiring teachers dynamically.

In Figure 8.8↓ class size in public schools is plotted against migration index with 95% confidence interval. The simulation is done ten times to remove the stochastic effect on the output. The migration index increases with the increase in the class size in the public school. To make the migration index more expressive and visualize the corresponding relationship of public and private schools' class size effect, a variation of private school size is also considered. The effect of class size on migration index is shown in Figure 4.8. To prove the hypothesis, we plotted variation of class size in public school against migration index while each color line corresponds to the class size in private school. The data values plotted in the Figure 4.9 are off 95% confidence interval collected from multiple repetitive experiments to remove the stochastic behavior.

Figure 8.9↓ proves the hypothesis in two folds. Firstly, the migration index increases, within the context of each line, with respect to increasing in class size of the public school. This is due to the fact that students prefer to study in school have smaller classes. As the class size in public school increases student migrate to private schools where class size is lesser as compared to public schools. Secondly, the value of migration index increases when the class size of the private school increases as represented by each line.

As discussed before in detail that class size of a school can only be reduced by hiring new teachers dynamically with an effective rate. Public schools hire teachers in static ways following static teacher induction policy while private schools hire teachers dynamically when the need arises. Due to this fact, private school can maintain smaller class size. This phenomenon leads to mass migration of student from public schools to private schools.

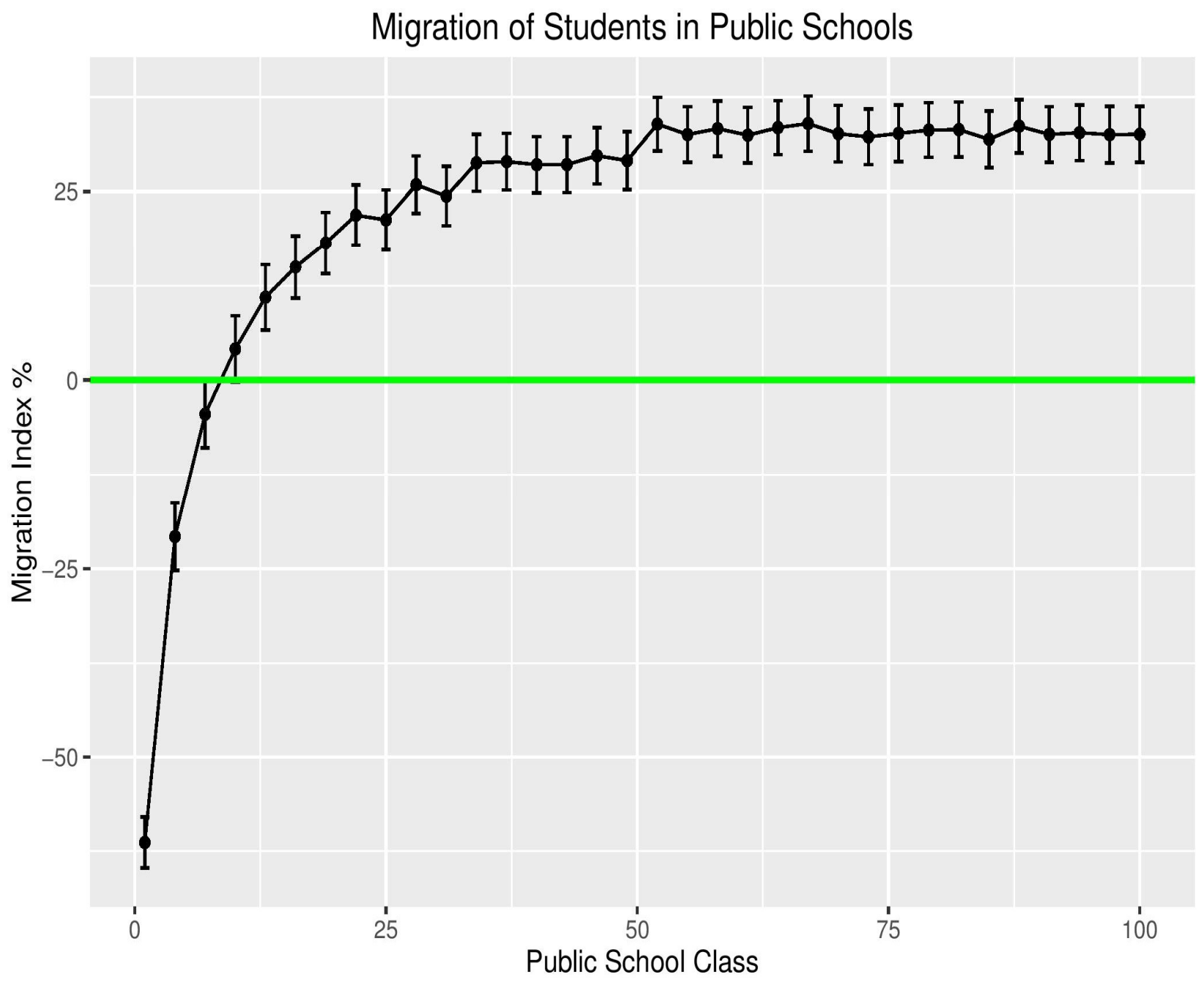

**Figure 4.8: Students migration in Public Schools**

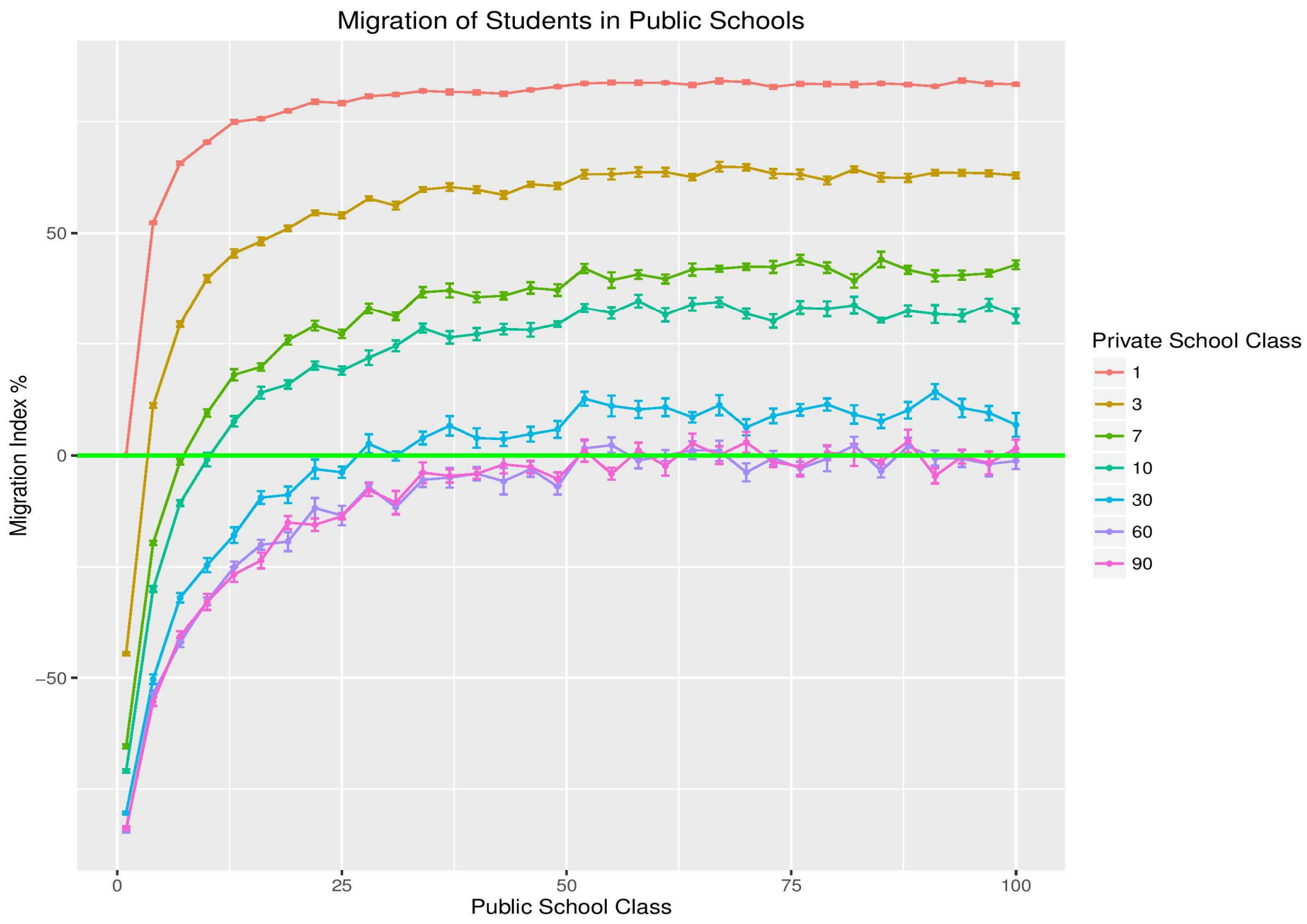

**Figure 4.9 : Migration index with 95% CI**

## 4.5 Segregation of Students in Public and Private School

### 4.5.1 Visual Output of the Migration of Students from Public to Private Schools' Due Class Size Difference

In this section, we give a detailed description of the visualization output of the simulation experiments and discussion on it. The model is represented as a two-dimensional grid of cells containing schools and students. Each cell can carry one student or a school. Schools are shown as building and students are represented by an Actor surrounding schools in which those are enrolled. Schools are colored with respect to the teacher induction policy they follow. For example, yellow colored school is private school following dynamic teacher induction policy and red colored schools are public school following relatively static teacher induction policy. Students are colored with respect to the output of each experiment.

We presented the ratio of students with respect to grades and wealth separately. In figure 4.10 students are colored according to their scores at the end of the simulation experiment. Darker colored actor denotes students with higher scores while lighter color actors denote the students with lower scores. It can be easily observed that in private-schools ratio of students with good scores is high. On the other hand, in public-schools ratio of students with poor scores is high.

In Figure 4.10 (a) students are colored according to the wealth they contain at the end of the simulation experiment. Darker black and red colored actors represent relatively wealthy students while white colored actors represent poor students with relatively lesser wealth. This figure clearly depicts that in private-schools the ratio of wealthy students is high while in public school the ratio of poor students is high.

The visual output of the experiments is evident that students become segregated dual: with respect to score and wealth. It means that students with better socioeconomic status are likely to study in private schools and poor students are forced to study in public school. It also shows that students in public schools (these are wealthy students also) are likely to get good results while students in public schools get poor scores.

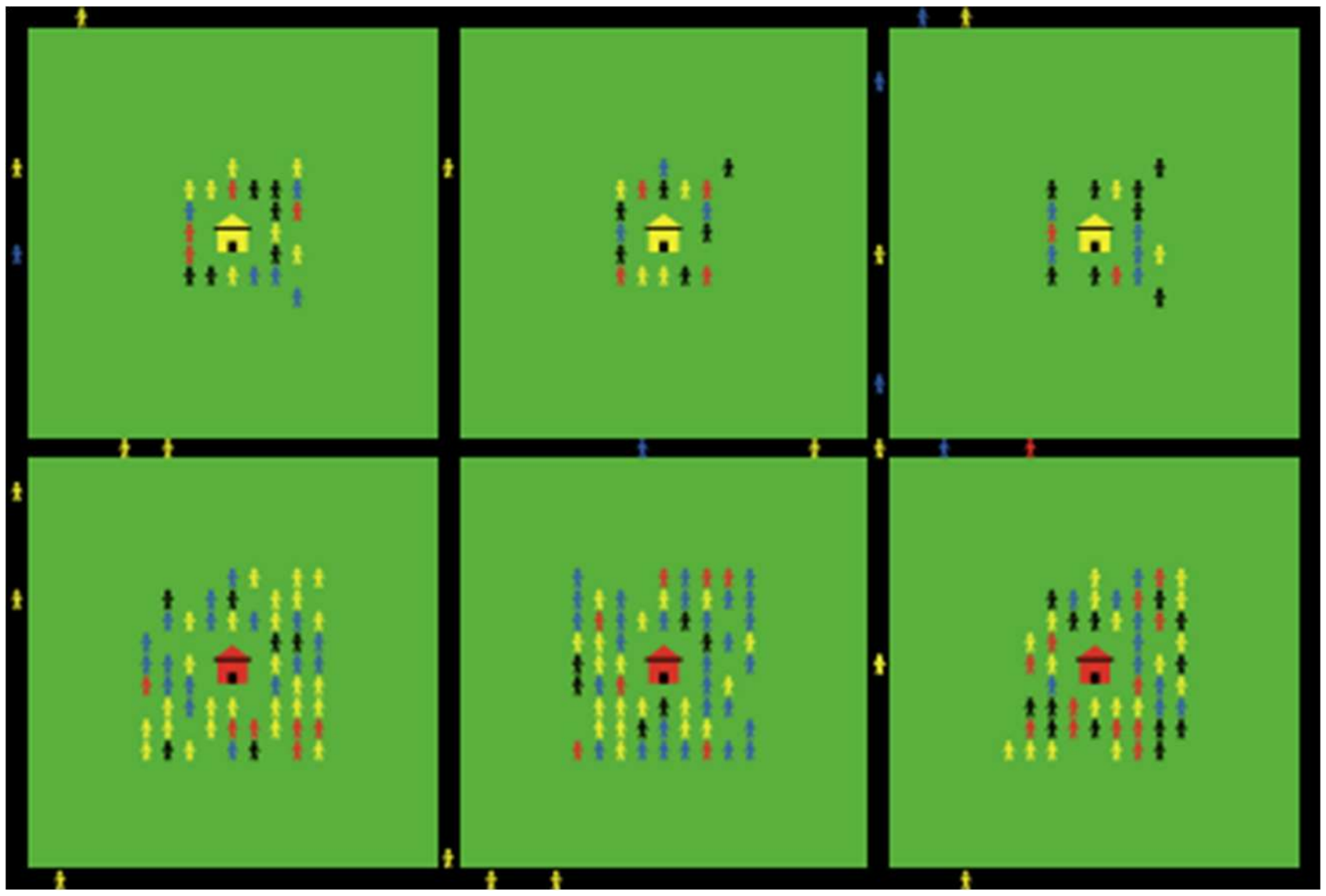

(a) Colorization of students by scores

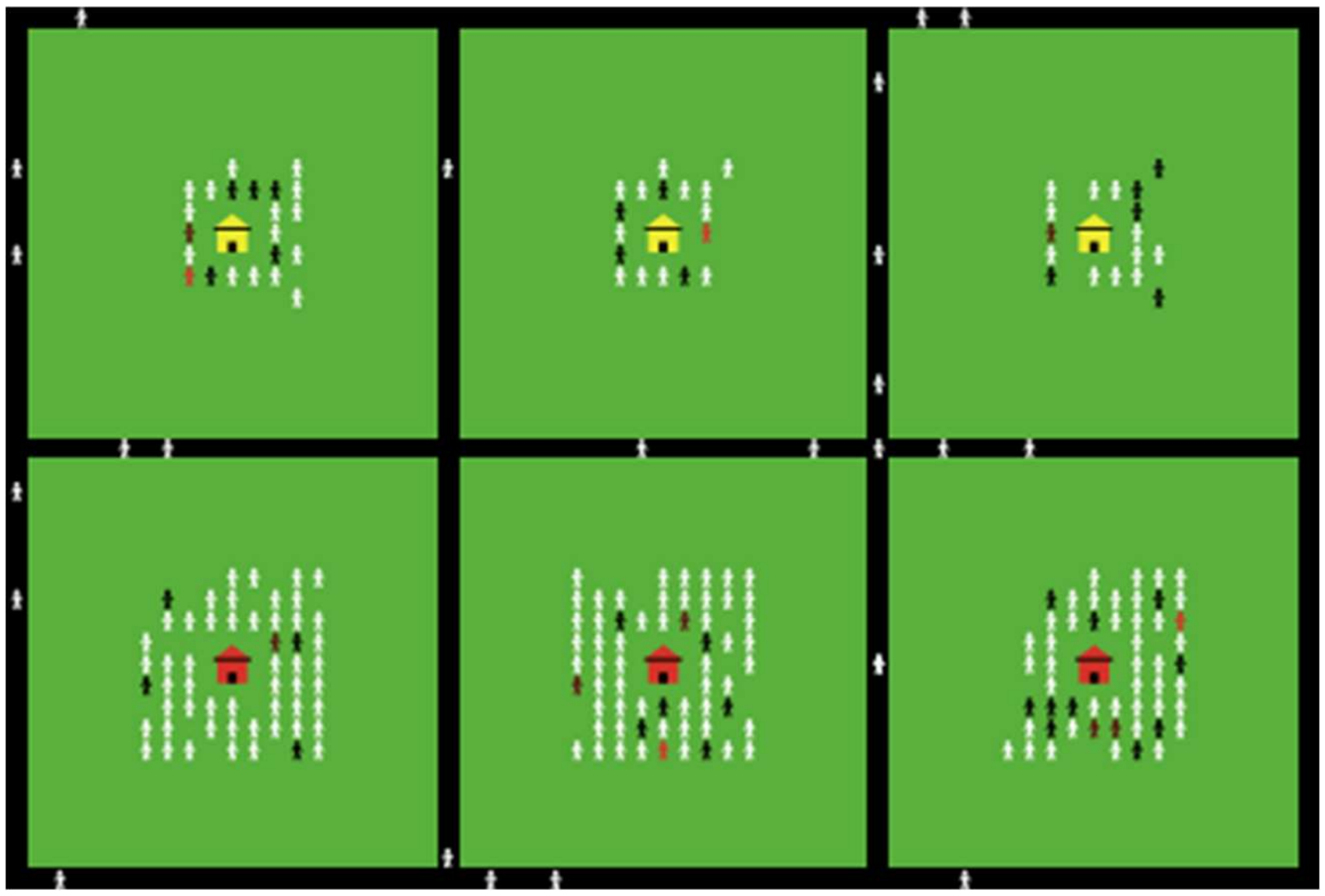

(b) Colorization of students by wealth

**Figure 4.10: Visual output of students' segregation**

### 4.5.2 Effect of Study Hours at Home on Segregation of Students in School

Result in this section would prove the following hypothesis

**Segregation Index (I)**

- Decreases with the increase in educational quality in public schools.
- Decreases with a decrease in the difference of educational quality of a school.

The relationship between required study hours and segregation index in public schools is depicted in Figure 4.11. The figure shows that as the required study hours in public schools increase, the segregation index decreases. This can be explained by the fact that school performance and educational quality are directly proportional to the study hours of students at home. To ensure the accuracy of the results, the simulation experiment was repeated ten times, reducing the impact of random fluctuations in the output. The results consistently showed that as the educational quality in public schools improved, students' migration to private schools decreased.

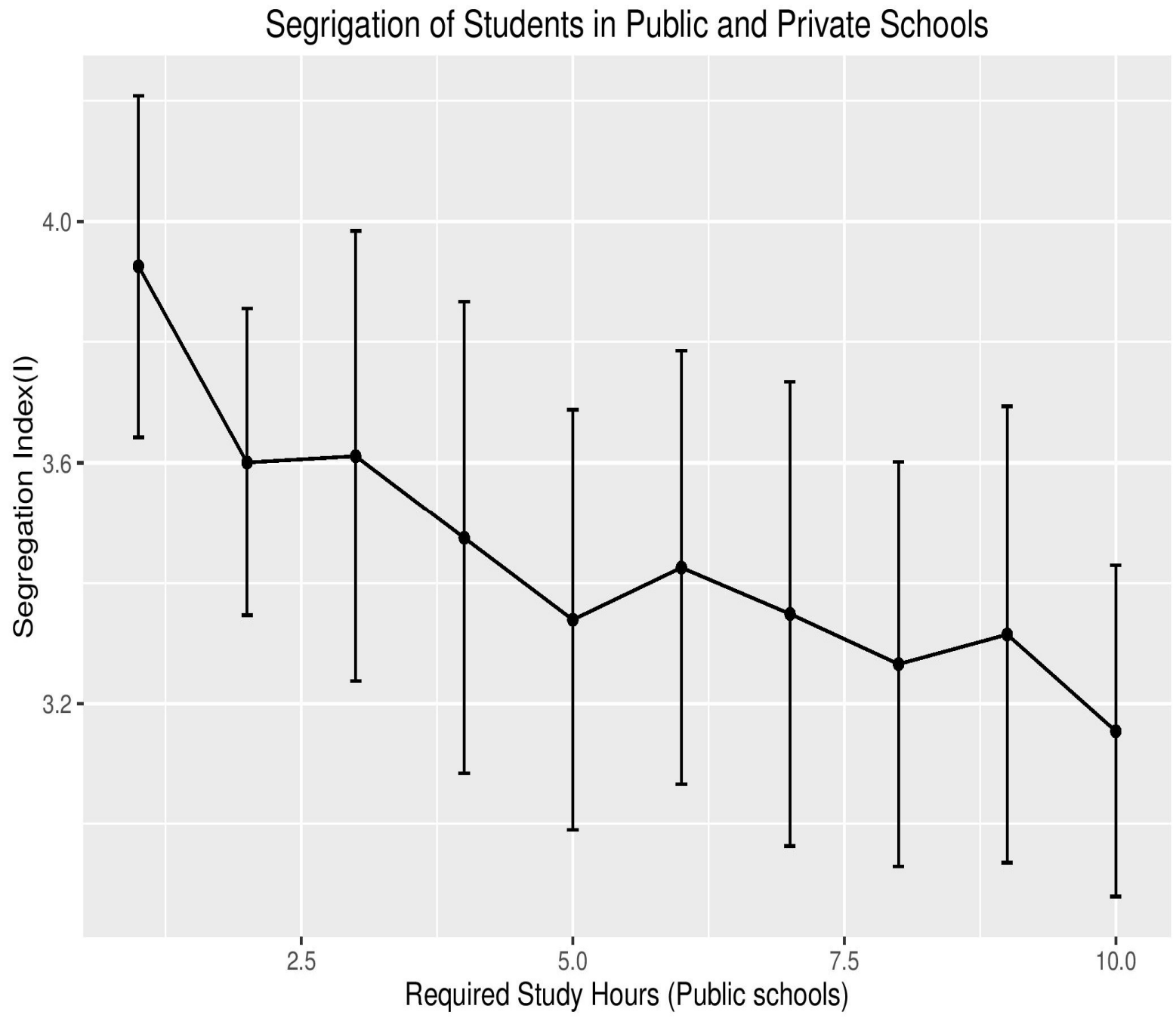

**Figure 4.11 : Segregation Index with 95% CI**

The hypothesis that the segregation of students in public and private schools decreases with the improvement in educational quality in public schools and the reduction in the difference in educational quality between public and private schools is further strengthened through the measurement of the required study hours at home in private schools. Figure 4.12 depicts the decrease in the segregation index of students in public and private schools as the educational quality of public schools in terms of the required study hours at home increases. It is evident that as the difference in educational quality between public and private schools decreases, the segregation also decreases. This supports the hypothesis and provides additional evidence for the relationship between educational quality, difference in educational quality and segregation of students in public and private schools.

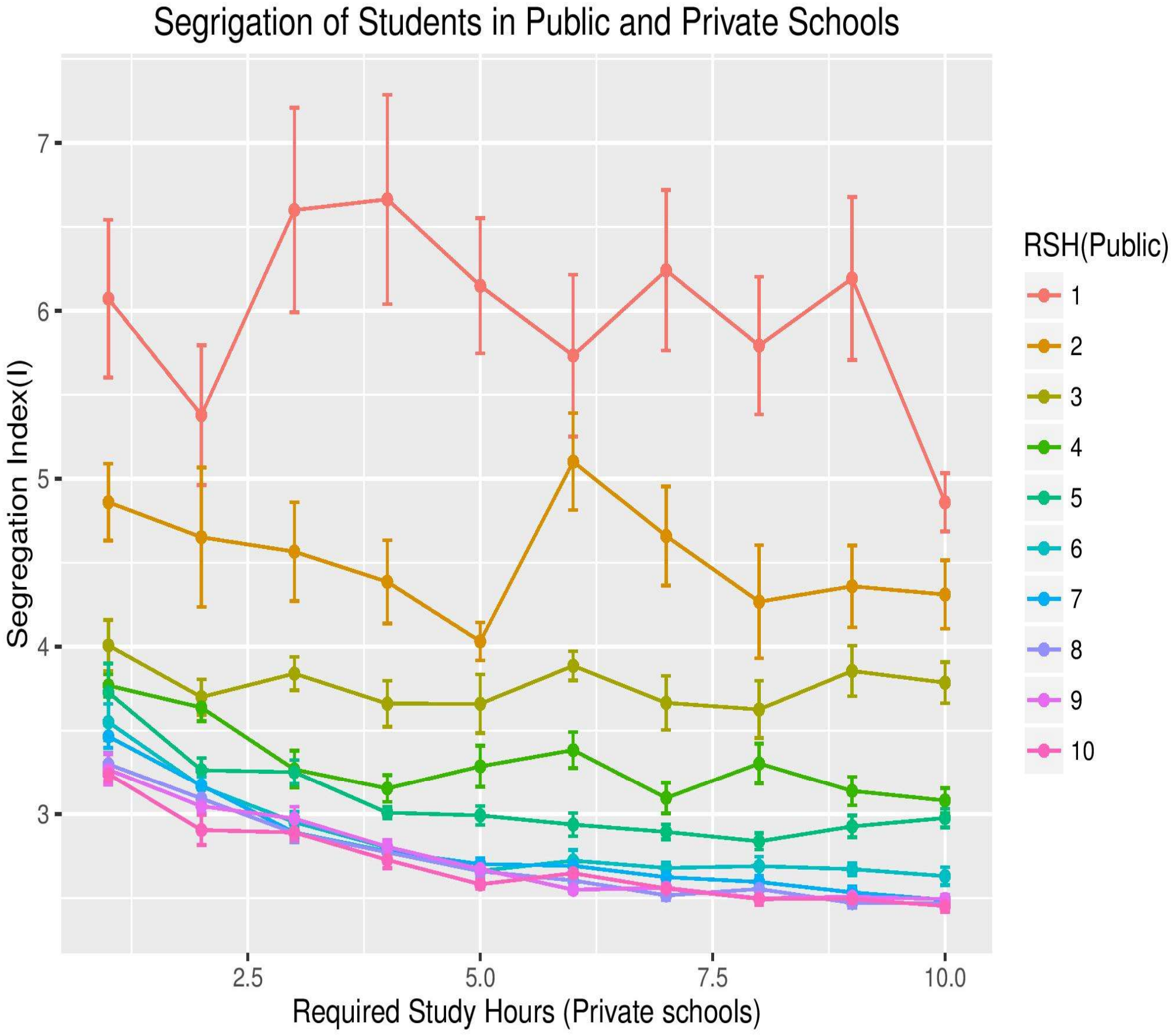

**Figure 4.12: Students' Segregation in Public and Private Schools**

### 4.5.3 Segregation of Student in Public and Private School Based on Wealth

Results in this section would prove the following hypothesis

- By enhancing the educational quality in the public schools, the socioeconomic gap between students in public and private schools decreases.
- Educational quality enhancement requires a cast in public school, so, enhancing educational quality in public school may increase the underclass student (out of school).

The unequal distribution of students in public and private schools is often a result of socioeconomic status. Students from better socioeconomic backgrounds tend to enroll in private schools that offer higher fees and better educational quality, while those from lower socioeconomic backgrounds would remain in public schools due to lower fees and comprehensible quality of education. This differentiation results in the segregation of students in the two types of schools. However, if public schools were to improve their educational quality, students with better socioeconomic status would also consider enrolling in public schools, thus reducing the segregation of students. The evidence supports the hypothesis that increasing the quality of education in public schools can lead to a reduction in the segregation of students based on their socioeconomic status.

The findings of Figure 8.13 indicate a strong correlation between the improvement of educational quality in public schools and the reduction of segregation among students of different socioeconomic status (SES). As the educational quality in public schools increases, due to factors such as the number of required study hours at home, students with better SES are more likely to enroll in public schools. This results in a decrease in segregation and a corresponding increase in the average wealth of students in public schools. Conversely, as the educational quality of public schools improves, the average wealth of students in private schools decreases. This suggests that there is a dynamic relationship between the educational quality of public and private schools and the distribution of students with different SES within these schools.

It is a positive development that the number of students in public schools with higher average wealth or better socioeconomic status (SES) is increasing. However, this may also lead to unintended consequences such as an increase in the number of underclass students. This is because the improvement in the quality of education in public schools also increases the cost of education, making it less accessible for lower-income students. Additionally, the enrollment of students with better SES in public schools increases the socioeconomic gap, further widening the divide between rich and poor students.

As shown in Figure 4.13, this trend is clearly evident. The increased number of underclass students in society is a cause for concern and must be addressed. It is important to ensure that educational opportunities are available and accessible to all students, regardless of their socioeconomic status. This requires a concerted effort to improve the quality of education in public schools while also making it more affordable for lower-income students. Only then can we truly create an equitable educational system that benefits all students.

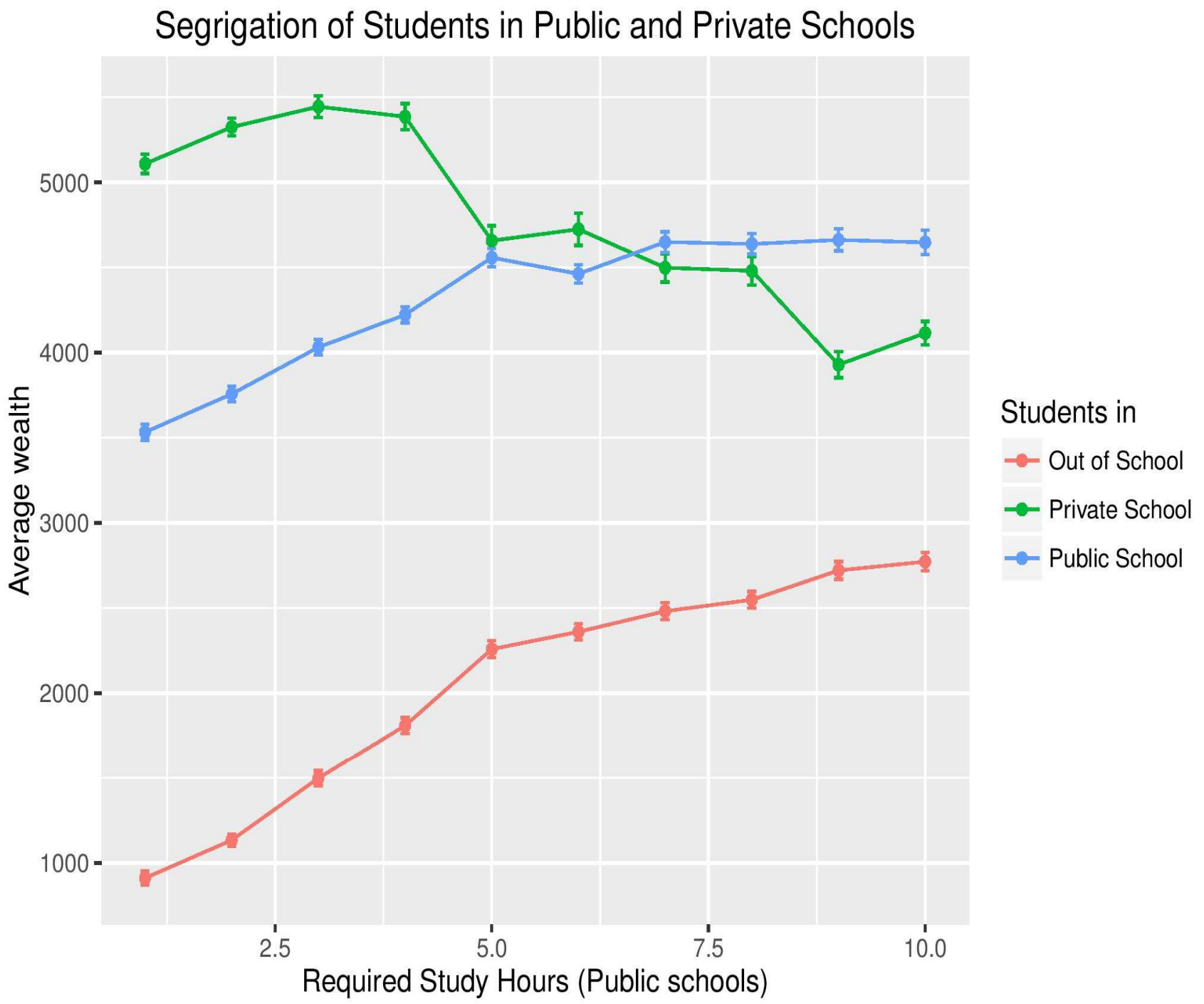

**Figure 4.13: Average wealth of students in the model**

### 4.5.4 Lorenz Curve Depiction of Inequality Due to SES in Public and Private School

Result in this section would prove the following hypothesis

- Due to unequal educational quality between public and private school, students are not evenly distributed in school with respect to socioeconomic status.

In the educational system under observation, educational quality in school differs from public to private sector. This difference in educational quality is caused by the educational the cost required to study in either educational sector. This difference in educational results in discrimination and even distribution of students with respect to the wealth they own and score attained as a result of the study in schools.

Lorenz Curve in figure 4.14 shows noticeable inequality in scores of students and socioeconomic status between students. Here skewed Lorenz curve depicts the inequalities in socioeconomic results in-discriminated in educational output. The skewed shape Lorenz curves evident that students' grades are not equally distributed.

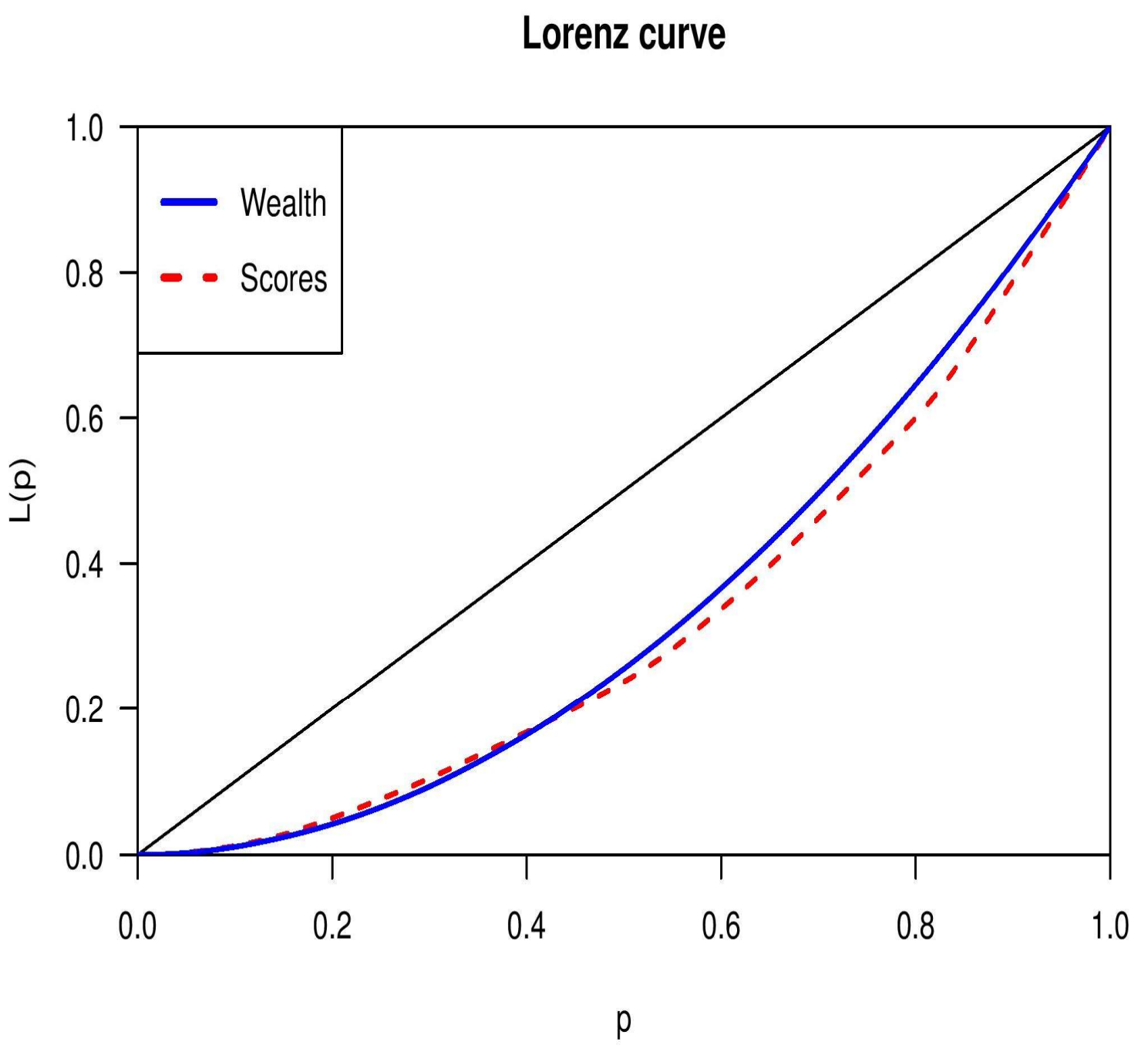

**Figure 4.14: Lorenz Curve for scores and Wealth of students**

## 4.6 Conclusion

Descriptive Agent based Modeling of agent-based models increase the comprehension, comparison, and replication of models. It contains the formal specification of the model, and complex network of the structure of the model. Both help a lot in understanding the underlying structure without knowing much the code syntax for cross domain researcher. Easy comprehension, replication, and completeness are the major differences that can be done while using descriptive agent-based modeling – DREAM, instead of exploratory agent-based modeling – ODD.

To mitigate the migration of students from public schools to private schools, a number of measures can be taken. One such measure is to reduce the class size in public schools by dynamically and instantly hiring teaching staff as per the need. This will help in improving the educational quality in public schools and reducing the gap between public and private schools in terms of education quality. The migration of students from public to private schools is often a result of segregation of students based on their socioeconomic status. Students who come from families with good socioeconomic status tend to prefer private schools, leaving behind students from families with weaker socioeconomic status to attend public schools. By reducing class size in public schools and improving the educational quality, it can help reduce the segregation of students and the migration from public to private schools.

Induction of teaching staff as per need is the basic parameter for quality education drive. In public schools, specifically in the primary sector, dynamic teacher induction is highly recommended. This would, certainly, increase the quality of education in public schools. Increase in educational quality, ultimately, boosts the confidence in students, and parents would enroll and retain pupils there. Hence, it would reduce the migration of students from public schools to private schools. This, lesser migration from public to private school, makes a healthy distribution of students in public school with respect to socioeconomic index.

# Chapter 5

# SUMMARY

## 5.1 Summary

The main objective of the thesis is to address two key issues related to software engineering of simulation and modeling. Firstly, the thesis aims to investigate the impact of formal specification on the comprehension and replication of simulation models. The hypothesis being tested is that a formal specification of the model, as opposed to a textual representation, will make the model easier to comprehend and replicate. The second aim of the thesis is to verify the hypothesis through a case study of complex adaptive behavior of student migration. To this end, the study employs simulation techniques to examine the migration patterns of students from one school to another. The factors contributing to this migration, such as differences in teacher induction policy, student-teacher ratio, and educational quality, are analyzed. The results of the simulation show that student migration leads to segregation based on socioeconomic status. By addressing these two issues, the thesis aims to contribute to the field of simulation and modeling by improving the understanding of the impact of formal specification on the comprehension and replication of simulation models.

The Descriptive agent-based modeling (DREAM) is a new and innovative method for describing complex adaptive behaviors in models and simulators. This approach utilizes semi-formal pseudocode specification and complex network analysis to provide a complete and clear demonstration of the model. The use of DREAM is beneficial as it allows for cross-domain comparison of models without the need to delve into the code specifically, making it a more efficient way to understand and replicate complex models. In contrast, the exploratory modeling technique (ODD) is a textual modeling method that tends to make the presentation more verbose and less specific. This approach can make the model more difficult to comprehend, replicate, and compare to models from other domains. The increased verbosity of the presentation makes it more challenging for the reader to understand the structure of the model, which can hinder its replication and comparison with other models.

## 5.2 Thesis Contributions

Our principal contribution to the software engineering of simulation and modeling is to provide a case study that proves formal methods as an easy tool for complete, clear, and concise demonstration of Agent-based models of complex adaptive system. Specifically, we prove via semi-formal specification and complex network analysis of a complex adaptive system that formal description is an effective, and efficient way of modeling and simulation. It makes a complete and clear presentation.

We investigated causes of migration and its outcomes using computer simulation. Results of simulation reveal that students' migration is because of teacher induction policy. Infect teacher induction policy affects the students-teacher ratio in schools. Schools with healthy students-teachers' ratio (smaller classes) have better quality education; however, it needs enough resources to be maintained. Students with better socioeconomic status prefer such schools' as they can afford. This causes the segregation of students based on socioeconomic status. Dynamic teacher induction policy, in public schools, results in an increase in quality of education in public school, hence, migration of student from public to private school reduces which causes reduction in segregation of students based on socioeconomic status.

## 5.3 Future Work

Formal methods are proven techniques for production of quality software. It is helpful in creating complete and concise requirements, and documentation for the software artifacts. We would use it further for modeling and simulation of complex adaptive system of schools with complete set of variables that was not possible in the current limited scenario of the thesis. We intend to provide a formal framework in this regard that provide efficiency, and effectiveness in modeling and simulation.

# Chapter 6 References